%% file: LRmm.tex
\documentclass[11pt]{article}
\pdfoutput=1

\usepackage[lmargin=50pt,rmargin=60pt,tmargin=60pt,bmargin=65pt]{geometry}
\usepackage{authblk} 
\usepackage{epsfig} 
\usepackage{amstext,amsmath,amssymb,amsfonts,amsmath,amsthm}
\numberwithin{equation}{section}
\usepackage{xcolor} 
\usepackage{tikz}
\usepackage{makecell}
\usepackage{braket} 
\usepackage{dsfont} 
\usepackage{graphicx} 
\usepackage[export]{adjustbox} 
\usepackage{float}  
\usepackage{subfig}
\usepackage{caption} 
\usepackage{color}
\usepackage[colorlinks, linkcolor=blue, citecolor=blue, urlcolor=blue]{hyperref}

\theoremstyle{definition}

\theoremstyle{remark}

\numberwithin{equation}{section}

\definecolor{cardinal}{rgb}{0.6,0,0}
\definecolor{darkgreen}{rgb}{0,0.5,0}
\definecolor{golden}{rgb}{0.92, 0.7, 0}
\definecolor{midnight}{rgb}{0, 0, 0.5}
\definecolor{darkblue}{rgb}{0.2, 0, 0.8}

\newcommand{\sqbracket}[1]{\left[#1\right]}
\newcommand{\abs}[1]{\left| #1 \right|}

\newcommand{\be}{\begin{equation}}
\newcommand{\ee}{\end{equation}} 
\newcommand{\ba}{\begin{equation}\begin{aligned}}
\newcommand{\ea}{\end{aligned}\end{equation}} 
\newcommand{\nn}{\nonumber} 
\newcommand{\f}{\frac}
\newcommand{\p}{\partial}
\newcommand{\la}{\langle}
\newcommand{\ra}{\rangle}
\newcommand{\dd}{{\rm d}^d}
\newcommand{\rmd}{{\rm d}}
\newcommand{\id}{\mathds{1}} 
\newcommand{\im}{\mathrm{i}}

\renewcommand{\a}{\alpha} 
\renewcommand{\b}{\beta}  
\newcommand{\g}{\gamma} 
\renewcommand{\d}{\delta}
\newcommand{\eps}{\epsilon}

\renewcommand{\l}{\lambda}
\newcommand{\m}{\mu}

\renewcommand{\r}{\rho}

\newcommand{\vph}{\varphi}
 
\newcommand{\D}{\Delta}

\newcommand{\zb}{\bar{z}}

\newcommand{\cA}{\mathcal{A}}

\newcommand{\cD}{\mathcal{D}}

\newcommand{\cM}{\mathcal{M}}

\newcommand{\cO}{\mathcal{O}}
\newcommand{\cP}{\mathcal{P}}
\newcommand{\cR}{\mathcal{R}}

\newcommand{\cT}{\mathcal{T}}

\newcommand{\N}{\mathbb{N}}

\newcommand{\phirs}[2]{\ensuremath{\phi_{#1,#2}}}

\begin{document}

\title{\bf Long-range minimal models}

\author[1,2]{Connor Behan}
\author[3]{Dario Benedetti}
\author[3]{Fanny Eustachon}
\author[4]{Edoardo Lauria}

\affil[1]{\normalsize \it 
 Perimeter Institute for Theoretical Physics, 31 Caroline St N, Waterloo, Ontario N2L 2Y5, Canada
  \authorcr \hfill}

\affil[2]{\normalsize \it 
 ICTP South American Institute for Fundamental Research, Instituto de F\'{i}sica Te\'{o}rica UNESP, Rua Dr. Bento Teobaldo Ferraz 271, S\~{a}o Paulo, SP 01140-070, Brazil
  \authorcr \hfill}

\affil[3]{\normalsize \it 
 CPHT, CNRS, \'Ecole polytechnique, Institut Polytechnique de Paris, 91120 Palaiseau, France
  \authorcr \hfill}

\affil[4]{\normalsize \it 
Laboratoire de Physique de l'\'Ecole Normale Sup\'erieure, Mines Paris, Inria, CNRS, ENS-PSL, Sorbonne Universit\'e, PSL Research University, Paris, France
  \authorcr \hfill}

\date{}
\maketitle

\hrule\bigskip

\begin{abstract}
We study a class of nonlocal conformal field theories in two dimensions which are obtained as deformations of the Virasoro minimal models. The construction proceeds by coupling a relevant primary operator $\phirs{r}{s}$ of the $m$-th minimal model to a generalized free field, in such a way that the interaction term has scaling dimension $2-\d$. Flowing to the infrared, we reach a new class of CFTs that we call \emph{long-range minimal models}. In the case $r=s=2$, the resulting line of fixed points, parametrized by $\d$, can be studied using two perturbative expansions with different regimes of validity, one near the mean-field theory end, and one close to the long-range to short-range crossover. This is due to a straightforward generalization of an infrared duality which was proposed for the long-range Ising model ($m = 3$) in 2017. We find that the large-$m$ limit is problematic in both perturbative regimes, hence nonperturbative methods will be required in the intermediate range for all values of $m$. For the models based on $\phirs{1}{2}$, the situation is rather different. In this case, only one perturbative expansion is known but it is well behaved at large $m$. We confirm this with a computation of infinitely many anomalous dimensions at two loops. Their large-$m$ limits are obtained from both numerical extrapolations and a method we develop which carries out conformal perturbation theory using Mellin amplitudes. For minimal models, these can be accessed from the Coulomb gas representations of the correlators. This method reveals analytic expressions for some integrals in conformal perturbation theory which were previously only known numerically.\\
\end{abstract}

\hrule\bigskip

\tableofcontents

\input{sections/introduction}

\input{sections/near_MFT}
\input{sections/phi22flow}
\input{sections/phi12flow}

\input{sections/large-m}
\input{sections/conclusions}

\paragraph{Acknowledgments.}

We thank A. Antunes, N. Bobev, F. Her\v{c}ek, L. Di Pietro, D. Maz\'{a}\v{c}, M. Paulos, T. Pochart, S. Rychkov, R. Santachiara, J. Silva, B. van Rees and P. van Vliet for discussions. EL is supported by the European Union (ERC, QFT.zip project, Grant Agreement no. 101040260). Views and opinions expressed are however those of the authors only and do not necessarily reflect those of the European Union or the European Research Council Executive Agency. Neither the European Union nor the granting authority can be held responsible for them. Research at Perimeter Institute is supported in part by the Government of Canada through the Department of Innovation, Science and Economic Development and by the Province of Ontario through the Ministry of Colleges and Universities.

\appendix

\input{sections/app_CPT}

\input{sections/app_MFT}
\input{sections/mmcorrelators}
\input{sections/coulomb}

\bibliographystyle{JHEP}
\bibliography{refs}
\addcontentsline{toc}{section}{References}


\end{document}

%% file: sections/introduction.tex
\section{Introduction and summary}
\label{sec:introduction}

Quantum field theory (QFT) in two dimensions (2d) abounds with exactly solvable models. 
In particular, in 2d, the algebra of local conformal transformations is infinite-dimensional, and its central extension, i.e.\ the Virasoro algebra, provides a powerful tool to solve many models describing the critical regime of two-dimensional statistical systems \cite{Belavin:1984vu}. Indeed, continuous phase transitions are associated to scale-invariant Euclidean QFTs, and for unitary local 2d QFTs, scale invariance is promoted to local conformal invariance \cite{Polchinski:1987}.
We refer the reader to the books \cite{Ginsparg:1988ui,Mussardo:2010mgq,DiFrancesco:1997nk}, for a detailed discussion on 2d CFTs and more references to the original work.

We have been careful to emphasize the locality assumption, because in this paper we will explore a class of models where some degree of nonlocality is introduced.
From the Lagrangian perspective, the notion of locality versus nonlocality is rather intuitive: Lagrangian densities that  depend only on some fields and a finite number of their derivatives, all at the same point, are the basic ingredient of a perturbative definition of local QFTs; Lagrangians that depend on fields at different points are instead the epitome of a nonlocal theory. 
However, solvable 2d CFTs do not typically come in Lagrangian form, they are rather obtained more abstractly from the representation theory of the Virasoro algebra.
In this framework, locality is intrinsically assumed in the fact of having a representation of the Virasoro algebra, and it manifests itself in particular via the existence of a local energy-momentum tensor.
Therefore, a natural definition of a nonlocal 2d CFT is as a 2d QFT that has no local energy-momentum tensor and that is invariant under just the global conformal group $SO(3,1)$.
The easiest example in such a class is provided by a bosonic  generalized free field (GFF), i.e.\ a Gaussian scalar field with conformal two-point function, and scaling dimension greater than zero.\footnote{In general dimension $d$, a GFF has scaling dimension different from $\D_{\rm free}=d/2-1$. The restriction to scaling dimensions greater than $\D_{\rm free}$ is equivalent to demanding reflection positivity (i.e.\ unitarity) of the GFF.}
Unlike the local case, we do not know any solvable nontrivial model in this class, therefore perturbation theory remains the main tool to approach nonlocal 2d CFTs.

The main motivation and source of examples for nonlocal CFTs comes from the study of \emph{long-range models}.
These are statistical models in which for example spins interact with each other not only via nearest-neighbour couplings, but all-to-all with a coupling having a power-law decay with respect to the distance.
The prototypical example is the long-range Ising model, that has a long history dating back to the 1960's \cite{d69a,d69b,t69}. In its Ginzburg-Landau formulation, the $d$-dimensional long-range Ising model is described by a GFF of scaling dimension $\D_{\varphi}=(d-s)/2$, perturbed by a quartic local interaction. For $d/2<s<s^\star$, with some fixed $d$-dependent $s^\star<2$, the interaction triggers a renormalization group (RG) flow to a nontrivial IR fixed point \cite{Fisher:1972zz,Sak:1973}, corresponding to a nonlocal CFT \cite{Paulos:2015jfa}. For $s>s^\star$, the model is instead in the short-range Ising universality class \cite{Sak:1977}, and this has been used in \cite{Behan:2017dwr,Behan:2017emf}\footnote{The case $d=1$ is special, as there is no nontrivial short-range Ising CFT. An alternative construction has been found recently in \cite{Benedetti:2024wgx,Benedetti:2025nzp}.} to construct a perturbative description of the long-range Ising CFT at $s\lesssim s^\star$.
The latter construction is built on the knowledge of a CFT, without the need of any Lagrangian, hence it can be generalized to a wide class of other cases, and this is the main idea that we will exploit here.

In this work, we will consider a class of nonlocal 2d CFTs that can be constructed as a perturbation of the simplest case of local 2d CFTs, namely the unitary and diagonal Virasoro minimal models. The 2d Ising CFT belongs to such class of models, and thus we will recover the 2d long-range Ising CFT of \cite{Behan:2017dwr,Behan:2017emf} as a special case. The perturbation will itself be local, but by coupling the minimal model to a GFF it will break local conformal invariance, leaving us with only a global $SO(3,1)$ invariance.

As the construction is very general, let us state it starting from a generic local CFT in general dimension, with expectation values denoted $\la\ldots\ra_0$.
Assume that we know the spectrum of such CFT, together with the operator product expansion (OPE) coefficients, i.e.\ we know its set of conformal data.
Given a relevant operator $\Phi_i(x)$, with scaling dimension $\D_i<d$, we construct a nonlocal model by introducing a GFF $\chi(x)$ with scaling dimension $\D_\chi=d-\D_i-\d$, with $0<\d\ll 1$, and coupling it to the CFT by deforming the expectation values as
\be \label{eq:deformCFT}
\la\ldots\ra_{g_0} \equiv   \int d\m_C[\chi] \, \la\ldots e^{-g_0\int \Phi_i \chi}\ra_0 \equiv  \la\ldots e^{-g_0\int \Phi_i \chi}\ra\,,
\ee
where the dots now stand for any operator insertion, and  $d\m_C[\chi]$ is a centered normalized Gaussian measure with covariance
\be
C(x,y) \equiv \int d\m_C[\chi] \, \chi(x) \chi(y)  
 = \f{1}{|x-y|^{2\D_{\chi}}}\,.
\ee
The perturbed model is always nonlocal, except in the special cases $\d=d/2+1-\D_i$ (the GFF $\chi$ is in this case a local free field theory) or $\d=d/2-\D_i$ (the GFF is ultralocal, i.e.\ it is a non-dynamical Hubbard-Stratonovich field for a ``double-trace" operator), which we will exclude from now on.

Notice that given the form of the perturbation in \eqref{eq:deformCFT}, for $\chi$-independent observables we could easily integrate out $\chi$, and obtain
\be  \label{eq:deformCFT_no-chi}
\la\ldots\ra_{g_0} =  \la\ldots e^{\f12 g_0^2\int \Phi_i C^{-1} \Phi_i}\ra\,,
\ee
making the nonlocal nature of the perturbation completely manifest. This form of the model is not very useful in practice, for at least two reasons. First, $\chi$-dependent observables are part of the spectrum of the theory.
Second, in the form \eqref{eq:deformCFT}, we have a (nonlocal) CFT at $g_0=0$, perturbed by a local operator, and we can thus apply (standard) conformal perturbation theory; on the other hand, in the integrated form \eqref{eq:deformCFT_no-chi}, the perturbation is nonlocal, so its perturbative treatment is not so straightforward.\footnote{
Nevertheless, the expression \eqref{eq:deformCFT_no-chi} can be helpful for the interpretation of the model, as it provides an action of the type more commonly encountered in long-range models, and therefore we will occasionally refer to it in the following.}

The choice of scaling dimension for $\chi$ is such that the operator $\cO=\Phi_i \chi$ has scaling dimension $\D_{\cO}=d-\d$, so that it is weakly relevant for $0<\d\ll 1$,
 and thus it induces a nontrivial flow towards the IR.\footnote{Notice that the GFF has a $\mathbb{Z}_2$ symmetry $\chi(x)\to-\chi(x)$, that is broken by the operator $\cO_i$. If the original CFT has also a $\mathbb{Z}_2$ symmetry, under which $\Phi_i$ is odd, then the perturbation preserves an invariance under the diagonal subgroup of $\mathbb{Z}_2\times\mathbb{Z}_2$.}
If there are no other (near-)marginal operators in the coupled model, then it suffices to express $g_0$ in terms of a renormalized coupling $g$, and thus the flow is one-dimensional. The beta function can be computed using conformal perturbation theory \cite{Zamolodchikov:1987ti}, as reviewed in \cite{Komargodski:2016auf,Behan:2017emf} (and here in Appendix~\ref{app:CPT}). Its general form is
\be \label{eq:beta-general}
\b(g) = -\d g + \b_3 g^3 + O(g^5) \,,
\ee
where the absence of a $g^2$ term is a consequence of the linearity in $\chi$ of $\cO$.
If $\b_3>0$, the beta function \eqref{eq:beta-general} has a real fixed point at $g=g_\pm\equiv \pm \sqrt{\d/\b_3}$, and we will assume  that at $g=g_\pm$ the deformed CFT \eqref{eq:deformCFT} has global conformal invariance, i.e.\ it defines a nonlocal CFT.\footnote{
We stress that Polchinski's result on scale and conformal invariance \cite{Polchinski:1987} does not hold for non-local CFTs. 
In the case of the long-range Ising model a strong argument for the conformal invariance of its fixed point has been given in \cite{Paulos:2015jfa}.
}

\

As anticipated, in this paper we will stick to $d=2$, and for the unperturbed local CFT we will choose any of the unitary and diagonal Virasoro minimal models, denoted $\cM_{m+1,m}$ with integer label $m\geq 3$.
We will consider different choices of $\Phi_i$, corresponding to different relevant Virasoro primaries $\phirs{r}{s}$.\footnote{Notice that taking $\Phi_i$ to be a scalar Virasoro descendant in $\cM_{m+1,m}$ will always lead to an irrelevant deformation.}
We refer to the theory at the IR fixed point of the flow driven by $\cO = \phirs{r}{s}\chi$ as \emph{long-range minimal model} (LRMM) of type $(m,r,s)$. 

We can slightly restrict the set of LRMMs by demanding that they stay unitary, which in particular requires $\b_3>0$, otherwise the fixed point is at imaginary coupling, and we will likely find either violations of unitarity bounds or complex OPE coefficients, or both. 
However, we do not know under which conditions we are bound to find $\b_3>0$, hence this has to be checked on a case by case basis.

This paper will treat three LRMMs in detail, namely $(m,2,2)$, $(m,1,2)$ and $(m,2,1)$. The first two happen to have $\beta_3 > 0$ for all $m$. Conversely, $(m,2,1)$ has $\beta_3 < 0$ but it is structurally very similar to $(m,1,2)$ in other respects. While many of our techniques will also apply to other LRMMs, there are several motivations for starting with these three.
\begin{enumerate}
\item At a technical level, perturbative analysis of $(m,r,s)$ has a lower barrier to entry when $rs$ is small. This is because $rs$ is the number of Virasoro primary families exchanged in the self-OPE of $\phirs{r}{s}$.
\item More conceptually, the $(m,2,2)$ LRMMs are important because they are all believed to cross over to mean-field behaviour for sufficiently large $\delta$. This makes it easy to conjecture a natural Landau-Ginzburg description. 
Indeed, following Zamolodchikov \cite{Zamolodchikov:1986db}, unitary minimal models $\cM_{m+1,m}$ have a Landau-Ginzburg description as the multicritical points of a scalar field $\vph$, that is identified with the most relevant Virasoro primary, i.e.\ $\phirs{2}{2}$. Therefore, this is the class of models that directly generalizes the Ising case of \cite{Behan:2017dwr,Behan:2017emf}, and therefore we expect their near-mean-field description to be provided by the IR fixed point of a GFF $\vph$ with perturbation $\vph^{2(m-1)}$.
\item Since the $(m,2,2)$ and $(m,1,2)$ families intersect at $m = 3$, they both appeared in a recent study of the 2d long-range Ising model using the numerical bootstrap \cite{Behan:2023ile} (see also \cite{Behan:2021tcn} for a closely-related study). Surprisingly, the bounds showed evidence that they are saturated by the LRMM of type $(m,1,2)$ for all $m$.
\item Studying at least two LRMM families presents an opportunity to learn about the different types of allowed large-$m$ behaviour. Indeed, we will find that the $(m,2,2)$ LRMMs differ greatly from $(m,1,2)$ at large $m$, even though they coincide at $m = 3$. This will be shown by computing a large amount of perturbative data and developing complementary techniques for extracting the large-$m$ limit. Perturbation theory at large $m$ goes back to the original work \cite{Zamolodchikov:1987ti,Cardy:1989da} about RG flows connecting minimal models and has been revisited many times since then \cite{Recknagel:2000ri,Graham:2001pp,dns01,Fredenhagen:2009tn,1303.3015,Behan:2021tcn,2205.05091,2211.16503,Lauria:2023uca,2412.21107}.\footnote{Another way to realize the long-range Ising model as part of a larger family is to consider the long-range $O(N)$ model \cite{Fisher:1972zz,Suzuki:1973a,Suzuki:1973b}. For recent explorations of the CFT data and free energy at large $N$, see \cite{Brezin:2014rkt,Benedetti:2020rrq,Benedetti:2024mqx,Chai:2021arp,Tarnopolsky:2016vvd,Fraser-Taliente:2025udk}.}
\end{enumerate}

\paragraph{Plan of the paper.}

Before coming to the details of coupling Virasoro minimal models to a GFF, we first consider a Lagrangian definition of long-range multicritical models in Sec.~\ref{sec:duality}. This is the nonlocal $\vph^{2(m - 1)}$ Landau-Ginzburg theory discussed above which is conjectured to be the near-mean-field description of the LRMM of type $(m,2,2)$. Our calculations in this section can be seen as a generalization of the approach to the long-range Ising model based on $\vph^4$ \cite{Fisher:1972zz}.

In Sec.~\ref{sec:phi22flow-nearSR}, we consider the case of LRMMs of type $(m,2,2)$ (i.e.\ with $\Phi_i=\phirs{2}{2}$) and compute anomalous dimensions for two types of operators. For operators that are UV Virasoro primaries, their anomalous dimensions need to be computed using standard two-loop conformal perturbation theory. This naturally leads to numerical integrals. For operators that are UV higher-spin currents, their anomalous dimensions can be found more easily with input from representation theory \cite{rt15,1601.01310}. We explain both types of calculations along with a brute-force method for extrapolating the results to large $m$.

In Sec.~\ref{sec:phi12flow}, we turn to the simpler LRMMs, $(m,1,2)$ and $(m,2,1)$, obtained with $\Phi_i=\phirs{1}{2}$ and $\Phi_i=\phirs{2}{1}$ respectively. As it turns out, their anomalous dimensions at large $m$ are related by a simple map, thus allowing us to focus on $(m,1,2)$ without loss of generality. The treatment of UV Virasoro primaries and UV currents parallels the $(m,2,2)$ case except for a newly encountered mixing problem. This leads to the appearance of some anomalous dimensions at one loop.

In Sec.~\ref{sec:large-m}, we find analytic explanations for why the numerical results in $(m,2,2)$ and $(m,1,2)$ LRMMs appear to take quite a simple form at large $m$. Our first approach is based on taking the large-$m$ limit first and works well for $(m,1,2)$. Our second approach, which takes the large-$m$ limit last, is more time consuming but it works for $(m,2,2)$ as well.

In Sec.~\ref{sec:conclusions}, we summarize the main lessons contained in the perturbative data for LRMMs and give an outlook on future directions.

Various technical results are collected in the appendices.
App.~\ref{app:CPT} contains a review of the type of conformal perturbation theory discussed above which is needed in most sections of the main text.
App.~\ref{app:MFT} contains extra detail on perturbative calculations using the Lagrangian theory in Sec.~\ref{sec:duality}.
App.~\ref{app:mmcorrelators} reviews some necesssary aspects from the solution of the diagonal Virasoro minimal models, which was completed in \cite{Dotsenko:1984nm,Dotsenko:1984ad,Dotsenko:1985hi}. Finally, App.~\ref{app:coulombgas} contains a crashcourse on a \texttt{Mathematica} package from \cite{c05} which is instrumental for being able to reproduce Sec.~\ref{sec:large-m}.

%% file: sections/near_MFT.tex

\section{Near mean field theory end -- the \texorpdfstring{$\varphi^{2(m - 1)}$}{phi2m-2} flow}
\label{sec:duality}

As we recalled in the introduction, one motivation for nonlocal CFTs comes from the study of long-range statistical models, of which long-range Ising is the most studied example. 
It is natural to generalize the long-range Ising model to the multicritical case by replacing the quartic operator with a higher even power, that is,
defining a new model with action:
\be \label{eq:LG}
S_{\text{LR},m} = \f{ \mathcal{N}}{2} \int \rmd^2 x_1 \rmd^2 x_2 \f{(\vph(x_1)-\vph(x_2))^2}{|x_1-x_2|^{2+s}} + \f{\l_0}{(2m-2)!}  \int \rmd^2 x  \, \vph(x)^{2(m-1)} \, .
\ee
Here, the $\mathcal{N}$ is a normalization factor, that we will choose such that the propagator is unit-normalized in direct space (see \cite{Behan:2023ile} for more details). 
The operator $\vph^{2(m-1)}$ has dimension $\D_{m} = (m-1) (2-s)$, with $m= 3$ corresponding to Ising and $m>3$ to the multicritical cases.  

The interaction is marginal when $\D_{m} = 2$, that is, at $s=\bar{s}$, where
\be
\bar{s} \equiv 2 \f{m-2}{m-1} \,.
\ee
For $s< \bar{s}$ the interaction is irrelevant, therefore we are in the regime of mean field theory (MFT).
For $s> \bar{s}$ the interaction is relevant, and as we show in Section \ref{beta22MFT} the model flows in the IR to an interacting fixed point, which can be studied perturbatively for $s\gtrsim \bar{s}$.
Such a fixed point defines (for each given $m$) a family of long-range multicritical CFTs that depend continuously on the parameter $s$.

Notably, despite the interacting character of such CFTs, the conformal dimension of the field $\vph$ sticks to its GFF value.
This is a standard feature of long-range models \cite{Fisher:1972zz,Lohmann:2017}, which can be understood as a consequence of the local nature of UV divergences, that therefore cannot lead to a renormalization of the nonlocal quadratic term in \eqref{eq:LG}.

\subsection{Crossover to short range}
As the distance $s-\bar{s}$ from the MFT regime increases, the CFT becomes more and more strongly coupled and eventually the perturbative treatment based on \eqref{eq:LG} becomes unreliable.
Eventually, we expect that, like in the Sak scenario for the Ising case \cite{Sak:1973,Sak:1977}, for some $s=s^\star>\bar{s}$ the anomalous dimension of the irrelevant operator $\p_\m\vph \p^\m \vph$ becomes so large to turn it into a relevant operator. That is, we expect that $\p_\m\vph \p^\m \vph$ will behave as a \emph{dangerously irrelevant} operator \cite{Amit:1982az}.
When this happens, the nonlocal CFT is no longer IR-stable, and thus the long-range multicritical model transitions to a different universality class, which is that of the corresponding short-range multicritical model. 
By continuity of the spectrum, one infers that such crossover must take place at a value of $s$ such $\D_\vph$ equals the conformal dimension of the  magnetization field in the corresponding short-range model.
Let us thus pause for a moment from the long-range domain, and discuss briefly what we know about the short-range case.

In two dimensions, short-range multicritical models have been related by Zamolodchikov  \cite{Zamolodchikov:1986db} (see also \cite{DiFrancesco:1997nk} or appendix A of \cite{Lencses:2022ira} for reviews) to the unitary a diagonal Virasoro minimal models, $\cM_{m+1,m}$. 
According to such correspondence, the Landau-Ginzburg description of the unitary minimal models $\cM_{m+1,m}$ is obtained by a $\mathbb{Z}_2$-even perturbation $\vph^{2(m-1)}$ of the free boson:
\be \label{eq:LG-SR}
S_{m} = \f12 \int \rmd^2 x \, \p_\m\vph(x) \p^\m \vph(x)  + \f{\l_0}{(2m-2)!}  \int \rmd^2 x  \, \vph(x)^{2(m-1)} \, .
\ee
The key argument for such correspondence is the identification of the relevant primaries of $\cM_{m+1,m}$ with scaling operators of the theory \eqref{eq:LG-SR} at its IR fixed point. The most relevant singlet operator in $\cM_{m+1,m}$, besides the identity, is $\phirs{2}{2}$, which is thus identified with the order parameter $\vph$ of the corresponding Landau-Ginzburg description. The relevant composite operators corresponding to powers of $\vph$ are then obtained by repeated use of the OPE, resulting in the identifications
\be \label{eq:phi-k}
:\vph^k : = \begin{cases} \phi_{(k+1,k+1)}  & k=1,\ldots,m-2 \,,\\
\phi_{(k+3-m,k+2-m)}  & k=m-1,\ldots,2m-4 \, .\end{cases}
\ee
Notice that $\vph(x)^{2m-3}$ is missing, because it is a descendant of $\vph$, as a consequence of the Schwinger-Dyson equations: $\vph(x)^{2m-3}\sim \p^2 \vph$.

Assuming the validity of Zamolodchikov's correspondence between $\cM_{m+1,m}$ and \eqref{eq:LG-SR}, and by the Sak scenario we discussed above, we expect that in the long-range model \eqref{eq:LG}, when  $\D_{\vph}\leq \D_{2,2}$, i.e. for $s\geq  s^\star$ with
\be
s^\star \equiv  2-2\D_{2,2} \,,
\ee
the model will fall into the short range universality class. However, continuity of the spectrum demands that the theory at $s\geq  s^\star$ will not be just $\cM_{m+1,m}$, but it will rather include also a decoupled sector.
Following the same line of reasoning as for the long-range Ising case \cite{Behan:2017dwr,Behan:2017emf}, and taking into consideration the identification between $\vph$ and $\phirs{2}{2}$, we are led to conclude that the CFT  associated to the IR fixed point of \eqref{eq:LG} will be described near $s^\star$ by a LRMM of type $(m,2,2)$, i.e.\ the IR fixed point of $\cM_{m+1,m}$ perturbed by $\cO= \phirs{2}{2}\chi$, and with the identification
\be
\d = s^\star - s \,.
\ee
In support of this IR duality, we have the following pieces of evidence:
\begin{itemize}
\item Integrating out $\chi$ as in equation \eqref{eq:deformCFT_no-chi}, with $\Phi_i=\phirs{2}{2}$, and using $\phirs{2}{2}\sim\vph$ one obtains precisely the quadratic term in  \eqref{eq:LG}. Therefore, the $\phirs{2}{2}\chi$ perturbation can be seen as being the analog of Sak's nonlocal perturbation \cite{Sak:1973,Sak:1977,Behan:2017dwr,Behan:2017emf}, rewritten via a Hubbard-Stratonovich trick;
\item One major difference between the long-range and the short-range case is that in the former $\vph(x)^{2m-3}$ is not a descendant of $\vph$. In the long-range case, the Schwinger-Dyson equations are nonlocal, hence they imply instead an exact shadow relation between the two operators, namely
\be
\D_{\vph^3} = 2 - \D_\vph \,.
\ee
By construction, this is also the dimension of $\chi$, hence we expect $\chi$ to be identifiable with the missing $\vph(x)^{2m-3}$ in \eqref{eq:phi-k}.
This field is present also at $s>s^\star$, but it decouples, thus reconciling continuity of the spectrum with the absence of such an operator in the short-range models;

\item At $\d>0$, the Schiwinger-Dyson equations for the deformed theory \eqref{eq:deformCFT} imply that at the IR fixed point, $\phirs{2}{2}$ is the shadow of $\chi$, and since the dimension of the latter is protected (for the same reason why $\vph$ in \eqref{eq:LG} needs no wave function renormalization), this means that $\phirs{2}{2}$ must acquire an anomalous dimension such that we can identify it with $\vph$.

\end{itemize}

In the absence of a full proof, it would be nice to test this duality between the long-range multiciritical models \eqref{eq:LG} and the LRMM of type $(m,2,2)$ by matching their predictions near $\bar{s}$ and near $s^\star$, respectively, with some nonperturbative (e.g.\ lattice) computation at intermediate values of $s$.
In view of such possible comparison, in the rest of this section we provide some perturbative results in the near-MFT regime, while in the next section we will study the near-crossover regime.

\subsection{Beta function}
\label{beta22MFT}

In order to make the interaction term of \eqref{eq:LG} weakly relevant we choose
\be\label{eq:sbarexpansion}
s = \overline{s} + \frac{\eps}{m-1} \,,\qquad \eps\ll 1\,,
\ee
such that the bare coupling has dimension $[\lambda_0]=2-(m-1)(2-s)=\eps$. We will treat $\eps$ as a UV regulator.
IR divergences are instead regulated by relegating the interaction to a  finite volume $V =\pi R^2$.

Exploiting the analytic regularization, and using an $\overline{\text{MS}}$ scheme, the bare coupling gets renormalized as $\lambda_0 = R^{-\eps} (\lambda+\delta_\lambda)$, with $\delta_\lambda$ the sum of appropriate counter-terms.
From the Callan–Symanzik equation, $R\, d\lambda_0/dR=0$, the beta function is found to be (see Appendix~\ref{app:MFT} for details):
\be\label{eq:beta_MFT}
\beta(\lambda) =R\, d\lambda/dR= -\eps \lambda + A \lambda^2 - 2 B_1 \lambda^3  + O\left(\lambda^4\right),
\ee
with
\be\label{eq:1loop_MFT}
A = \pi \frac{\sqbracket{2(m-1)}!}{(m-1)!^3} \,,
\ee
and
\be\label{eq:hard_sums}
\begin{split}
B_1 =& \frac{\sqbracket{2(m-1)}!}{3!} \pi^2 \sum_{\substack{a+b+c=2(m-1)\\ a,\,b,\,c\,\neq 0\\a,\,b,\,c\,\neq m-1}}\frac{1}{(a!b!c!)^2}\frac{\Gamma\left(1-\frac{a}{m-1}\right)\Gamma\left(1-\frac{b}{m-1}\right)\Gamma\left(1-\frac{c}{m-1}\right)}{\Gamma\left(\frac{a}{m-1}\right)\Gamma\left(\frac{b}{m-1}\right)\Gamma\left(\frac{c}{m-1}\right)}\\
&-\frac{\sqbracket{2(m-1)}!}{2(m-1)!^2}\pi^2\sum_{\substack{a+b=m-1\\ a,\,b\,\neq 0}}\frac{1}{(a!b!)^2}\sqbracket{\psi_0\left(\frac{a}{m-1}\right)+\psi_0\left(\frac{b}{m-1}\right)}\\
&- \left(\frac{\sqbracket{2(m-1)}!^2}{(m-1)!^6 }- 2\frac{\sqbracket{2(m-1)}!}{(m-1)!^4}\right)\gamma_E\pi^2 \,,
\end{split}
\ee
with $\gamma_E$ Euler's constant.

The beta function \eqref{eq:beta_MFT} admits two perturbative zeros: the trivial fixed point $\lambda_*=0$ and the nontrivial fixed point
\be\label{eq:FP}
\lambda_* = \f{\eps}{A} + \f{2 B_1 \eps^2}{A^3} + O\left(\eps^3\right) \,.
\ee
From the derivative of the beta function at this fixed point, we find the conformal dimension $\varphi^{2(m - 1)}$:
\be\label{eq:perturbingDim}
\Delta_{\varphi^{2(m - 1)}} = 2 + \beta'(\lambda_*) = 2 + \eps - 2 \frac{B_1}{A^2} \eps^2 + O(\eps^3)\,.
\ee

\subsection{Anomalous dimensions of $ \varphi^\alpha$ operators}

Consider a bare operator $\Phi$, of dimension $\Delta_\Phi^{\text{UV}}$, in the unperturbed theory. In the perturbed theory, we compute the anomalous dimension of $\Phi$ by renormalising its two-point correlation function, i.e. we define $\Phi = Z_\Phi\Phi_r$ and require
\be
\la\Phi(\infty)\Phi_r(0)\ra = Z_\Phi^{-1} \la\Phi(\infty)\Phi(0)\ra\,, %
\ee
to be finite from UV divergences (in the $\eps\to0$ limit), by choosing appropriately $Z_\Phi$. Note that the operator at infinity does not get renormalized, because it is inserted outside the (finite) region of integration.

The anomalous dimension of $\Phi$ is found from the Callan–Symanzik equation to be:
\be
\gamma(\lambda) = -\frac{1}{Z_\Phi}R\frac{d Z_\Phi}{dR}\,.
\ee
The conformal dimension of $\Phi$ at the IR fixed point is
\be
\Delta_\Phi = \Delta_\Phi^{\text{UV}}+\gamma(\lambda_*)\,.
\ee

For monomial operators $\Phi = \varphi^\alpha$, in the $\overline{\text{MS}}$ we find:
\be\label{eq:gamma}
\gamma(\lambda) = \tilde{A}\lambda - \tilde{B}_1\lambda^2 + O(\lambda^3)\,,
\ee
where
\be
\tilde{A} = \frac{2\pi \alpha!^2}{(m - 1)!^2 (\alpha - m + 1)!} \Theta(\alpha - m + 1)\,,
\ee
and $\Theta$ is the Heaviside function with $\Theta(0) = 0$.
The coefficient $\tilde{B}_1$ depends more sensitively on $\alpha$, in particular (we define $c \equiv 2m - 2 - a - b$):
\begin{itemize}
	\item If $\alpha < m$,
	\ba
	\tilde{B}_1 &= \pi^2 \sum_{a = 1}^{\alpha - 2} \sum_{b = 2m - 1 - \alpha}^{2m - 3 - a} \frac{1}{(b + \alpha - 2m + 2)!} \frac{\alpha!^2}{a!^2 b! c!^2} \frac{\Gamma\left(1-\frac{a}{m-1}\right)\Gamma\left(1-\frac{b}{m-1}\right)\Gamma\left(1-\frac{c}{m-1}\right)}{\Gamma\left(\frac{a}{m-1}\right)\Gamma\left(\frac{b}{m-1}\right)\Gamma\left(\frac{c}{m-1}\right)} \\
	&+ \pi^2 \sum_{a = 1}^{\alpha - 1} \frac{\alpha!^2}{a!^2 (2m - 2 - \alpha)! (\alpha - a)!^2} \frac{\Gamma\left(1-\frac{a}{m-1}\right)\Gamma\left(1-\frac{2m - 2 - \alpha}{m-1}\right)\Gamma\left(1-\frac{\alpha - a}{m-1}\right)}{\Gamma\left(\frac{a}{m-1}\right)\Gamma\left(\frac{2m - 2 - \alpha}{m-1}\right)\Gamma\left(\frac{\alpha - a}{m-1}\right)}\,.
	\ea
	\item If $\alpha = m$,
	\ba
	\tilde{B}_1 &= -\pi^2 \sum_{a = 1}^{m - 2} \frac{m!^2}{a!^2 (m - 1)! c!^2} \left ( H_{-\frac{a}{m-1}} + H_{-\frac{c}{m-1}} \right ) \\
	&+ \pi^2 \sum_{a = 1}^{m - 2} \sum_{b = m}^{2m - 3 - a} \frac{1}{(b - m + 2)!} \frac{m!^2}{a!^2 b! c!^2} \frac{\Gamma\left(1-\frac{a}{m-1}\right)\Gamma\left(1-\frac{b}{m-1}\right)\Gamma\left(1-\frac{c}{m-1}\right)}{\Gamma\left(\frac{a}{m-1}\right)\Gamma\left(\frac{b}{m-1}\right)\Gamma\left(\frac{c}{m-1}\right)} \\
	&+ \pi^2 \sum_{\substack{a = 1 \\ a \neq m - 1}}^{m - 1} \frac{m!^2}{a!^2 (m - 2)! (m - a)!^2} \frac{\Gamma\left(1-\frac{a}{m-1}\right)\Gamma\left(1-\frac{m - 2}{m-1}\right)\Gamma\left(1-\frac{m - a}{m-1}\right)}{\Gamma\left(\frac{a}{m-1}\right)\Gamma\left(\frac{m - 2}{m-1}\right)\Gamma\left(\frac{m - a}{m-1}\right)} \\
	&- \pi^2 \frac{m!^2}{(m - 1)!^2 (m - 2)!} \left ( H_{\frac{2 - m}{m - 1}} + H_{\frac{a - m}{m - 1}} \right )\,.
	\ea
	\item If $m < \alpha < 2m - 2$,
	\ba
	\tilde{B}_1 &= \pi^2 \sum_{a = 1}^{\alpha - 2} \sum_{\substack{b = 2m - 1 - \alpha \\ a \neq m - 1, b \neq m - 1}}^{2m - 3 - a} \frac{1}{(b + \alpha - 2m + 2)!} \frac{\alpha!^2}{a!^2 b! c!^2} \frac{\Gamma\left(1-\frac{a}{m-1}\right)\Gamma\left(1-\frac{b}{m-1}\right)\Gamma\left(1-\frac{c}{m-1}\right)}{\Gamma\left(\frac{a}{m-1}\right)\Gamma\left(\frac{b}{m-1}\right)\Gamma\left(\frac{c}{m-1}\right)} \\
	&- \pi^2 \sum_{a = 1}^{m - 2} \frac{1}{(\alpha - m + 1)!} \frac{\alpha!^2}{a!^2 (m - 1)! c!^2} \left ( H_{-\frac{a}{m-1}} + H_{-\frac{c}{m-1}} \right ) \\
	&- \pi^2 \sum_{b = 2m - 1 - \alpha}^{m - 2} \frac{1}{(b + \alpha - 2m + 2)!} \frac{\alpha!^2}{(m - 1)!^2 b! c!^2} \left ( H_{-\frac{b}{m-1}} + H_{-\frac{c}{m-1}} \right ) \\
	&+ \pi^2 \sum_{\substack{a = 1 \\ a \neq m - 1}}^{\alpha - 1} \frac{\alpha!^2}{a!^2 (2m - 2 - \alpha)! (\alpha - a)!^2} \frac{\Gamma\left(1-\frac{a}{m-1}\right)\Gamma\left(1-\frac{2m - 2 - \alpha}{m-1}\right)\Gamma\left(1-\frac{\alpha - a}{m-1}\right)}{\Gamma\left(\frac{a}{m-1}\right)\Gamma\left(\frac{2m - 2 - \alpha}{m-1}\right)\Gamma\left(\frac{\alpha - a}{m-1}\right)} \\
	&- \pi^2 \frac{\alpha!^2}{(m - 1)!^2 (2m - 2 - \alpha)! (\alpha - m + 1)!^2} \left ( H_{\frac{\alpha - 2m + 2}{m - 1}} + H_{\frac{a - \alpha}{m - 1}} \right )\,.
	\ea
	\item If $\alpha \geq 2m - 2$,
	\ba
	\tilde{B}_1 &= \pi^2 \sum_{a = 1}^{2m - 4} \sum_{\substack{b = 2m - 1 - \alpha \\ a \neq m - 1, b \neq m - 1}}^{2m - 3 - a} \frac{1}{(b + \alpha - 2m + 2)!} \frac{\alpha!^2}{a!^2 b! c!^2} \frac{\Gamma\left(1-\frac{a}{m-1}\right)\Gamma\left(1-\frac{b}{m-1}\right)\Gamma\left(1-\frac{c}{m-1}\right)}{\Gamma\left(\frac{a}{m-1}\right)\Gamma\left(\frac{b}{m-1}\right)\Gamma\left(\frac{c}{m-1}\right)} \\
	&- \pi^2 \sum_{a = 1}^{m - 2} \frac{1}{(\alpha - m + 1)!} \frac{\alpha!^2}{a!^2 (m - 1)! c!^2} \left ( H_{-\frac{a}{m-1}} + H_{-\frac{c}{m-1}} \right ) \\
	&- \pi^2 \sum_{b = 2m - 1 - \alpha}^{m - 2} \frac{1}{(b + \alpha - 2m + 2)!} \frac{\alpha!^2}{(m - 1)!^2 b! c!^2} \left ( H_{-\frac{b}{m-1}} + H_{-\frac{c}{m-1}} \right )\,.
	\ea
\end{itemize}

\subsection{Example results}
While we do not have a closed expression for the sums in the coefficient $B_1$ at general $m$, the latter can be easily evaluated at each finite $m$. For the for the long-range Ising model ($m = 3$) we find:
\be
\beta_3(\lambda) = -\eps \lambda +3\pi\lambda^2 -24\pi^2\log(2)\lambda^3 + O(\lambda^4) \,,
\ee
in agreement with the results of \cite{Benedetti:2020rrq}. For the long-range tricritical Ising model ($m = 4$) we find:
\be
\beta_4(\lambda) = -\eps \lambda +\frac{10\pi}{3}\lambda^2 - \sqbracket{30\pi^2\log(3) + \frac{45\sqrt{3}}{32\pi}\left(\Gamma\left(\frac{1}{3}\right)^6-3\Gamma\left(\frac{2}{3}\right)^6\right)}\lambda^3 + O(\lambda^4) \,,
\ee
which is a new result.

We can also evaluate the coefficients in the large-$m$ expansion, as we now explain. Considering the second line in \eqref{eq:hard_sums} first, $[a! (m - 1 - a)!]^{-2}$ is sharply peaked around $a = \frac{m - 1}{2}$. We can therefore approximate the sum by expanding $\psi_0 \left ( \frac{a}{m - 1} \right )$ around this value with the leading term producing
\ba
\sum_{\substack{a+b=m-1 \\ a,b \neq 0}} \frac{1}{(a! b!)^2} \psi_0 \left ( \frac{a}{m - 1} \right ) &\sim \psi_0 \left ( \frac{1}{2} \right ) \sum_{a = 1}^{m - 2} [a! (m - 1 - a)!]^{-2} \\
&= \psi_0 \left ( \frac{1}{2} \right ) \frac{\Gamma(2m - 1) - 2\Gamma(m)^2}{\Gamma(m)^4}\,.
\ea
For the first line in \eqref{eq:hard_sums}, the same approach tells us to evaluate the gamma functions at $a = b = c = \frac{2}{3}(m - 1)$ resulting in an overall factor of $[\Gamma \left ( \frac{1}{3} \right ) / \Gamma \left ( \frac{2}{3} \right )]^3$. In this case, we are still left with a nontrivial sum, which we approximate as a Gaussian integral. In particular, we can write
\ba
\frac{1}{(a! b! c!)^2} &= \exp \left [ f(a, b) \right ] \\
&\sim \exp \left [ f \left ( \frac{2m - 2}{3}, \frac{2m - 2}{3} \right ) + \frac{1}{2} \left ( a - \frac{2m - 2}{3} \right )^2 \frac{\partial^2 f}{\partial a^2} \bigl |_{a = b = \frac{2m - 2}{3}} \right. \\
&\left. + \left ( a - \frac{2m - 2}{3} \right ) \left ( b - \frac{2m - 2}{3} \right ) \frac{\partial^2 f}{\partial a \partial b} \bigl |_{a = b = \frac{2m - 2}{3}} + \frac{1}{2} \left ( b - \frac{2m - 2}{3} \right )^2 \frac{\partial^2 f}{\partial b^2} \bigl |_{a = b = \frac{2m - 2}{3}} \right ]\,.
\ea
The leading term is of course $f \left ( \frac{2m - 2}{3}, \frac{2m - 2}{3} \right ) = -6 \log \Gamma \left ( \frac{2m + 1}{3} \right )$ while the subleading term becomes tractable after using Stirling's approximation. As a result,
\ba
& \sum_{\substack{a+b+c=2(m-1)\\ a,\,b,\,c\,\neq 0\\a,\,b,\,c\,\neq m-1}}\frac{1}{(a!b!c!)^2}\frac{\Gamma\left(1-\frac{a}{m-1}\right)\Gamma\left(1-\frac{b}{m-1}\right)\Gamma\left(1-\frac{c}{m-1}\right)}{\Gamma\left(\frac{a}{m-1}\right)\Gamma\left(\frac{b}{m-1}\right)\Gamma\left(\frac{c}{m-1}\right)} \\
&\sim \left [ \Gamma \left ( \frac{1}{3} \right ) / \Gamma \left ( \frac{2}{3} \right ) \right ]^3 \Gamma \left ( \frac{2m + 1}{3} \right )^{-6} \int \text{d}^2 \, \textbf{x} \exp \left [ -\frac{3}{2m - 2} \textbf{x}^{\intercal} \begin{pmatrix} 2 & 1 \\ 1 & 2 \end{pmatrix} \textbf{x} \right ] \\
&= \left [ \Gamma \left ( \frac{1}{3} \right ) / \Gamma \left ( \frac{2}{3} \right ) \right ]^3 \Gamma \left ( \frac{2m + 1}{3} \right )^{-6} \frac{2\pi (m - 1)}{\sqrt{27}}\,.
\ea

These expressions make it clear that the double sum dominates over the other terms in \eqref{eq:hard_sums} at large $m$. It is therefore valid to take
\be
A = \pi \frac{(2m - 2)!}{(m - 1)!^3}, \quad B_1 = \pi^3 \frac{\Gamma \left ( \frac{1}{3} \right )^3}{\Gamma \left ( \frac{2}{3} \right )^3} \frac{(m - 1) (2m - 2)!}{3^{5/2} \left [ \frac{2}{3}(m - 1) \right ]!^6}\,,
\ee
in this limit. One can then plug this into \eqref{eq:perturbingDim} for instance and see that the coefficient of $\eps^2$ diverges exponentially as $m \to \infty$. This leads to a breakdown of perturbation theory when using small but finite values of $\eps$. In consequence, while the near mean-filed and near short-range ends become close in this limit as $\abs{\bar{s}-s^\star} = 2/m+O(1/m^2)$, the acceptable range for perturbation theory ($\eps \ll A^2/2B_1$) shrinks at a faster speed.\footnote{In particular, one cannot set $\eps = (m - 1)(2 - 2\Delta_{2,2} - \bar{s}) \approx 2$ to try to approximate the large-$m$ minimal model $\mathcal{M}_{m, m + 1}$ using the $\varphi^{2(m - 1)}$ deformation.}
%

%% file: sections/phi22flow.tex

\section{Near short-range minimal model end -- the \texorpdfstring{$\phirs{2}{2}\chi$}{phi22chi} flow}
\label{sec:phi22flow-nearSR}
When the scaling dimension of $\varphi$ takes the value
\begin{equation}
	\D_{2,2} = 2\,h_{2,2} = \frac{3}{2 m (m+1)}\,,
\end{equation}
we expect that \eqref{eq:LG} flows to a product theory between the short-range minimal model $\mathcal{M}_{m,m+1}$ and a GFF $\chi$ of dimension $2 - \D_{2,2}$. From that product theory, we can induce a weakly-relevant flow into the LRMM of type $(m,2,2)$ via:
\be\label{eq:flow22}
S'_{\text{LR},m} = S_{\text{SR},m} + \f{ \mathcal{N}}{2} \int \rmd^2 x_1 \rmd^2 x_2 \f{(\chi(x_1)-\chi(x_2))^2}{|x_1-x_2|^{2-s}} + g_0 \int \rmd^2 x  \, \phirs{2}{2} \chi\,,
\ee
where $\chi$ has dimension $2 - \D_{2,2} - \delta$ for $0 < \delta \ll 1$, and $\mathcal{N}$ is a normalization factor.  We recall that the dimension of $\chi$ is protected because UV divergences cannot renormalize its nonlocal kinetic term. Furthermore, the Schwinger-Dyson equations imply the following shadow relation at the IR fixed point:
\be\label{eq:shadow2}
\Delta_{\phi_{2,2}} = 2 - \Delta_\chi.
\ee

In this section, we compute the beta function for the coupling at leading order in conformal perturbation theory, to show that the flow \eqref{eq:flow22} leads to unitary and interacting IR fixed points for all integer $m\geq 3$. We also compute several anomalous dimensions of operators at the IR fixed points.
%

\subsection{Beta function}
Finding the beta function to non-trivial order in $g$ means computing the $\beta_3$ coefficient defined in \eqref{eq:beta-general}. The method we use to compute $\beta_3$, adapted from \cite{Komargodski:2016auf} and \cite{Behan:2017emf}, is reviewed in Appendix \ref{app:CPT}. We start from the (regularized) integral:
\be\label{eq:beta3R}
\beta_3 = -\pi \int_{\mathcal{R}} \text{d}^2z \la \mathcal{O}(0) \mathcal{O}(z, \bar{z}) \mathcal{O}(1) \mathcal{O}(\infty) \ra \biggl |_{\text{finite}}\,,
\ee
with $\mathcal{O}=\phirs{2}{2}\chi$, the integration domain is $\mathcal{R}=\{z\,:\, |z|<1, |z|<|z-1|\}$, and the integrand is evaluated in the unperturbed theory. The latter can be computed recursively (to arbitrary precision) using the BPZ differential equation \cite{Belavin:1984vu}, and we find (see Appendix \ref{app:mmcorrelators} for a derivation):
\begin{align}
\la \mathcal{O}(0) \mathcal{O}(z, \bar{z}) \mathcal{O}(1) \mathcal{O}(\infty) \ra &= \la \chi(0) \chi(z, \bar{z}) \chi(1) \chi(\infty) \ra \la \phirs{2}{2}(0) \phirs{2}{2}(z, \bar{z}) \phirs{2}{2}(1) \phirs{2}{2}(\infty) \ra \label{eq:phi22betaBlocks} \\
&= \frac{1 + |z|^{-2\D_\chi} + |1 - z|^{-2\D_\chi}}{|z|^{2\D_{2,2}}} \sum_{r \in \{ 1,3 \}} \sum_{s \in \{ 1,3 \}} C_{(2,2)(2,2)(r,s)}^2 \left | \sum_{n = 0}^\infty a_n^{(r,s)} \rho^{h_{r,s} + n} \right |^2 \nonumber\,,
\end{align}
where $C$'s are known OPE coefficients, the $a_n^{(r,s)}$ are coefficients of the Virasoro blocks for $\phirs{r}{s}$ exchange in the radial expansion as obtained from BPZ equation, and $(\rho, \bar{\rho})$ are the radial coordinates defined in \cite{Hogervorst:2013sma}
\be\label{eq:rhocoord}
\rho = \frac{z}{(1 + \sqrt{1 - z})^2}, \quad \bar{\rho} = \frac{\bar{z}}{(1 + \sqrt{1 - \bar{z}})^2}\,.
\ee
The regularization in \eqref{eq:beta3R} is achieved by cutting off a ball of radius $a$ around the origin in the complex plane, integrating term-by-term in \eqref{eq:phi22betaBlocks}, and dropping power-law singular contributions (as $a\rightarrow0$) afterwards. In practice, we truncate the sum over $n$ to some higher order $n_{\text{max}}$, and perform integration by combining both numerical and analytic strategies, as detailed in appendix \ref{app:CPT}.

We list in Table \ref{tab:phi22beta3} several chosen values of $\beta_3$ computed this way. Note that the $m = 3$ result matches that of \cite{Behan:2017emf} for the 2d long-range Ising model. 
\begin{table}[h]
\centering\setlength\doublerulesep{1em}\setlength\extrarowheight{0.2em}%
\begin{tabular}{|c|ccc|}
\hline
 $(m,2,2)$ & $m=3$ & $m=4$ & $m=5$ \\
\hline
$\beta_3$ & $1.268404308939(1\pm4)$ & $0.451640757399(67\pm10)$ & $0.241916794937(59\pm10)$ \\
\hline
\hline
$(m,2,2)$ & $m=10$ & $m=15$ & $m=20$\\ 
\hline
$\beta_3$ & $0.04955840966(37\pm27)$ & $0.02151175317(88\pm10)$ & $0.01205562419(35\pm10)$ \\
\hline
\end{tabular}
\caption{Values of $\beta_3$ for LRMM of type $(m, 2,2)$. For practical purposes, we have truncated the sum over $n$ to $n_{\text{max}}=20$, and checked that the results are stable against increasing $n_{\text{max}}$. The series truncation order is taken high enough such that the numerical error is dominated by the precision of the numerical integration scheme.}
\label{tab:phi22beta3}
\end{table}
We have computed $\beta_3$ for a few hundred integer values $m\geq 3$, and for all of them we have found that $\beta_3$ is positive. This is also  true for the analytic results in the $1/m$ expansion discussed in \ref{largem22}. We conjecture that all LRMM $(m,2,2)$, with (at least) $m\geq 3$ and integer have $\beta_3 > 0$.

%
\subsection{Anomalous dimensions of Virasoro primaries}
\label{sec:phi22flow-anodim}
Next, we compute anomalous dimensions of Virasoro primaries $\phirs{r}{s}$. As reviewed in Appendix \ref{app:CPT}, we have that $\gamma_{r,s}(g) = -\frac{1}{2} B_{r,s} g^2 + O(g^4)$ to leading order in conformal perturbation theory, where
\be\label{eq:phi22Bano}
B_{r,s} = 2\pi \int_{\mathcal{R}} \text{d}^2z \left [ 2 \la \phirs{r}{s}(0) \mathcal{O}(z,\bar{z}) \mathcal{O}(1) \phirs{r}{s}(\infty) \ra + \la \mathcal{O}(0) \mathcal{O}(z,\bar{z}) \phirs{r}{s}(1) \phirs{r}{s}(\infty) \ra \right ] \biggl |_{\text{finite}}\,,
\ee
with $\mathcal{R}$ is the same as before and the four-point functions are those of the unperturbed theory.
We will carry out all integrations using the series method once again. This involves writing
\begin{align}
\la \phirs{r}{s}(0) \mathcal{O}(z,\bar{z}) \mathcal{O}(1) \phirs{r}{s}(\infty) \ra &= \la \chi(z, \bar{z}) \chi(1) \ra \la \phirs{r}{s}(0) \phirs{2}{2}(z,\bar{z}) \phirs{2}{2}(1) \phirs{r}{s}(\infty) \ra \nonumber\\
&= \frac{|1 - z|^{-2\D_\chi}}{|z|^{\D_{2,2} + \D_{r,s}}} \sum_{r' = r \pm 1} \sum_{s' = s \pm 1} C_{(2,2)(r,s)(r',s')}^2 \left | \sum_{n = 0}^\infty a_n^{(r',s')} \rho^{h_{r',s'} + n} \right |^2 \nonumber\,,\\
\la \mathcal{O}(0) \mathcal{O}(z,\bar{z}) \phirs{r}{s}(1) \phirs{r}{s}(\infty) \ra &= \la \chi(0) \chi(z, \bar{z}) \ra \la \phirs{2}{2}(0) \phirs{2}{2}(z,\bar{z}) \phirs{r}{s}(1) \phirs{r}{s}(\infty) \ra \nonumber\\
&= |z|^{-4} \sum_{r' \in \{ 1,3 \}} \sum_{s' \in \{ 1,3 \}} C_{(2,2)(2,2)(r',s')} C_{(r,s)(r,s)(r',s')} \left | \sum_{n = 0}^\infty a_n^{(r',s')} \rho^{h_{r',s'} + n} \right |^2\,,
\end{align}
where the coefficients $a_n^{(r',s')}$ are again computed recursively from BPZ differential equation  (see Appendix \ref{app:mmcorrelators}) and $C$'s are known OPE coefficients.

Upon truncating the series in $n$ to some high $n_{\text{max}}$, and performing the integration for fixed $m$, it is not difficult to generate tables of anomalous dimensions along the lines of Table \ref{tab:phi22ano}.
\begin{table}[h]
\centering\setlength\doublerulesep{1.43em}\setlength\extrarowheight{0.2em}%
\begin{tabular}[t]{|c|c|}
\hline
 $(m,2,2)$& $m=3$ \\
\hline
$\gamma_{1,2}$ & $1.000000000000(06\pm25) \,\delta$ \\
\hline
$\gamma_{1,3}$ & $(0.0\pm2.5)\cdot 10^{-13}\,\delta$ \\
\hline
\hline
 $(m,2,2)$& $m=4$ \\
\hline
$\gamma_{2,2}$ & $0.999999999999(98\pm25)\,\delta$ \\
\hline
$\gamma_{1,2}$ & $1.853972502143(2\pm7)\,\delta$ \\
\hline
$\gamma_{2,1}$ & $-1.49448378662(13\pm12)\,\delta$ \\
\hline
$\gamma_{1,3}$ & $2.387848466943(8\pm7)\,\delta$ \\
\hline
$\gamma_{1,4}$ & $(0\pm3)\cdot10^{-10}\,\delta$ \\
\hline
\end{tabular}\quad
\begin{tabular}[t]{|c|c|}
\hline
 $(m,2,2)$& $m=5$ \\
\hline
$\gamma_{2,2}$ & $1.00000000000(02\pm13)\,\delta$ \\
\hline
$\gamma_{2,3}$ & $1.47226190793(71\pm13)\,\delta$ \\
\hline
$\gamma_{1,2}$ & $2.71621107485(0\pm5)\,\delta$ \\
\hline
$\gamma_{2,1}$ & $-2.70495211922(48\pm23)\,\delta$ \\
\hline
$\gamma_{2,4}$ & $0.64046753713(04\pm13)\,\delta$ \\
\hline
$\gamma_{1,3}$ & $4.70461558572(07\pm23)\,\delta$ \\
\hline
$\gamma_{2,5}$ & $-5.5457825655(85\pm15)\,\delta$ \\
\hline
$\gamma_{1,4}$ & $4.157465930(9\pm8)\,\delta$ \\
\hline
$\gamma_{1,5}$ & $(0\pm3)\cdot10^{-6}\,\delta$ \\
\hline
\end{tabular}
\caption{
	Leading order in the anomalous dimensions at $g=g_*$ for LRMM of type $(m, 2,2)$, with $m=3,4,5$. We have truncated the sum over $n$ to $n_{\text{max}}=30$, and checked that the results are stable against increasing $n_{\text{max}}$. The uncertainty is given by the maximal errors between the absolute error $\abs{\gamma_{r,s}(n_{\text{max}}=30)-\gamma_{r,s}(n_{\text{max}}=25)}$ and the precision of the numerical integration at each point.}
\label{tab:phi22ano}
\end{table}

The $m = 3$ result is consistent with the analytic results of \cite{Behan:2017emf} for the 2d long-range Ising model, i.e. $\gamma_{1,2}=\delta$ (from shadow relation \eqref{eq:shadow2}) and $\gamma_{1,3}=0$.

\subsection{Large-\texorpdfstring{$m$}{m} analysis}\label{largem22}

\begin{figure}[h]
\centering
\includegraphics[scale=0.73, valign=t]{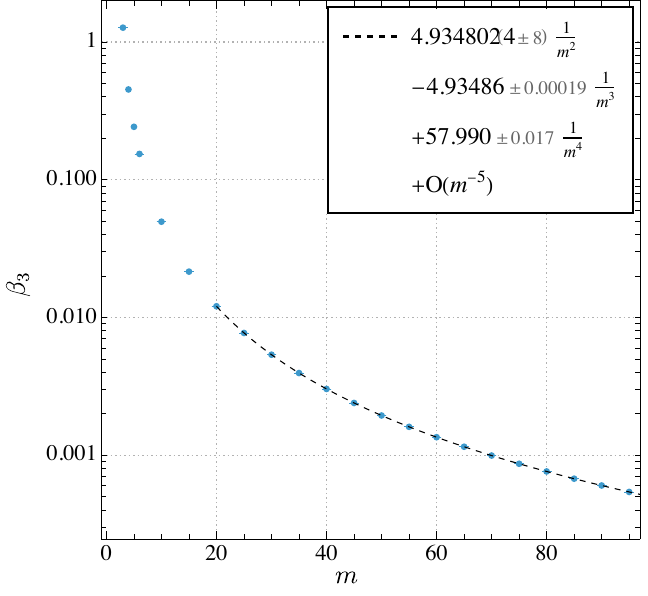}\quad
\includegraphics[scale=0.75, valign=t]{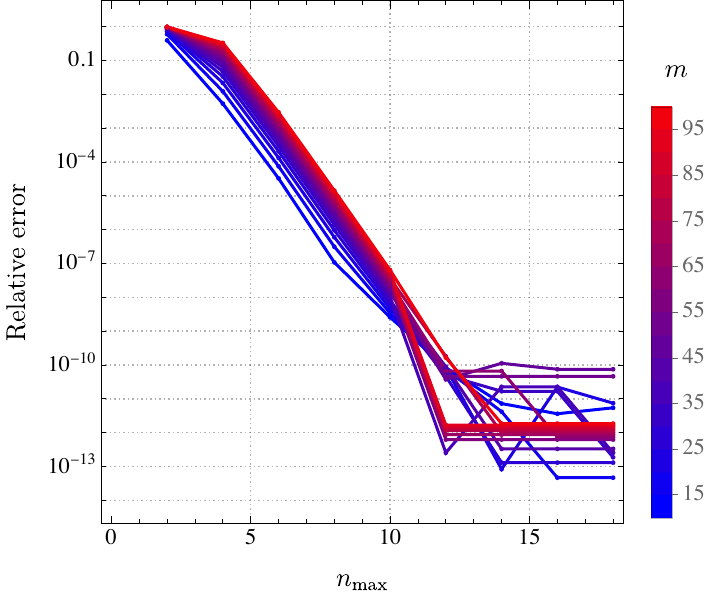}
\caption{Left: Polynomial fit (up to $m^{-8}$) for numerical values of $\beta_3$ and $m\ge20$. For each data point, the error is taken to be the max between $\abs{\beta_3(n_{\text{max}}=20)-\beta_3(n_{\text{max}}=18)}$ and the precision of the numerical integration at each point. The uncertainty on the fit corresponds to the 95\% confidence interval. Right: Error fluctuations, for several choices of $n \leq n_{\text{max}}$ and $m$. After some value $n_{\text{max}}<20$, the relative error get saturated by the numerical error.}
\label{fig:phi22beta3Fit}
\end{figure}

After enlarging Table \ref{tab:phi22beta3} to several more values of $m$, it is possible to make conjectures about the large-$m$ expansion of $\beta_3$ using a numerical fit. Figure \ref{fig:phi22beta3Fit} shows the result of this fit along with evidence that the sum over $n$ has converged sufficiently well.
A conclusion we can draw from the numerics is that
\be
\beta_3 = \frac{\pi^2}{2m^2} - \frac{\pi^2}{2m^3} + O(m^{-4})\,.
\ee
We will prove this result analytically in section \ref{sec:large-m}, using an entirely different method based on the Coulomb gas formalism.
We find a perturbative (for $m^2\delta\ll 1$) IR fixed point at:
\be
g_*^2/\delta = \frac{2m^2}{\pi^2} + \frac{2m}{\pi^2} + O(1)\,. \label{eq:phi22FPlargem}
\ee

\begin{figure}[H]
\centering
\begin{tabular}[c]{l}\includegraphics[scale=0.75]{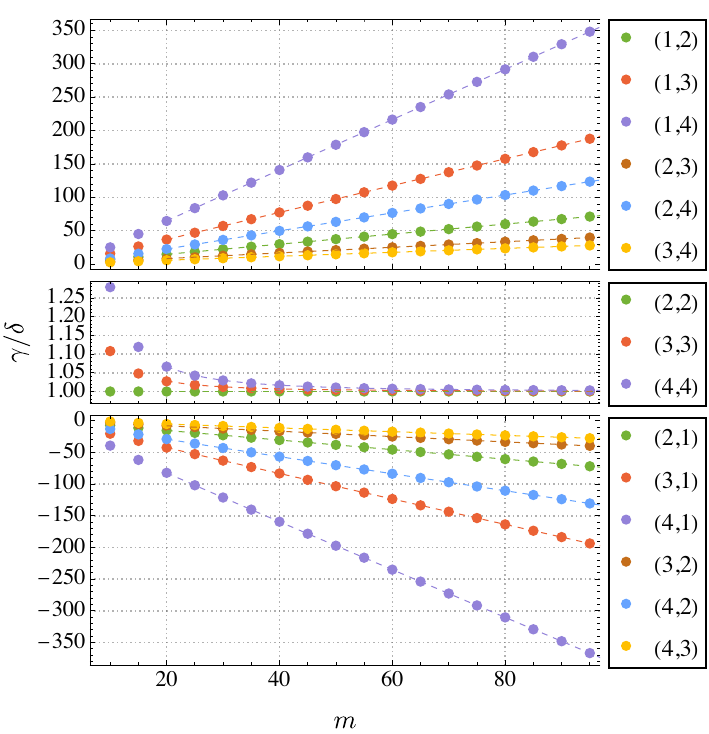}\end{tabular}\,
\setlength\extrarowheight{1em}
\begin{tabular}[c]{l}
$\gamma_{2,2}/\delta= 1.00000(0\pm8)+O(m^{-1})$\\
$\gamma_{3,3}/\delta= 1.00000(0\pm8)+O(m^{-1})$\\
$\gamma_{1,2}/\delta= 0.75000(01\pm13)m+O(1)$\\
$\gamma_{2,1}/\delta=-0.75000(01\pm13)m+O(1)$\\
$\gamma_{1,3}/\delta= 2.00000(04\pm34)m+O(1)$\\
$\gamma_{3,1}/\delta=-2.00000(04\pm35)m +O(1)$\\
\\
\end{tabular}
\caption{Left: Polynomial fits (up to $m^{-6}$) for several anomalous dimension for $m\ge20$. Here $n_{\text{max}}=50$. For each data point, the error is taken to be the max between $\abs{\beta_3(n_{\text{max}}=50)-\beta_3(n_{\text{max}}=45)}$ and the precision of the numerical integration at each point. The uncertainty on the fit corresponds to the 95\% confidence interval. The expansion order was increased compared to figure \ref{fig:phi22beta3Fit}, in order to improve convergence. Right: Leading non trivial order of the fit, for some operators.}
\label{fig:phi22anoFit}
\end{figure}

For anomalous dimensions, we have performed the fits with \eqref{eq:phi22FPlargem} plugged in from the start. An example plot which scans over different values of $(r,s)$ is shown in Figure \ref{fig:phi22anoFit}. Based on data from approximately 100 operators, we conjecture:
\be\label{eq:phi22gammaFitResult}
\gamma_{r,s}(g_*) = \begin{cases}
\frac{m}{2} \frac{(r - s)^2 (r + s)}{(r + 1) (s - 1)} \delta + O(1), & r \leq s \\
-\frac{m}{2} \frac{(r - s)^2 (r + s)}{(s + 1) (r - 1)} \delta + O(1), & r > s
\end{cases}\,.
\ee
Two comments are now in order.

First, the formula is anti-symmetric under $r \leftrightarrow s$. This is a consequence of a transformation applied to $m$, which formally exchanges Kac table weights:
\be
m \leftrightarrow -1-m, \quad h_{r,s} \leftrightarrow h_{s,r} \label{eq:formalm}\,.
\ee
Since the perturbing operator $\phirs{2}{2} \chi$ is symmetric in the Kac labels, the transformation \eqref{eq:formalm}  commutes with this flow. In particular, $(r,s,m) \leftrightarrow (s,r,1-m)\simeq (s,r,-m)$ should be a symmetry of the leading term in the large-$m$ expansion.

Second, the fixed point \eqref{eq:phi22FPlargem} and anomalous dimensions \eqref{eq:phi22gammaFitResult} are plagued with the same issue as the fixed point and anomalous dimensions discussed already at the end of section \ref{sec:duality}, in the sense that it is not valid to consider small finite values of $\delta$ when $m$ is large. This is a further indication that, for unprotected quantities, one cannot make the $\eps$ and $\delta$ expansions perturbative at the same time.

%

\subsection{Anomalous dimensions of higher-spin currents}

In this section, we consider the global primaries in the Virasoro identity multiplet, i.e the stress tensor $T$ and its higher-spin composites $T^I$. These cease to be conserved as we turn on the nonlocal interaction, and at the IR fixed point can generically acquire anomalous dimension via a multiplet recombination \cite{rt15,1601.01310}.

Taking the stress tensor as a simple example, the broken conformal Ward identities imply
\be
\bar{\partial} T = g_* b \mathcal{V} + O(g_*^2)\,, \label{eq:recombinationEOM}
\ee
where $b$ is a numerical coefficient.
As $g_* \to 0$, and assuming spectrum continuity, $\mathcal{V}$ needs to become an independent Virasoro primary with quantum numbers $(h, \bar{h}) = (2, 1)$ and there is only one such candidate which can be built from the minimal model $\mathcal{M}_{m,m + 1}$ and a GFF of dimension $2 - \D_{2,2}$. We can therefore use the recombination to compute the anomalous dimension of $T$ at the IR fixed point.\footnote{
In the case of the long-range Ising model, the anomalous dimension of the broken stress tensor was computed in \cite{Behan:2017dwr,Behan:2017emf}. This was followed by a similar computation for the (unique) spin-4 current in \cite{Behan:2018hfx}. The calculations here will extend these results to higher long-range minimal models of type $(m, 2, 2)$. They will also be new for the long-range Ising model because we will go up to spin 6.}

For higher-spin currents $T^I$, we have many degenerate operators, and so \eqref{eq:recombinationEOM} should be modified as follow:
\be
\bar{\partial} T^I = g_* b^I_{\;\;J} \mathcal{V}^J + O(g_*^2)\,, \label{eq:recomb}
\ee
and note that the matrix $b^I_{\;\;J}$ needs not be square. (Recently, \cite{2211.16503,2412.21107} used \eqref{eq:recomb} to study the breaking of extended chiral symmetry when perturbing away from a rational CFT. In that application, the discussion was geared towards the question of whether the anomalous dimensions were zero or non-zero.)

The explicit indices in \eqref{eq:recomb} indicate that we have chosen bases for two degenerate spaces of operators. The elements do not need to be orthogonal so we will refer to the Gram matrices $G_T^{IJ}$ and $G_\mathcal{V}^{IJ}$ defined by the inner products
\be
\left < T^{\dagger I} T^J \right > = G_T^{IJ}, \quad \left < \mathcal{V}^{\dagger I} \mathcal{V}^J \right > = G_\mathcal{V}^{IJ}\,,
\ee
in radial quantization.
The only perturbation theory enters when fixing the coefficient matrix $b^I_{\;\;J}$. This can be done by writing the mixed two-point function between $\bar{\partial} T^I$ and $\mathcal{V}^J$ in two different ways. Fixing the spin to be $\ell$,
\ba
& \la \bar{\partial} T^I(z_1, \bar{z}_1) \mathcal{V}^J(z_2, \bar{z}_2) \ra = g_* b^I_{\;\;K} \la \mathcal{V}^K(z_1, \bar{z}_1) \mathcal{V}^J(z, \bar{z}) \ra_0 = g_* b^I_{\;\;K} \frac{G_\mathcal{V}^{KJ}}{z_{12}^{2\ell} \bar{z}_{12}^2} \\
& \la \bar{\partial} T^I(z_1, \bar{z}_1) \mathcal{V}^J(z_2, \bar{z}_2) \ra = g_* \int \text{d}^2z_3 \la \bar{\partial} T^I(z_1, \bar{z}_1) \mathcal{V}^J(z_2, \bar{z}_2) \mathcal{O}(z_3, \bar{z}_3) \ra_0 = \pi g_* \frac{\lambda^{IJ}}{z_{12}^{2\ell} \bar{z}_{12}^2}\,,
\ea
where unimportant labels on the OPE coefficient matrix $\lambda^{IJ}$ have been suppressed. The integral has been evaluated using the identity $\bar{\partial}_1 z_{13}^{-1} = 2\pi \delta(z_{13}, \bar{z}_{13})$ and the fact that $\text{d}^2z_3 = \frac{1}{2} \text{d}z_3 \text{d}\bar{z}_3$.\footnote{Acting with $\bar{\partial}_1$ on $z_{12}^{-1}$ is not relevant for this calculation because these insertions can always be considered a finite distance apart.} Equating these two rows,
\be
b^I_{\;\;J} = \pi \lambda^{IK} \left [ G_\mathcal{V}^{-1} \right ]_{KJ}\,.
\ee

This solution can now be plugged into the equation of motion \eqref{eq:recomb} and its conjugate to get
\be
\bar{L}_{-1} T^I = \pi g_* \lambda^{IK} \left [ G_\mathcal{V}^{-1} \right ]_{KJ} \mathcal{V}^J, \quad T^{\dagger I} \bar{L}_1 = \pi g_* \mathcal{V}^{\dagger J} \left [ G_\mathcal{V}^{-1} \right ]_{JK} \left [ \lambda^\dagger \right ]^{KI}\,.
\ee
The key will be to use both inside $\la T^{\dagger I} T^J \ra$ since $[\bar{L}_1, \bar{L}_{-1}] = 2\bar{L}_0$ gives access to the scaling dimension. We will also write $T^{\dagger I}$ and $T^J$ as linear combinations of dilation eigenstates $\tilde{T}^J$ using
\be
T^I = M^I_{\;\;J} \tilde{T}^J = (HU)^I_{\;\;J} \tilde{T}^J\,.
\ee
Note that we have also taken the polar decomposition of $M$ into a Hermitian matrix $H$ and unitary matrix $U$. Putting the ingredients together,
\ba
(\pi g_*)^2 \lambda^{IJ} \left [ G_\mathcal{V}^{-1} \right ]_{JK} \left [ \lambda^\dagger \right ]^{KL} &= \la T^{\dagger L} \bar{L}_1 \bar{L}_{-1} T^I \ra = \left [ HU \text{diag}(\gamma) U^\dagger H \right ]^{LI}\,. \label{eq:degenRecomb}
\ea
Since the original basis of operators determines all OPE coefficients and Gram matrices, we just need to strip off $H$ from $HU \text{diag}(\gamma) U^\dagger H$ in order to compute the anomalous dimension matrix. This is easily done by realizing that
\be
M M^\dagger = H^2 = G_T\,.
\ee
Therefore, $H$ is the unique positive-definite Hermitian matrix squaring to $G_T$: it may be found by taking the positive square roots of all eigenvalues in the spectral decomposition of $G_T$.

Let us now see how this works for examples with increasing spin. At spin 2, the current which breaks is the stress tensor and there is only one operator with which it can possibly recombine. For later convenience, we will write these operators with the notation
\be
T = L_{-2}, \quad \mathcal{V} = \D_{2,2} \phirs{2}{2} L_{-1} \chi - \D_\chi L_{-1} \phirs{2}{2} \chi\,. \label{eq:uniqueTVex}
\ee
OPE coefficients involving them are fixed in terms of the central charge. It is therefore a simple exercise to apply \eqref{eq:degenRecomb} and find
\be
\gamma_2 = (\pi g_*)^2 c^{-1} \D_{2,2} \D_\chi = (\pi g_*)^2 \frac{3(2m + 3)(2m - 1)}{4m(m + 1)(m + 3)(m - 2)}\,. \label{eq:uniqueTVanom}
\ee

At spin 4, there is again one current to break and it is the well known quasiprimary\footnote{This standard terminology in the context of 2d CFTs refers to operators that are Virasoro descendants, only primaries under the global conformal group.}
\be
T^1 = L_{-4} - \frac{5}{3} L_{-2}^2\,.
\ee
Conversely, suitable candidates for the divergence of $T^1$ span a nontrivial subspace. The dimension of it depends on the minimal model under consideration. For $m = 3$, we have found that
\ba
\mathcal{V}^1 &= L_{-2} L_{-1} \phirs{2}{2} \chi - \frac{76}{45} L_{-1}^2 \phirs{2}{2} L_{-1} \chi + \frac{76}{115} L_{-1} \phirs{2}{2} L_{-1}^2 \chi - \frac{76}{10695} \phirs{2}{2} L_{-1}^3 \chi \\
\mathcal{V}^2 &= L_{-1}^3 \phirs{2}{2} \chi - \frac{17}{5} L_{-2}^2 \phirs{2}{2} L_{-1} \chi + \frac{153}{115} L_{-1} \phirs{2}{2} L_{-1}^2 \chi - \frac{51}{3565} \phirs{2}{2} L_{-1}^3 \chi \label{v3-m3}\,,
\ea
is a basis for the kernel of $L_1$ among $(h, \bar{h}) = (4, 1)$ operators. At higher values of $m$, there are fewer null states and a third operator is needed. In other words, the $L_{-3}$ action on $\phirs{2}{2}$ is independent and no longer a linear combination of $L_{-2}L_{-1}$ and $L_{-1}^3$.\footnote{Even though it is commonly said that $\phirs{2}{2}$ for $m = 3$ only has null states at level 2 and 4, this is colloquially referring to null Verma modules. At level 3, there is a null state which lives in the level 2 null Verma module.} For the $m = 4$ example, we have found
\begin{align}
\mathcal{V}^1 &= L_{-3} \phirs{2}{2} \chi - \frac{160}{77} L_{-2} \phirs{2}{2} L_{-1} \chi + \frac{3200}{3003} L_{-1} \phirs{2}{2} L_{-1}^2 \chi - \frac{3200}{471471} \phirs{2}{2} L_{-1}^3 \chi \nonumber \\
\mathcal{V}^2 &= L_{-2} L_{-1} \phirs{2}{2} \chi - \frac{3}{77} L_{-2} \phirs{2}{2} L_{-1} \chi - \frac{120}{77} L_{-1}^2 \phirs{2}{2} L_{-1} \chi + \frac{1780}{3003} L_{-1} \phirs{2}{2} L_{-1}^2 \chi - \frac{1780}{471471} \phirs{2}{2} L_{-1}^3 \chi \nonumber \\
\mathcal{V}^3 &= L_{-1}^3 \phirs{2}{2} \chi - \frac{240}{77} L_{-1}^2 \phirs{2}{2} L_{-1} \chi + \frac{3569}{3003} L_{-1} \phirs{2}{2} L_{-1}^2 \chi - \frac{3569}{471471} \phirs{2}{2} L_{-1}^3 \chi\,. \label{v3-m4}
\end{align}
Although it would take more space to write out, it is clear that the quasiprimary basis $\{ \mathcal{V}^1, \mathcal{V}^2, \mathcal{V}^3 \}$ can be computed for general $m \geq 4$. Using these to compute OPE coefficients and Gram matrices, we find
\be\label{eq:phi22Spin4Current}
\gamma_4 = (\pi g_*)^2 \frac{3(2m - 1)(2m + 3)(112m^4 + 224m^3 - 148m^2 - 260m + 75)}{32(m - 2)m^2(m + 1)^2(m + 3)(3m - 2)(3m + 5)}\,,
\ee
for the $m \geq 4$ anomalous dimension. Even though the $m = 3$ anomalous dimension should in principle require a separate calculation, it happens to be given by \eqref{eq:phi22Spin4Current} once again.

\begin{table}[h]
\centering\setlength\extrarowheight{0.2em}
\begin{tabular}{|c|c|}
\hline
$m$ & $\gamma_6/(\pi g_*)^2$ \\
\hline
3 & $97455/131072$ \\
4 & $11 (21567259 \pm \sqrt{16239267339481})/557056000$ \\
5 & $3 (1030002563 \pm \sqrt{34772932634594569})/12058624000$ \\
6 & $55 (138204789 \pm \sqrt{607495480660345})/43724570624$ \\
10 & $207 (66360398593 \pm \sqrt{134889823876370114049})/222258003968000$ \\
15 & $319 (304817158333 \pm \sqrt{2817314386305624849289})/3508717748224000$ \\
20 & $43 (41876782359 \pm \sqrt{52995318467200907281})/114592350208000$ \\
25 & $3339 (7507982225219 \pm \sqrt{1700866497134624314157961})/2477904412672000000$ \\
\hline
\end{tabular}
\caption{Anomalous dimensions of the two-loop dilation eigenstates among spin-6 operators in the Virasoro identity multiplet. There are two of these for $m \geq 4$ but only one for $m = 3$.}
\label{tab:phi22HScurrents}
\end{table}
At spin 6, the new ingredient that appears is that the $(h, \bar{h}) = (6, 0)$ quasiprimaries can be degenerate. This is not the case for $m = 3$ which just has
\begin{align}
T^1 &= L_{-4} L_{-2} - \frac{5}{2} L_{-3}^2 + \frac{5}{3} L_{-2}^3 \label{eq:t6m3}\,,
\end{align}
but all $m \geq 4$ additionally have
\begin{align}
T^2 &= L_{-6} + \frac{21}{4} L_{-3}^2 - \frac{14}{3} L_{-2}^3 \label{eq:t6m4}\,.
\end{align}
The reason of course has to do with null states again. The many null states of $\phi_{1,1}$ from the level 1 Verma module are joined by one from the level 6 Verma module in the case of $m = 3$. We will not write expressions for the operators available for recombination but the number of them is 3 for $m = 3$ and 6 for $m \geq 4$. A general $m$ formula for the anomalous dimensions is likely to be very cumbersome and we have not attempted to find it. Even the square root of the Gram matrix
\begin{align}
G_T &= \frac{(m - 2)(m + 3)}{4m^3(m + 1)^3} \\
&\times \begin{pmatrix}
7(1271m^4 + 2542m^3 - 941m^2 - 2212m + 336) & -28(131m^4 + 262m^3 - 107m^2 - 238m + 30) \\
-28(131m^4 + 262m^3 - 107m^2 - 238m + 30) & 1517m^4 + 3034m^3 - 1403m^2 - 2920m + 300
\end{pmatrix} \nonumber\,,
\end{align}
is already quite long. Instead, we have computed anomalous dimensions individually for several values of $m$ and listed the results in Table \ref{tab:phi22HScurrents}.

%% file: sections/phi12flow.tex

%
\section{Another long-range minimal model -- the \texorpdfstring{$\phirs{1}{2}\chi$}{phi12chi} flow}
\label{sec:phi12flow}

In this section, we study the LRMM of type $(m,1,2)$. Because the analysis closely parallels that of section \ref{sec:phi22flow-nearSR}, we will omit most intermediate steps and concentrate on the main results.

%

\subsection{Beta function}
The computation of $\beta_3$ get slightly simplified such that the change of variable to radial coordinates \eqref{eq:rhocoord} is not necessary anymore. Indeed, for $\cO=\phirs{1}{2}\chi$, the integrand is known in closed form and thus converges everywhere including the integration region boundary:
\be\label{phi12chi4pt}
\la\cO(0) \cO(z, \bar{z}) \cO(1) \cO(\infty) \ra = \frac{1 + |z|^{-2\D_\chi} + |1 - z|^{-2\D_\chi}}{|z|^{2\D_{1,2}}}\sum_{s \in \{ 1,3 \}} C_{(1,2)(1,2)(1,s)}^2 \left | \mathcal{F}_{(1,s)}^{(1,2)(1,2)(1,2)(1,2)}(z,\bar{z}) \right |^2\,,
\ee
where $C$'s are known OPE coefficients and blocks (see Appendix \ref{app:mmcorrelators})
\begin{align}
	\mathcal{F}^{(1,2)(1,2)(1,2)(1,2)}_{(1,1)}(z)&=(1-z)^{\frac{m}{2m+2}} \, _2F_1\left(\frac{1}{m+1},\frac{m}{m+1};\frac{2}{m+1};z \right)\,,\nonumber\\
	\mathcal{F}^{(1,2)(1,2)(1,2)(1,2)}_{(1,3)}(z)&=z ^{h_{1,3}} (1-z)^{\frac{m}{2m+2}} \, _2F_1\left(\frac{m}{m+1},\frac{2m-1}{m+1};\frac{2 m}{m+1};z \right)\,.
\end{align}

Due to the complicated domain of integration, it is still hard to find analytic expressions for $\beta_3$. We proceed as in section \ref{sec:phi22flow-nearSR}, re-expanding \eqref{phi12chi4pt} around $z=\bar{z}=0$ to some high order $n_{\text{max}}$ only to remove IR divergencies and combining both analytic and numerical integration strategies. Table \ref{tab:phi12beta3} shows several values for $\beta_3$.

\begin{table}[h]
\centering\setlength\doublerulesep{1em}\setlength\extrarowheight{0.2em}%
\begin{tabular}{|c|cccc|}
\hline
$(m,1,2)$ & $m=3$ & $m=4$ & $m=5$ & $m=6$ \\
\hline
$\beta_3$ & $1.268404308(9\pm6)$ & $3.169321538(4\pm 6)$ & $5.475808980(0\pm6)$ & $8.062529231(4\pm6)$\\
\hline
\hline
$(m,1,2)$& $m=10$ & $m=15$ & $m=20$ & $m=25$ \\
\hline
$\beta_3$ & $20.000315421(0\pm6)$ & $36.615640393(7\pm6)$ & $54.016175065(8\pm6)$ & $71.796653176(4\pm6)$ \\
\hline
\end{tabular}
\caption{Values of $\beta_3$ for LRMM of type $(m, 1,2)$. For practical purposes, we have truncated the sum over $n$ to $n_{\text{max}}=45$, and checked that the results are stable against increasing $n_{\text{max}}$. The series truncation order is taken high enough such that the numerical error is dominated by the precision of the numerical integration scheme.}
\label{tab:phi12beta3}
\end{table}

Two comments are in order. First, for $m=3$ we find the same result as for the LRMM of type $(3,2,2)$. This is of course expected, since for $m=3$ the operator $\phirs{1}{2}$ is the same as $\phirs{2}{2}$. Second, we have checked that $\beta_3>0$ for many values of (integer) $m$. Since this remains true in the $1/m$ expansion, as we will see in Section \ref{largem22}, we conjecture that it must be true all LRMM $(m,1,2)$, with (at least) $m\geq 3$ and integer.

\subsection{Anomalous dimensions of Virasoro primaries}
\label{sec:mixing}
Anomalous dimensions of Virasoro primaries are obtained from an integral like \eqref{eq:phi22Bano}. The integrand is again known in closed form (see Appendix \ref{app:mmcorrelators} for a derivation), and we have
\begin{align}\label{phi12chi4asdpt}
	\la \phirs{r}{s}(0) \mathcal{O}(z,\bar{z}) \mathcal{O}(1) \phirs{r}{s}(\infty) \ra &= \frac{|1 - z|^{-2\D_\chi}}{|z|^{\D_{1,2} + \D_{r,s}}} \sum_{r' = r \pm 1} \sum_{s' = s \pm 1} C_{(1,2)(r,s)(r',s')}^2 \left | \mathcal{F}_{(1,s)}^{(r,s)(1,2)(1,2)(r,s)}(z,\bar{z}) \right |^2 \nonumber\,,\\
	\la \mathcal{O}(0) \mathcal{O}(z,\bar{z}) \phirs{r}{s}(1) \phirs{r}{s}(\infty) \ra 	&= |z|^{-4} \sum_{r' \in \{ 1,2 \}} \sum_{s' \in \{ 1,3 \}} C_{(1,2)(1,2)(r',s')} C_{(r,s)(r,s)(r',s')} \left | \mathcal{F}_{(1,s)}^{(1,2)(1,2)(r,s)(r,s)}(z,\bar{z}) \right |^2\,,
\end{align}
where $C$'s are known OPE coefficients and
\begin{align}
		\begin{split}
		\mathcal{F}^{(r,s)(1,2)(1,2)(r,s)}_{(r,s+1)}(z)&=(1-{z} )^{\frac{m}{2 m+2}} {z} ^{h_{r,s+1}} \, _2F_1\left(\frac{m}{m+1},\frac{-r m+s m+m-r}{m+1};-r+\frac{m s}{m+1}+1;{z} \right)\,,\\
	\mathcal{F}^{(r,s)(1,2)(1,2)(r,s)}_{(r,s-1)}(z)&=(1-{z} )^{\frac{m}{2 m+2}} {z} ^{h_{r,s-1}} \, _2F_1\left(\frac{m}{m+1},\frac{r m-s m+m+r}{m+1};r-\frac{m s}{m+1}+1;{z} \right)\,,\\
		\mathcal{F}^{(1,2)(1,2)(r,s)(r,s)}_{(1,1)}(z)&=(1-{z} )^{\frac{m r-m s+r+1}{2 m+2}} \, _2F_1\left(\frac{1}{m+1},\frac{m r+r-m s+1}{m+1};\frac{2}{m+1};{z} \right)\,,\\
	\mathcal{F}^{(1,2)(1,2)(r,s)(r,s)}_{(1,3)}(z)&={z} ^{h_{1,3}} (1-{z} )^{\frac{m r-m s+r+1}{2 m+2}} \, _2F_1\left(\frac{m}{m+1},\frac{r m-s m+m+r}{m+1};\frac{2 m}{m+1};{z} \right)\,.
		\end{split}
\end{align}
 
 As for the integration, we proceed as before expanding \eqref{phi12chi4asdpt} around $z=\bar{z}=0$ to some high order $n_{\text{max}}$ only to remove IR divergencies and combining both analytic and numerical integration strategies. Several computed anomalous dimensions are listed in table \ref{tab:phi12ano}.
\begin{table}[h]
\centering\setlength\extrarowheight{0.2em}\setlength\doublerulesep{1.43em}%
\begin{tabular}[t]{|c|c|}
\hline
 $(m,1,2)$& $m=3$ \\
\hline
$\gamma_{1,2}$ & $1.00000000(00\pm10) \,\delta$ \\
\hline
$\gamma_{1,3}$ & $(0\pm5)\cdot 10^{-10}\,\delta$ \\
\hline
\hline
 $(m,1,2)$& $m=4$ \\
\hline
$\gamma_{2,2}$ & $0.264198357(6\pm4)\,\delta$ \\
\hline
$\gamma_{1,2}$ & $1.000000000(0\pm4)\,\delta$ \\
\hline
$\gamma_{2,1}$ & $(0.0\pm1.9)\cdot 10^{-10}\,\delta$ \\
\hline
$\gamma_{1,3}$ & $0.444444444(45\pm31)\,\delta$ \\
\hline
$\gamma_{1,4}$ & $\gamma_{3,1}$ see eq.~\eqref{eq:31-1loop} \\
\hline
\end{tabular}\quad
\begin{tabular}[t]{|c|c|}
\hline
 $(m,1,2)$& $m=5$ \\
\hline
$\gamma_{2,2}$ & $0.120000000(00\pm13)\,\delta$ \\
\hline
$\gamma_{2,3}$ & $0.369504172(28\pm22)\,\delta$ \\
\hline
$\gamma_{1,2}$ & $1.00000000(00\pm21)\,\delta$ \\
\hline
$\gamma_{2,1}$ & $(0.0\pm1.1)\cdot 10^{-10}\,\delta$ \\
\hline
$\gamma_{2,4}$ & $0.360000000(00\pm18)\,\delta$ \\
\hline
$\gamma_{1,3}$ & $0.577350269(19\pm20)\,\delta$ \\
\hline
$\gamma_{2,5}$ & $\gamma_{3,1}$ see eq.~\eqref{eq:31-1loop} \\
\hline
$\gamma_{1,4}$ & $17.06825153(96\pm10)\,\delta$ \\
\hline
$\gamma_{1,5}$ & $(0.0\pm1.1)\cdot 10^{-10}\,\delta$ \\
\hline
\end{tabular}
\caption{Leading order in the anomalous dimensions for LRMM of type $(m, 1,2)$, with $m=3,4,5$. We have truncated the sum over $n$ to $n_{\text{max}}=45$, and checked that the results are stable against increasing $n_{\text{max}}$. The uncertainty is given by the maximum between the absolute error $\abs{\gamma_{r,s}(n_{\text{max}}=45)-\gamma_{r,s}(n_{\text{max}}=40)}$ and the precision of the numerical integration at each point.}
\label{tab:phi12ano}
\end{table}

A few comments are in order. As expected, the anomalous dimensions in LRMM of type $(3,1,2)$ match perfectly with those of $(3,2,2)$ LRMM. Furthermore, we see that $\gamma_{1,2}=\delta$, consistently with the shadow relation (analogous to \eqref{eq:shadow2}). We also observe that, at least for all integer $m\leq 100$, $\gamma_{r,1}=\gamma_{m-r,m}$ appears to be compatible with zero for even $r$. For $m=3$, we know from \cite{Behan:2017emf} that $\gamma_{2,1}=\gamma_{1,3}=0$, but this observation for higher $r$ and $m$ is new. An analytic explanation for it will be given in section \ref{sec:large-m}. 

\paragraph{A mixing problem.\\}

One could try to apply the same procedure as before to compute anomalous dimensions of operators $\phirs{r}{1}$, with $r$ odd. It is not difficult to see that the integral would feature additional $\log a$ divergences, coming from the negative integer values of $h_{r,2}-h_{r,1}-h_{1,2}$ (for all values of $m$) in the first line of \eqref{phi12chi4asdpt}.  As no lower-order terms are available to cancel those divergencies, they are signs of operator mixing.

The long-range interaction breaks the local Virasoro symmetry while keeping the global conformal symmetry untouched. This allows $\phirs{r}{1}$ to mix with quasiprimary operators, as well as with operators made of $\chi$'s powers and other Virasoro primaries. The mixing happens when two or more operators form a degenerate subspace: their UV dimensions are equal.
Due to the fusion rules, the operators that can mix with $\phirs{r}{1}$ are combinations of $\chi$, $\phirs{r}{2}$ and their descendants. Their dimensions obey the relation: 
\be
\D_{r,2}+\D_{\chi}-\D_{r,1} = 3-r \in\mathbb{Z}\,.
\ee
Acting on the operator with a quasiprimary or descendant combination of $L_{-k}$, $k\in\N$ raises the dimension by $k$.
Thus, $\phirs{r}{1}$ mixes with the quasiprimary operator
\be
M_{(r,2)} := \chi \mathcal{L}_{\frac{3-r}{2}}\overline{\mathcal{L}_{\frac{3-r}{2}}}\phirs{r}{2}, \ r>2 \text{ odd}\,,
\ee
where $\mathcal{L}_{-k}$ denotes a specific combination of Virasoro generators forming a quasiprimary at level $k$ and normalized with respect to its two-point function.
When $r$ is even, it is not possible to form a spinless operator with the appropriate dimension, and its contribution vanishes upon angular integration.\medskip

Under the mixing, the renormalization of the operators get slightly modified to
\be
Q_i = \sum_j Z^{(r,1)}_{i,j} \sqbracket{Q}_j, \quad Q = \begin{pmatrix}
   \phirs{r}{1}\\
   M_{(r,2)}
\end{pmatrix}\,.
\ee
This means that instead of only computing the \eqref{eq:phi22Bano} contribution, one should compute the full two-point function in the degenerate subspace
\be
Z^{(r,1)} = \begin{pmatrix}
	1 + g^2\,\left(I^{\text{div}(1)}_{\phi \cO\cO\phi}+I^{\text{div}(2)}_{\phi \cO\cO\phi}\right) & -g\, I^{\text{div}(1)}_{\phi \cO M}\\
   -g\, I^{\text{div}(1)}_{M \cO\phi} & 1 + g^2\, \left(I^{\text{div}(1)}_{M \cO\cO M}+I^{\text{div}(2)}_{M \cO\cO M}\right)
\end{pmatrix}+O(g^3)\,,
\ee
where $I^{\text{div}(k)}_{\cO_1\dots \cO_n}$ denotes the pole term of order $k$ in $1/\delta$ of the regulated $n$-point function integral, see Appendix \ref{app:CPT}. 
The renormalized operators and their associated anomalous dimension are obtained by solving the eigensystem of $\rmd Z^{(r,1)}/\rmd \log(1/R)$. Already at this level, it is clear that the one-loop contribution, allowed by the OPE, will compensate for the double poles at two-loops, by consistency of the RG flow. \medskip

Here we choose to not compute the anomalous dimension of $\phirs{r}{1}$, for all $r$ odd, as the computation of the four-point function involving descendants placed at infinity requires a more subtle handling.
Instead, we report the special case of $\phirs{3}{1}$, which mixes with $M_{3,2} = \chi \phirs{3}{2}$.
There, the one-loop contribution does not vanish anymore:
\be\begin{split}\label{eq:31-1loop}
\gamma_{3,1}(g) &= - \pi \, \frac{m+2}{m+3}g+\gamma_{(3,1)}^{(2)}g^{2}+O(g^{3})\\
\gamma_{M(3,2)}(g) &= \pi \, \frac{m+2}{m+3}g+\gamma_{(3,1)}^{(2)}g^{2}+O(g^{3})\,.
\end{split}\ee
The two-loop order $\gamma_{(3,1)}^{(2)}$ contribution is equal for both operators and scheme dependent. The four-point integral contributions are computed numerically following the procedure described in section \ref{sec:phi22flow-anodim}. Several values are given in table \ref{tab:phi12ano31}. The consistency condition for the RG flow was verified up to machine precision. Note that $\phirs{3}{1}$ does not belong to the Kac table \eqref{eq:kactable} of the long-range Ising model $(3,1,2)$ LRMM.

\begin{table}[h]
\centering\setlength\doublerulesep{1em}\setlength\extrarowheight{0.2em}%
\begin{tabular}{|c|ccc|}
\hline
$(m,1,2)$ &$m=3$ & $m=4$ & $m=5$\\
\hline
$\gamma_{3,1}^{(2)}$  & / & $-0.993807196(68\pm21)$ & $-3.046624486(44\pm19)$\\
\hline
\hline
$(m,1,2)$ & $m=10$ & $m=15$ & $m=20$\\
\hline
$\gamma_{3,1}^{(2)}$ & $-11.716079789(22\pm15)$ & $-20.455101516(67\pm14)$ & $-29.371518375(64\pm14)$\\
\hline
\end{tabular}
\caption{Two-loops coefficient in the anomalous dimensions for  $\phirs{3}{1}$ in LRMM of type $(m, 1,2)$. We used $n_{\text{max}}=45$. The uncertainty is given by the maximum between the absolute error $\abs{\gamma_{3,1}(n_{\text{max}}=45)-\gamma_{3,1}(n_{\text{max}}=40)}$ and the precision of the numerical integration at each point.}
\label{tab:phi12ano31}
\end{table}

\subsection{Large-\texorpdfstring{$m$}{m} analysis}
\label{largem12}
\begin{figure}[H]
\centering
\includegraphics[scale=0.70, valign=t]{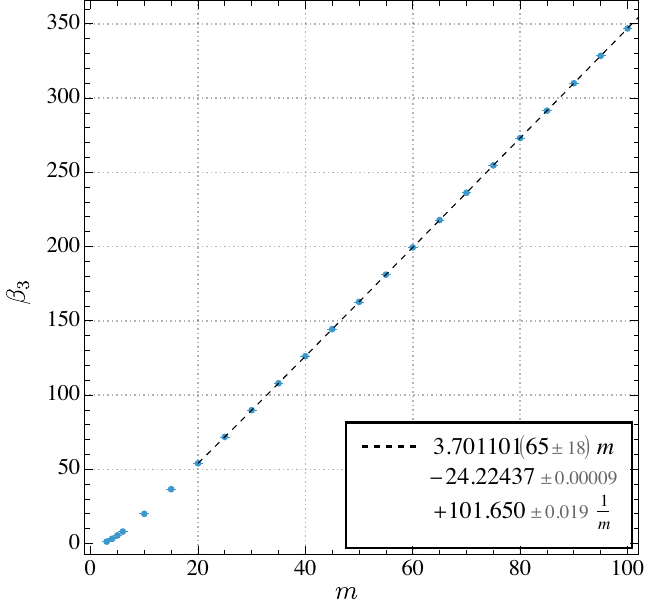} \quad 
\includegraphics[scale=0.74, valign=t]{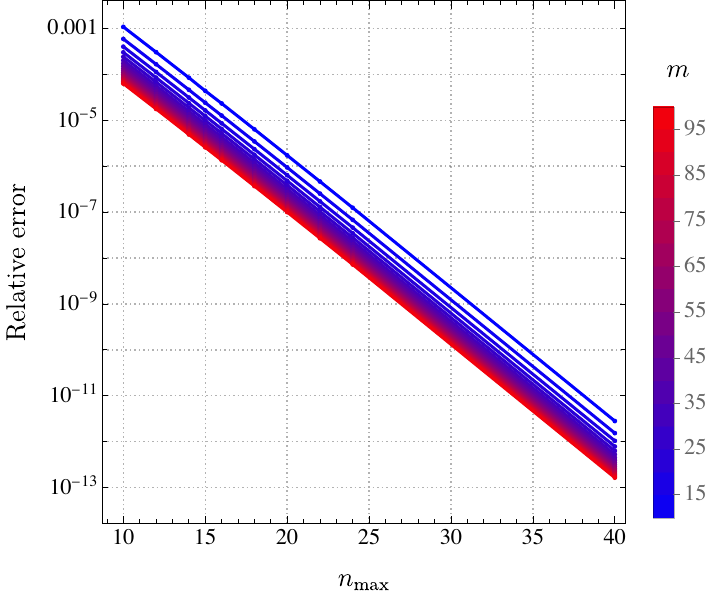}
\caption{Left: Polynomial fit (up to $m^{-8}$) for numerical values of $\beta_3$ and $m\ge20$. For each data point, the error is taken to be the max between $\abs{\beta_3(n_{\text{max}}=50)-\beta_3(n_{\text{max}}=45)}$ and the precision of the numerical integration at each point. The uncertainty on the fit corresponds to the 95\% confidence interval. Right: Error fluctuations, for several choices of $n \leq n_{\text{max}}$ and $m$.}
\label{fig:phi12beta3Fit}
\end{figure}

As we will prove in section \ref{sec:large-m}, at large $m$ we have that:
\be\label{eq:phi12betaFitResult}
\beta_3 = \frac{3\pi ^2}{8}  (m-1-8 \log 2) + O(m^{-1})\,.
\ee
Figure \ref{fig:phi12beta3Fit} shows the result of this fit (after extending table \ref{tab:phi12beta3} to several more values of $m$), along with evidence that the sum over $n$ has converged sufficiently well. We find a family of perturbative (for $\delta/m\ll 1$) IR fixed points at:
\be\label{eq:phi12FP}
g_*^2/\delta = \frac{8}{3\pi ^2} \left ( \frac{1}{m}+\frac{1+8 \log 2}{m^2} \right )+ O(m^{-3})\,.
\ee

\begin{figure}[H]
\centering
\begin{tabular}[c]{l}\includegraphics[scale=0.75]{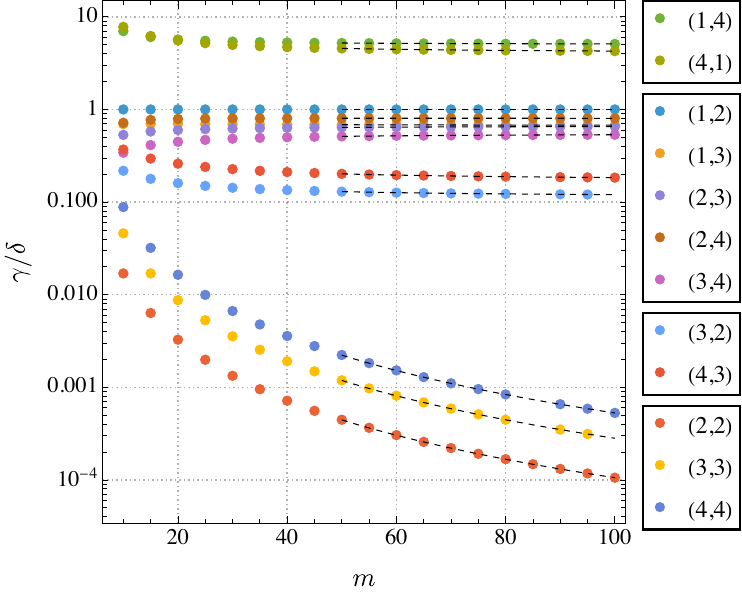}\end{tabular}\,%
\setlength\extrarowheight{1em}
\begin{tabular}[c]{ll}
$\gamma_{2,2}/\delta= (1.0000\pm 0.0014)\frac{1}{m^2}+O(m^{-3})$\\
$\gamma_{3,3}/\delta=(2.6667\pm 0.0026)\frac{1}{m^2}+O(m^{-3})$\\
$\gamma_{1,2}/\delta= 1.0000000(00\pm21)+O(m^{-1})$\\
$\gamma_{1,3}/\delta=0.6666666(67\pm23)+O(m^{-1})$\\
$\gamma_{2,3}/\delta=0.6666666(65\pm23)+O(m^{-1})$\\
$\gamma_{3,2}/\delta=0.1111111(1\pm4)+O(m^{-1})$\\
\\
\end{tabular}
	\caption{Left: Polynomial fits (up to $m^{-6}$) for several anomalous dimension. Here $n_{\text{max}}=45$. For each data point, the error is taken to be the max between $\abs{\beta_3(n_{\text{max}}=45)-\beta_3(n_{\text{max}}=40)}$ and the precision of the numerical integration at each point. The uncertainty on the fit corresponds to the 95\% confidence interval. Right: Leading non trivial order of the fit, for some operators.}
\label{fig:phi12anoFit}
\end{figure}

Figure \ref{fig:phi12anoFit} shows the result for several anomalous dimensions of $\phirs{r}{s}$, along with their large $m$ fit. For operators that do not mix (see the discussion in section \ref{sec:mixing}), we conjecture that:
\be\label{eq:phi12gammaFitResult}
\gamma_{r,s}(g_*) = \begin{cases}
\frac{1}{3}(r-s)^2\frac{s+(-1)^{s+r}}{s+(-1)^{s+r+1}}\delta+O(m^{-1}), & r> s >1\\
\frac{1}{3}(r-s)^2\frac{s+(-1)^{s+r+1}}{s+(-1)^{s+r}}\delta+O(m^{-1}), & 1\le r \le s
\end{cases}\,.
\ee

\subsection{Anomalous dimensions of higher-spin currents}
Since the $\phirs{1}{2}\chi$ flow breaks Virasoro symmetry, quasiprimaries in the Virasoro identity multiplet (which become degenerate above the lowest spins) acquire anomalous dimensions which can be found with the recombination method. This refers to an $(h, \bar{h}) = (\ell, 1)$ operator $\mathcal{V}^I$ becoming a descendant of an $(h, \bar{h}) = (\ell, 0)$ operator $T^I$ when the coupling is switched on. As explained around \eqref{eq:degenRecomb}, we will use
\be
\left [ U \text{diag}(\gamma) U \right ]_{IN} = (\pi g_*)^2 \left [ G_T^{-1/2} \right ]_{IJ} \lambda^{JK} \left [ G_\mathcal{V}^{-1} \right ]_{KL} \left [ \lambda^\dagger \right ]^{LM} \left [ G_T^{-1/2} \right ]_{MN}\,,
\ee
as a master formula where
\be
\la T^{\dagger I} T^J \ra = G^{IJ}_T, \quad \la T^I(0) \mathcal{V}^J(1) \mathcal{O}(\infty) \ra = \lambda^{IJ}, \quad \la \mathcal{V}^{\dagger I} \mathcal{V}^J \ra = G^{IJ}_\mathcal{V} \label{eq:gramOPE}\,,
\ee
and the notation $G_T^{-1/2}$ means we are taking the square root of $G_T^{-1}$ which has all positive eigenvalues.

At spin 2, the difference between $\phirs{1}{2}$ and $\phirs{2}{2}$ plays no role and one can trivially change \eqref{eq:uniqueTVex} to
\be
T = L_{-2}, \quad \mathcal{V} = \D_{1,2}\phirs{1}{2}L_{-1}\chi - \D_\chi L_{-1}\phirs{1}{2}\chi\,.
\ee
Similarly, \eqref{eq:uniqueTVanom} becomes
\be\label{anom_spin2_RG12}
\gamma_2 = (\pi g_*)^2 c^{-1} \D_{1,2} \D_\chi = (\pi g_*)^2 \frac{3m(m + 2)}{4(m + 1)(m + 3)}\,.
\ee

At spin 4, we start to see results that are more sensitive to this particular LRMM. Even though the current is still
\be
T^1 = L_{-4} - \frac{5}{3} L_{-2}^2\,,
\ee
the subspace of operators which can become its divergence is two-dimensional for any $m$. Picking $m = 4$ for example, a valid basis is
\ba
\mathcal{V}^1 &= L_{-2} L_{-1} \phirs{1}{2} \chi - \frac{65}{36} L_{-1}^2 \phirs{1}{2} L_{-1} \chi + \frac{65}{84} L_{-1} \phirs{1}{2} L_{-1}^2 \chi - \frac{65}{4788} \phirs{1}{2} L_{-1}^3 \chi \\
\mathcal{V}^2 &= L_{-1}^3 \phirs{1}{2} \chi - \frac{11}{3} L_{-1}^2 \phirs{1}{2} L_{-1} \chi + \frac{11}{7} L_{-1} \phirs{1}{2} L_{-1}^2 \chi - \frac{11}{399} \phirs{1}{2} L_{-1}^3 \chi\,.
\ea
Using operators that are quasiprimary for general $m$, the correlators in \eqref{eq:gramOPE} lead to
\be\label{anom_spin4_RG12}
\gamma_4 = (\pi g_*)^2 \frac{3m(m + 2)(29m^2 + 94m + 80)}{32(m + 1)^2(m + 3)(3m + 5)}\,.
\ee
Notice that after plugging in the fixed point, both $\gamma_2$ and $\gamma_4$ are $O(1/m)$ this time instead of $O(1)$.

\begin{table}[h]
\centering\setlength\extrarowheight{0.2em}%
\begin{tabular}{|c|c|}
\hline
$m$ & $\gamma_6/(\pi g_*)^2$ \\
\hline
3 & $97455/131072$ \\
4 & $(12317 \pm \sqrt{575089})/14875$ \\
5 & $35(103841 \pm \sqrt{60739393})/4194304$ \\
6 & $1760/1813, 132/161$ \\
10 & $90(24622 \pm \sqrt{6505053})/2301299$ \\
15 & $17(40596377 \pm \sqrt{21350564024689})/687865856$ \\
20 & $55(29233 \pm \sqrt{12123049})/1567657$ \\
25 & $2025(8280345 \pm \sqrt{1026132436081})/16126050304$ \\
\hline
\end{tabular}
\caption{Anomalous dimensions for the broken quasiprimary currents of spin 6 analogous to table \ref{tab:phi22HScurrents}.}
\label{tab:phi12HScurrents}
\end{table}
At spin 6, there is one current for $m = 3$ and two for $m \geq 4$. These are the same as \eqref{eq:t6m3} and \eqref{eq:t6m4}. A basis of operators which can recombine with these currents has three elements for $m = 3$ and four for $m \geq 4$. The building blocks again have unwieldy expressions for general $m$ so we have computed a representative set of anomalous dimensions which are in table \ref{tab:phi12HScurrents}.

\subsection{What about \texorpdfstring{$\phirs{2}{1}\chi$}{phi21chi}?}

Just as $\mathcal{O} = \phirs{1}{2} \chi$ defines a LRMM of type $(m,1,2)$, we can use $\mathcal{O} = \phirs{2}{1} \chi$ to define a LRMM of type $(m,2,1)$. It should be no surprise that these models are formally quite similar. In fact, without any calculation, large-$m$ results can be written down for one model as soon as they are known for the other. The reason for this is the transformation \eqref{eq:formalm} which implements the swapping of Kac indices. We saw in section \ref{sec:phi22flow-nearSR} that the anomalous dimensions \eqref{eq:phi22gammaFitResult} had definite symmetry properties under $(r,s,m) \leftrightarrow (s,r,-m)$ due to the fact that $\phirs{2}{2}$ is mapped to itself. In this section, the same transformation instead takes anomalous dimensions associated with $\phirs{1}{2}\chi$ to those associated with $\phirs{2}{1}\chi$. From \eqref{eq:phi12gammaFitResult}, we therefore conclude that the formula
\be\label{eq:phi21gammaFitResult}
\gamma_{r,s}(g_*) = \begin{cases}
\frac{1}{3}(r-s)^2\frac{r+(-1)^{s+r+1}}{r+(-1)^{s+r}}\delta+O(m^{-1}), & r\ge s \\
\frac{1}{3}(r-s)^2\frac{r+(-1)^{s+r}}{r+(-1)^{s+r+1}}\delta+O(m^{-1}), & r< s
\end{cases},
\ee
is valid for the $(m,2,1)$ LRMM.

Despite these formal similarities, it is worth pointing out that the $(m,2,1)$ LRMM is conceptually more exotic. This can be seen by using the same trick to compute the large-$m$ beta function. The result is
\be
\beta_3 = -\frac{3\pi ^2}{8}  (m+2+8 \log 2) + O(m^{-1})\,,
\ee
with the fixed point
\be
g_*^2/\delta = -\frac{8}{3\pi^2} \left ( \frac{1}{m} - \frac{2 + 8 \log 2}{m^2} \right ) + O(m^{-3})\,.
\ee
This shows that the fixed point is complex, and thus the corresponding CFT is nonunitary.\footnote{The anomalous dimensions of weakly broken currents \eqref{anom_spin2_RG12} and \eqref{anom_spin4_RG12} are positive for $\phi_{1,2}\chi$. They would be negative in $\phi_{2,1}\chi$ therefore giving a clear sign of unitarity violation. This is because they are proportional to an odd power of $m$ in the large-$m$ limit.} Although the result above  was obtained at large $m$, we have checked numerically that this fixed point is still not real at $m = 4$. When $m = 3$, $\D_{(2,1)} = 1$ and thus also $\chi^2$ becomes marginal at $\d=0$; moreover, the OPE gives $\phirs{2}{1}\times\phirs{2}{1}=\id$, implying $\cO\times\cO\supset \chi^2$, and thus an RG analysis involving two couplings is required.

%% file: sections/large-m.tex

\section{Analytic approaches to large \texorpdfstring{$m$}{m}}
\label{sec:large-m}

So far, our asymptotic results \eqref{eq:phi22FPlargem}, \eqref{eq:phi22gammaFitResult} have been obtained from a fit with input data computed at several large values of $m$. In each case, the closed form expression we quote for the slope is clearly conjectural since the fit has finite precision. In this section, we develop two methods for bypassing the fit and proving the asymptotic results directly. The first, in subsection \ref{sec:CGfirst}, introduces a multi-coupling RG flow which flows to the same fixed point as the $\phirs{1}{2}\chi$ flow. The second, in the remaining subsections, is applicable to the $\phirs{2}{2}\chi$ flow as well and involves a new mathematical toolkit.

\subsection{Multiple couplings}
\label{sec:CGfirst}

As discussed, for any perturbation of the type $\cO= \phirs{r}{s}\chi$, computing the two-loop beta function boils down to evaluating integrals of the type in \eqref{eq:beta3R}. We are interested in taking $m$ to be large but a subtlety arises if we do this before evaluating the integral.
This procedure is not valid, as the integral exhibits a UV divergence $\propto\log a$, and therefore needs further regularization.

Let us consider the simple case of $\cO= \phirs{1}{2}\chi$, for which we know that $\beta_3$ is of order $m$, in the large $m$ limit.
The four-point correlation function of $\cO$ is written in closed form in \eqref{phi12chi4pt}, and in the large $m$ limit we find:
\be
\la\cO(0) \cO(z, \bar{z}) \cO(1) \cO(\infty) \ra = \frac{(2+2 z \bar{z}-z-\bar{z}) \left(\frac{1}{(z \bar{z})^{3/2}}+\frac{1}{((1-z) (1-\bar{z}))^{3/2}}+1\right)}{2|z| \sqrt{1-z} \sqrt{1-\bar{z}}}+O(m^{-1})\,.
\ee
Along with the power-law divergent terms as $a\to 0$ (which we drop, as explained earlier), the integrated large-$m$ four-point function exhibits a logarithmic UV divergence, as announced:
\begin{align}
\beta_3 = -\pi \int_{\mathcal{R}} \text{d}^2z \la \mathcal{O}(0) \mathcal{O}(z, \bar{z}) \mathcal{O}(1) \mathcal{O}(\infty) \ra  =& -\frac{3\pi}{2} \log{a} +  \text{regular terms as $a\to 0$}\,.
\end{align}
(We recall that $\beta_3$ is itself the coefficient of a logarithmic divergence, so $\log a$ above really means $(\log a)^2$ in the perturbative expansion.) Before any discussion of the finite remainder can be meaningful, this divergence must be removed by a consistent renormalization of couplings at lower order, which is possible only if we turn on more couplings in the UV. (Analogous logarithmic divergences --- with somewhat more complicated numerical coefficients --- will generically appear for other choices of $\phirs{r}{s}\chi$ as well.)

In the present case, the consistent RG flow must involve turning on in the UV both $\phirs{1}{2}\chi$ and $\phirs{1}{3}$, the latter being weakly relevant at large $m$. We thus consider:\footnote{If we take $\delta=\frac{3}{2m}$, this RG flow furnishes an interesting class of conformal boundary conditions for the theory of a 3d free massless scalar field \cite{Behan:2021tcn}.}
\be\label{eq:flow13largem}
S'_{\text{LR},m} = S_{\text{SR},m} + \f{ \mathcal{N}}{2} \int \rmd^2 x_1 \rmd^2 x_2 \f{(\chi(x_1)-\chi(x_2))^2}{|x_1-x_2|^{2-s}} + g_0 \int \rmd^2 x  \, \phirs{1}{2} \chi+ h_0 \int \rmd^2 x  \, \phirs{1}{3}\,.
\ee

The beta functions at leading order are found to be:
\begin{align}\label{beta_minimal}
	\begin{split}
		\beta(h) =&-\frac{4}{m} h+\pi  h^2 C_{(1,3)(1,3)}^{(1,3)}+\pi  g^2  C_{(1,2)(1,2)}^{(1,3)}+O(h^3, h g^2)\,,\\
		\beta(g)=&-\delta g+2\pi  h g C_{(1,2)(1,3)}^{(1,2)}+O(g^3, h^2 g)\,,
	\end{split}
\end{align}
where, for the OPE coefficients, we use special cases of the formula from \eqref{bulkOPEcoeffrsrs13}
\be\label{structureC}
C_{(r,s)(r,s)}^{(1,3)} = \frac{(r - s)^2}{2\sqrt{3}} \frac{s + 1}{s - 1} + O(m^{-1})\,.
\ee

Our goal is to reproduce the results of the flow in section \ref{sec:phi12flow} which had a single coupling constant. In this case, we took $\delta \to 0$ first and then $m \to \infty$. For this reason, the limit we are interested in is
\be\label{double_scaling}
\delta \ll \frac{1}{m} \ll 1\,.
\ee
Solving for the zeros at one-loop and then taking the leading order for small $\delta m$,
we find two families of IR fixed points related by the $\chi \leftrightarrow -\chi$ (bulk) $\mathbb{Z}_2$ symmetry at:
\begin{align}\label{minimal_fixed_points}
	h_* =-\frac{\delta }{\sqrt{3} \pi } +O(\delta^2)\,, \quad  g_* = \pm \frac{1}{\pi}\sqrt{\frac{8 \d}{3 m}}+O(\delta^{3/2} m^{1/2})\,.
\end{align}
Another solution, besides $g_* = h_* = 0$, is the one that corresponds to the IR fixed point of the staircase flow between consecutive minimal models \cite{Zamolodchikov:1987ti,Cardy:1989da}, where $g_*=0$ and $h_* = -\frac{\sqrt{3}}{\pi m}$.

It is tempting to conjecture that \eqref{minimal_fixed_points} is the same as the IR fixed point of the finite-$m$ RG flow (induced by $\phirs{1}{2}\chi$) when \eqref{double_scaling} is satisfied. Indeed, by linearizing around the fixed point \eqref{minimal_fixed_points}, we find the following IR scaling dimensions for the leading UV-marginal singlet operators:
\be\label{stability_evs}
\D_{O}=2+2\delta+O(\delta^2)\,,\quad \D_{O'}=2-\f{4}{m}+\frac{2\delta}{3}+O(\delta^2)\,.
\ee
These are equal to the IR scaling dimensions of $\phirs{1}{2}\chi$ and $\phirs{1}{3}$ respectively, along the finite-$m$ RG flow induced by $\phirs{1}{2}\chi$ in the limit \eqref{double_scaling}. For $\phirs{1}{2}\chi$, this follows by computing $\beta'(g_*)$ for the beta function \eqref{eq:beta-general}, with coefficient $\b_3$ given in \eqref{eq:phi12betaFitResult}, and fixed point \eqref{eq:phi12FP}. For $\phirs{1}{3}$, this can be seen by using \eqref{eq:phi12gammaFitResult}.

As a further check, we will show that infinitely many anomalous dimensions of Virasoro primaries match between the two fixed points.\footnote{Checking that anomalous dimensions of spin-$\ell$ currents match should also be straightforward.} Since $\phirs{1}{3}$ can appear in the self-OPE of $\phirs{r}{s}$ but $\phirs{1}{2}\chi$ cannot, the one-loop anomalous dimension of a non-degenerate Virasoro primary is:
\be
\gamma_{r,s} = -2\pi h_* C^{(1,3)}_{(r,s)(r,s)} = \frac{\delta}{3} (r - s)^2 \frac{s + 1}{s - 1}\,.
\ee
This agrees with \eqref{eq:phi12gammaFitResult} when $s > r$ and $r + s$ is odd.\footnote{In particular, $\phirs{1}{2}$ gets an anomalous dimension of $\delta$ which means that the shadow relation holds.} Based on \eqref{stability_evs}, we should already expect that the disagreement for $r + s$ even occurs because of mixing. Indeed, $\phirs{r}{s}$ appears in the OPE of $\phirs{1}{2} \chi$ and $\phirs{r}{s-1} \chi$ and their dimensions obey the relation
\be
\D_{r,s-1} + \D_\chi - \D_{r,s} = 2 + r - s + O(m^{-1})\,.
\ee
If this integer is even, we can build the scalar
\be
M_{(r,s-1)} = \chi \mathcal{L}_{\frac{2 + r - s}{2}} \overline{\mathcal{L}_{\frac{2 + r - s}{2}}} \phirs{r}{s-1}\,,
\ee
and find degeneracy with $\phirs{r}{s}$.

It is possible, but more difficult, to write a multi-coupling RG flow --- like the one just discussed for $\phirs{1}{2}\chi$ --- to investigate systematically the large-$m$ behavior of other $\phirs{r}{s}\chi$ minimal models. This is partly because there is more large-$m$ degeneracy in the UV. For example, in the $\phirs{2}{2}\chi$ case, one would need to consider at least two more couplings at one loop: $\phirs{1}{3}$ and $\phirs{3}{1}$. There is also the possibility that higher loops will be needed since some of the OPE coefficients entering the beta function are suppressed at large $m$ (see Appendix \ref{app:mmcorrelators}).  In the following subsections, we will develop an alternative approach which sticks to a single-coupling RG flow, at the price of computing integrals at finite $m$ first.

\subsection{The advantages of Mellin space}
\label{sec:CGMellin}

Based on the discussion above, one might assume that our task is now very difficult.
This is because
integrated four-point functions at finite $m$ appear not to have closed form expressions. The way out is to be more flexible about which types of expressions we consider to be closed form. In particular, we will exploit the Mellin representation
\be
\mathcal{F}_{(r,s)(r',s')(r',s')(r,s)}(z, \bar{z}) = \int_{-i\infty}^{i\infty} \frac{\text{d}x \text{d}y}{(2\pi i)^2} |z|^x |1 - z|^y \mathcal{M}_{(r,s)(r',s')(r',s')(r,s)}(x, y)\,. \label{eq:mellin-schematic}
\ee
As long as we are careful about the contour, it is possible to first integrate $|z|^x |1 - z|^y$ over all of space, which is an easier calculation than integrating it over the region $\mathcal{R}$. Well developed Mellin space methods then exist to handle the remaining contour integrals.

Going back to \cite{m09,p10,fkprv11}, Mellin amplitudes are perhaps best known for the role they have played in studies of holographic CFTs in a large $N$ expansion. Nevertheless, they exist for general CFT correlators as well. This was rigorously established in \cite{psz19}, which also analyzed Mellin amplitudes for minimal model four-point functions in some detail. To obtain the ones we need, it will be important to use the so called Coulomb gas formalism of \cite{Dotsenko:1984nm,Dotsenko:1984ad,Dotsenko:1985hi}, which provides a convenient integral representation of either a multi-valued block or a single-valued correlator. The former are analogous to open string amplitudes, while the latter are analogous to closed string amplitudes. For all of the propagators that appear, the identities
\be
\frac{\Gamma(\D)}{(t_i - t_j)^\D} = \int_{-i\infty}^{i\infty} \frac{\text{d}\gamma}{2\pi i} \Gamma(-\gamma) \Gamma(\gamma + \D) t_i^\gamma (-t_j)^{-\D - \gamma}\,,
\ee
and
\be
\int \frac{\text{d}^2 t}{\pi} \prod_{i = 1}^n \frac{\Gamma(\D_i)}{|t - z_i|^{2\D_i}} = \prod_{i < j} \int_{-i\infty}^{i\infty} \frac{d \gamma_{ij}}{2\pi i} \Gamma(\gamma_{ij}) |z_{ij}|^{-2\gamma_{ij}}\,,
\ee
(subject to various conditions) may be used repeatedly in order to arrive at expressions that take the form of \eqref{eq:mellin-schematic}.

By doing conformal perturbation theory in this way, the large-$m$ limit becomes accessible. The step-by-step prescription is schematically summarized as follows:
\begin{enumerate}
	\item Write the integrand in the Mellin-Barnes representation
	\ba
	\lim_{m\rightarrow\infty}\int \rmd^2 z\, &\la\cO_{r',s'}(0)\cO_{r,s}(z,\zb)\cO_{r,s}(1)\cO_{r',s'}(\infty)\ra\\ 
	&\sim \lim_{m\rightarrow\infty}\int \rmd^2 z\, \int_{-i\infty}^{i\infty} \prod_k \dd x_k\, \mathcal{\tilde{M}}_{(r,s)(r',s')(r',s')(r,s)}(\abs{z},\abs{1-z},x_k).
	\ea
	\item Commute the contour integrals to the left, introducing regulators and deforming the contours when necessary to avoid colliding poles
	\be
	\lim_{m\rightarrow\infty}\int \rmd^2 z\, \textcolor{cyan!70!black}{\int_{-i\infty}^{i\infty} \prod_k \rmd x_k} \mapsto \left[\textcolor{cyan!70!black}{\int_{-i\infty}^{i\infty} \prod_k \rmd x_k}+\textcolor{cyan!70!black}{\sum \text{Res}}\right]\,\lim_{m\rightarrow\infty}\int \rmd^2 z.
	\ee 
	\item Compute first the integral over space, then expand in the large-$m$ limit, and finally compute the remaining contour integrals.
\end{enumerate}
Note that in contrast to the previous section, the limit is never commuted with the integral over space, avoiding the issue of the flow with multiple couplings. However, while commuting the contour integrals, the contour can be pinched by colliding poles. A systematic regulating procedure is thus necessary to separate those poles and keep track of their contributions.
The rest of this section will explain the technical steps involved with a focus on the coefficient $\beta_3$ for the $\phirs{1}{2}\chi$-flow and (with significantly greater complexity) the $\phirs{2}{2}\chi$-flow. We have also examined anomalous dimensions of the $\phirs{1}{2}\chi$-flow. Although it remains a challenge to treat all operators uniformly, our approach is sufficient to prove our conjectured $\gamma_{r,s}$ formula for any particular value of $(r, s)$.

\subsection{Coulomb gas review}
\label{sec:CGreview}

The Coulomb gas formalism was first used by \cite{Dotsenko:1984nm,Dotsenko:1984ad} to compute the structure constants of diagonal Virasoro minimal models $C_{(r_1, s_1)(r_2, s_2)(r_3, s_3)}$. While some expressions for these were used in sections \ref{sec:phi22flow-nearSR} and \ref{sec:phi12flow}, we will instead focus on the four-point functions entering conformal perturbation theory.
Our conventions will be those of \cite{Esterlis:2016psv} in the case $\alpha' = 4$. For more about the history and subtleties of the linear dilaton theory which underlies this approach, see \cite{km20}.

Consider a massless free scalar field in two dimensions which can be written as $\phi(z) + \bar{\phi}(\bar{z})$ on shell. With an additional topological term in the action, it is easy to modify this theory so that conformal transformations are generated by
\be
T(z) = -\frac{1}{4} (\partial \phi)^2(z) + i \alpha_0 \partial^2 \phi(z)\,. \label{stress-tensor}
\ee
This stress tensor determines the central charge
\be
c = 1 - 24\alpha_0^2\,,
\ee
and the conformal weights of operators involving exponentials of $\phi(z)$ and $\bar{\phi}(\bar{z})$. In this work, we will indicate normalization factors separately and use the notation
\be
V_\alpha(z) = e^{i\alpha \phi(z)}, \quad V_\alpha(z, \bar{z}) = e^{i\alpha [\phi(z) + \bar{\phi}(\bar{z})]}\,.
\ee
A straightforward calculation then shows that $V_{\alpha}(z)$ and $V_{2\alpha_0 - \alpha}(z)$ both have
\be
h = \alpha(\alpha - 2\alpha_0)\,,
\ee
while $V_{\alpha}(z, \bar{z})$ and $V_{2\alpha_0 - \alpha}(z, \bar{z})$ both have
\be
\D = 2\alpha(\alpha - 2\alpha_0)\,.
\ee
When using these exponentials (informally vertex operators) to represent Virasoro primaries of minimal models, it will be convenient that this can be done in two ways.

It is now important to discuss the neutrality condition, which is familiar from the free scalar CFT. The linear dilaton CFT having \eqref{stress-tensor} as a stress tensor is similar, except it requires us to place a background charge  $-2\alpha_0$ at infinity. As such, correlation functions of vertex operators are given by
\ba
\la V_{\alpha_1}(z_1) \dots V_{\alpha_n}(z_n) \ra &= \prod_{i < j} z_{ij}^{2\alpha_i \alpha_j} \delta_{\sum_k \alpha_k - 2\alpha_0, 0} \\
\la V_{\alpha_1}(z_1, \bar{z}_1) \dots V_{\alpha_n}(z_n, \bar{z}_n) \ra &= \prod_{i < j} |z_{ij}|^{4\alpha_i \alpha_j} \delta_{\sum_k \alpha_k - 2\alpha_0, 0}\,. \label{vertex-correlators}
\ea
This neutrality condition is rather limiting, but it can be relaxed if we deform the theory once more by inserting in correlators additional charged operators, integrated over some contour. Such nonlocal operators are known as screening charges. In order to preserve conformal invariance, the integrated operators must be exactly marginal and this fixes their charges to have one of two values,
\be
h = 1 \Rightarrow \alpha \in \left \{ \alpha_0 \pm \sqrt{\alpha_0^2 + 1} \right \}\,.
\ee
These solutions are denoted by $\alpha_\pm$, which satisfy
\be
\alpha_+ + \alpha_- = 2\alpha_0, \quad \alpha_+ \alpha_- = -1 \label{amp-relations}\,,
\ee
or
\be
\alpha_+ = \sqrt{\frac{m + 1}{m}}, \quad \alpha_- = -\sqrt{\frac{m}{m + 1}}\,,
\ee
in terms of the usual label for a minimal model. As a result, correlators in the Coulomb gas theory are non-zero if their total charge differs from $2\alpha_0$ by a linear combination of $\alpha_\pm$ with non-negative integer coefficients. Importantly, this means that all of the exponentials involved must have charges given by $\alpha_{r, s}$ or $2\alpha_0 - \alpha_{r,s}$ for
\be
\alpha_{r,s} = \frac{1 - r}{2}\alpha_+ + \frac{1 - s}{2}\alpha_-, \quad r, s \in \mathbb{N}\,,
\ee
which is precisely the statement that their conformal weights belong to the Kac table. To actually compute a correlation function, we integrate the positions of $V_{\alpha_\pm}(z)$ or $V_{\alpha_\pm}(z, \bar{z})$ insertions and then use \eqref{vertex-correlators}. We will see various examples in the following subsections.

For now, let us note that a Virasoro block can be computed from
\ba
& z^{h_{r_1, s_1} + h_{r_2, s_2}} \la V_{\alpha_{r_1,s_1}}(0) V_{\alpha_{r_2,s_2}}(z) V_{\alpha_{r_3,s_3}}(1) V_{2\alpha_0 - \alpha_{r_4,s_4}}(\infty) \ra \label{chiral-ex} \\
&= z^{h_{r_1, s_1} + h_{r_2, s_2} + 2\alpha_{r_1,s_1} \alpha_{r_2,s_2}} (1 - z)^{2\alpha_{r_2,s_2} \alpha_{r_3,s_3}} \int \prod_{i = 1}^{N_+} \text{d}s_i s_i^{2\alpha_+ \alpha_{r_1,s_1}} (s_i - z)^{2\alpha_+ \alpha_{r_2,s_2}} (1 - s_i)^{2\alpha_+ \alpha_{r_3,s_3}} \\
&\quad \times\int \prod_{j = 1}^{N_-} \text{d}t_j t_j^{2\alpha_- \alpha_{r_1,s_1}} (t_j - z)^{2\alpha_- \alpha_{r_2,s_2}} (1 - t_j)^{2\alpha_- \alpha_{r_3,s_3}} \prod_{i < k} s_{ik}^{2\alpha_+^2} \prod_{j < l} t_{jl}^{2\alpha_-^2} \prod_{m,n} (s_m - t_n)^{-2}\,,
\ea
in a two-step process. The first step is to choose a contour which consists of line segments running between the branch points of the integrand. Specifically, \cite{Dotsenko:1984nm,Dotsenko:1984ad} defined two bases for the homology of \eqref{chiral-ex} and showed that one yields four-point blocks in the $s$-channel while the other yields four-point blocks in the $t$-channel.\footnote{The change of basis matrix (or crossing kernel) was found in \cite{fgp90} using a sequence of contour deformations. One of its interesting properties is the appearance of a 6j-symbol for the quantum group $U_{q_+}(\mathfrak{sl}(2)) \times U_{q_-}(\mathfrak{sl}(2))$ with $q_\pm = e^{\pi i \alpha_\pm^2}$. This has a non-trivial origin since the global symmetry group of a generic minimal model is simply $\mathbb{Z}_2$.} The second step is to divide by a certain integral independent of $z$ so that the leading $z \to 0$ or $z \to 1$ asymptotics of \eqref{chiral-ex} become unit-normalized. Fortunately, this integral admits the closed-form expression
\ba
& \int_0^1 \prod_{i = 1}^{N_+} \frac{\text{d}s_i}{s_i^{-A} (1 - s_i)^{-B}} \int_0^1 \prod_{j = 1}^{N_-} \frac{\text{d}t_j}{t_j^{AC^{-1}} (1 - t_j)^{BC^{-1}}} \prod_{i < k} s_{ik}^{2C} \prod_{j < l} t_{jl}^{2C^{-1}} \prod_{m,n} (s_m - t_n)^{-2} \label{chiral-selberg} \\
&= \prod_{i = 0}^{N_+ - 1} \prod_{j = 0}^{N_- - 1} \frac{\Gamma((i + 1)C + 1) \Gamma((j + 1)C^{-1} + 1) (A + iC - j)^{-1} (B + iC - j)^{-1}}{\Gamma(C + 1) \Gamma(C^{-1} + 1) [A + B + C(N_+ + i - 1) - (N_- + j - 1)] [(i + 1)C - j - 1]} \\
&\quad\times \frac{\Gamma(1 + A + iC) \Gamma(1 + B + iC)}{\Gamma(2 - 2N_- + (N_+ + i - 1)C + A + B)} \frac{\Gamma(1 - AC^{-1} + jC^{-1}) \Gamma(1 - BC^{-1} + jC^{-1})}{\Gamma(2 - 2N_+ + (N_- + j - 1)C^{-1} - AC^{-1} - BC^{-1})}\,,
\ea
where the relation between $A, B, C$ and the parameters in \eqref{chiral-ex} depends on the choice of contour. In the special cases of $N_+ = 0$ or $N_- = 0$, \eqref{chiral-selberg} becomes the Selberg integral for $\mathfrak{sl}(2)$. Selberg integrals for higher-rank Lie algebras can be defined but almost none of them are known in closed form. This is the reason why most of the minimal model CFTs for $W_N$-algebras remain unsolved \cite{p09}.\footnote{Although W-algebra minimal models are most often studied with higher-rank generalizations of the Coulomb gas formalism, there are newer methods which appear to be promising. In particular, a semi-analytic bootstrap technique first developed for loop models in \cite{nrj23} was recently used to solve the Virasoro minimal models with partition functions given by the $E_6$ modular invariant \cite{nr25}.}

To compute a single-valued correlation function
\be
\la V_{\alpha_{r_1, s_1}}(z_1, \bar{z}_1) \dots V_{\alpha_{r_{n - 1}, s_{m - 1}}}(z_{n - 1}, \bar{z}_{n - 1}) V_{2\alpha_0 - \alpha_{r_n, s_n}}(z_n, \bar{z}_n) \ra\,,
\ee
there is no longer a contour ambiguity since the requisite number of screening charges $V_{\alpha_\pm}(z, \bar{z})$ are integrated over all of space. To normalize one of these integrals we divide it by factors of the two-point function\footnote{Notice that $V_{2\alpha_0}$ is degenerate with the identity and it is indeed a two-point function.}
\ba
& \la V_{\alpha_{r,s}}(0) V_{\alpha_{r,s}}(1) V_{2\alpha_0}(\infty) \ra = \int \prod_{i = 1}^{r - 1} \text{d}^2u_i (|u_i| |1 - u_i|)^{4\alpha_+ \alpha_{r,s}} i^{-1} \label{nonchiral-selberg} \\
& \quad \times \int \prod_{j = 1}^{s - 1} \text{d}^2v_j (|v_j| |1 - v_j|)^{4\alpha_- \alpha_{r,s}} j^{-1} \prod_{i < k} |u_{ik}|^{4\alpha_+^2} \prod_{j < l} |v_{jl}|^{4\alpha_-^2} \prod_{m, n} |u_m - v_n|^{-4} \\
&= \frac{\pi^{r + s - 2}}{\alpha_+^{8(r - 1)(s - 1)}} \prod_{i = 0}^{r - 2} \prod_{j = 0}^{s - 2} \frac{\Gamma(1 - \alpha_+^2) \Gamma(1 - \alpha_-^2) \Gamma((i + 1)\alpha_+^2) \Gamma((j + 1)\alpha_-^2 - s + 1)}{\Gamma(\alpha_+^2) \Gamma(\alpha_-^2) \Gamma(1 - (i + 1)\alpha_+^2) \Gamma(1 - (j + 1)\alpha_-^2 + s - 1)} \\
&\quad \times \frac{\Gamma(2 - s + 2\alpha_- \alpha_{r,s} + i\alpha_-^2)^2 \Gamma(s - 4\alpha_- \alpha_{r,s} - (r + i - 2)\alpha_-^2)}{\Gamma(s - 1 - 2\alpha_- \alpha_{r,s} - i\alpha_-^2)^2 \Gamma(3 - s + 4\alpha_- \alpha_{r,s} + (r + i - 2)\alpha_-^2)} \\
&\quad \times \frac{\Gamma(1 + 2\alpha_+ \alpha_{r,s} + j\alpha_+^2)^2 \Gamma(2r - 3 - 4\alpha_+ \alpha_{r,s} - (s + j - 2)\alpha_+^2)}{\Gamma(-2\alpha_+ \alpha_{r,s} - j\alpha_+^2)^2 \Gamma(4 - 2r + 4\alpha_+ \alpha_{r,s} + (s + j - 2)\alpha_+^2)}\,,
\ea
where the relations \eqref{amp-relations} hide the symmetry under $(\alpha_+, r) \leftrightarrow (\alpha_-, s)$. Once again the existence of a closed-form expression here is crucial. A numerical computation of \eqref{nonchiral-selberg} would not work because the integral is divergent and needs to be defined through analytic continuation.

Before moving on, it is worth noting that various authors have introduced modified Coulomb gas formalisms since the work of \cite{Dotsenko:1984nm,Dotsenko:1984ad,Dotsenko:1985hi}. The motivation for doing so is to enable exact calculations in other classes of CFTs. These include the aformentioned minimal models for W-algebras and also critical $q$-state Potts models for general $q$ \cite{dsz87}. Very recently, \cite{ggqzz24} developed a Coulomb gas formalism which can be used to solve a family of CFTs having the quantum group $U_q(\mathfrak{sl}(2))$ as a genuine global symmetry.

\subsection{Warm-up in the chiral case}
\label{sec:CGeasy}

As a first application of these techniques, let us consider the problem of expanding Virasoro blocks for minimal models at large $m$. The simplest ones to study are the two blocks that appear in a four-point function of $\phirs{1}{2}$ operators, namely $\mathcal{F}_{(1,1)}^{(1,2)(1,2)(1,2)(1,2)}(z)$ and $\mathcal{F}_{(1,3)}^{(1,2)(1,2)(1,2)(1,2)}(z)$. As reviewed in Appendix \ref{app:mmcorrelators}, we already know that both of these blocks may be expressed as hypergeometric functions but the point we would like to illustrate is that one does not need to rely on this fact. The Coulomb gas formalism applies to all Virasoro blocks and it makes the large-$m$ expansion transparent once it is combined with the Mellin representation. The identity we will use to go to Mellin space is
\be
\frac{1}{(t_i - t_j)^\D} = \frac{1}{\Gamma(\D)} \int_{c -i\infty}^{c +i\infty} \frac{\text{d}\gamma}{2\pi i} \Gamma(-\gamma) \Gamma(\gamma + \D) t_i^\gamma (-t_j)^{-\D - \gamma} \label{chiral-mellin}\,,
\ee
where the contour separates the poles of $\Gamma(-\gamma)$ from those of $\Gamma(\gamma + \D)$. This is possible with a straight contour of real part $c\in(-\D,0)$ if and only if $\D > 0$.
The case of noninteger $\D<0$ can be reached by analytic continuation, which corresponds to taking a contour slaloming between the intertwined poles of the two gamma functions. Lastly, $\D\in-\mathbb{N}_0$ can be reached as a limit from one of the previous cases.

Since the external operators of interest are all $\phirs{1}{2}$, \eqref{chiral-ex} reduces to a single integral and we can consider the contours $t \in (0, z)$ and $t \in (1, \infty)$. These contours correspond to $\mathcal{F}_{(1,1)}^{(1,2)(1,2)(1,2)(1,2)}(z)$ and $\mathcal{F}_{(1,3)}^{(1,2)(1,2)(1,2)(1,2)}(z)$, respectively, because the first line of
\ba
\mathcal{F}_{(1,1)}^{(1,2)(1,2)(1,2)(1,2)}(z) &\sim z^{2\alpha_-^2 - 1} (1 - z)^{\alpha_-^2/2} \int_0^z \text{d}t [t(z - t)(1 - t)]^{-\alpha_-^2} \\
\mathcal{F}_{(1,3)}^{(1,2)(1,2)(1,2)(1,2)}(z) &\sim z^{2\alpha_-^2 - 1} (1 - z)^{\alpha_-^2/2} \int_1^\infty \text{d}t [t(t - z)(t - 1)]^{-\alpha_-^2}\,,
\ea
has $O(z^{h_{1,1}})$ scaling as $z \to 0$, while the second line has $O(z^{h_{1,3}})$ scaling. However, the $z \to 0$ limit of these integrals tells us more than just the scaling. It also tells us the prefactor. The prefactor of the first line is \eqref{chiral-selberg} for $A = B = 1, C = \alpha_-^{-2}$ while that of the second line is \eqref{chiral-selberg} for $A = 2\alpha_-^{-2} - 3, B = 1, C = \alpha_-^2$. We can therefore be careful about normalizations, to find
\ba
\mathcal{F}_{(1,1)}^{(1,2)(1,2)(1,2)(1,2)}(z) &= z^{2\alpha_-^2 - 1} (1 - z)^{\alpha_-^2/2} \frac{\Gamma(2 - 2\alpha_-^2)}{\Gamma(1 - \alpha_-^2)^2} \int_0^z \text{d}t [t(z - t)(1 - t)]^{-\alpha_-^2} \label{vir-blocks-normalized} \\
\mathcal{F}_{(1,3)}^{(1,2)(1,2)(1,2)(1,2)}(z) &= z^{2\alpha_-^2 - 1} (1 - z)^{\alpha_-^2/2} \frac{\Gamma(2\alpha_-^2)}{\Gamma(1 - \alpha_-^2) \Gamma(3\alpha_-^2 - 1)} \int_1^\infty \text{d}t [t(t - z)(t - 1)]^{-\alpha_-^2}\,.
\ea
Note that the limit $m \to \infty$, which corresponds to $\alpha_\pm \to \pm 1$, turns the integrals of \eqref{vir-blocks-normalized} into local Feynman integrals.

\begin{figure}[h]
\centering
\includegraphics[scale=0.7]{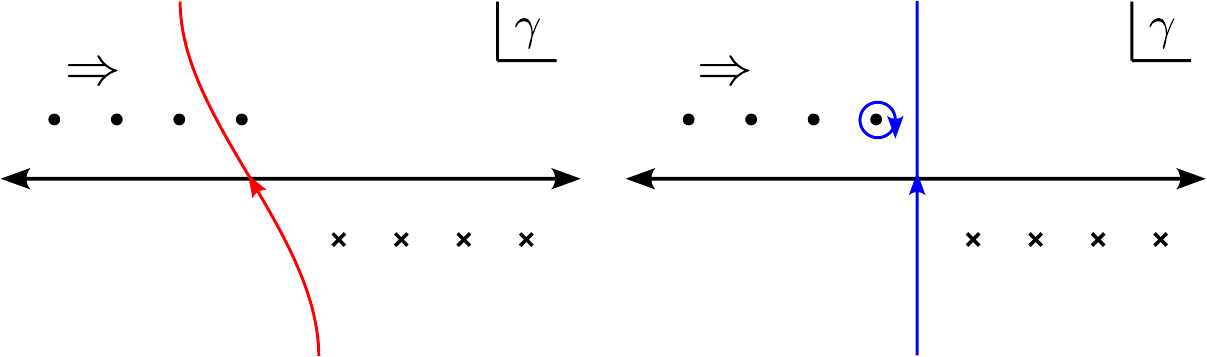}
\caption{Sequences of poles for both lines of \eqref{vir-blocks-mellin} which are offset from the real axis for clarity. The rightmost dot is $\alpha_-^2 - 1$ and the leftmost cross is $0$. The red contour is completely free of singular behaviour in the $\alpha_-^2 \to 1$ limit (double arrow) but it is not the natural Mellin-Barnes contour (the blue one). The difference is a simple residue at the rightmost dot.}
\label{fig:deform}
\end{figure}

The key step now is passing to the representation \eqref{chiral-mellin} which trivializes the integral over $t$. The proper values of $t_i$ and $t_j$ here are chosen so as to avoid branch points. For the first line of \eqref{vir-blocks-normalized}, we take $t_i = z - t$ and $t_j = z - 1$. This turns $t_{ij}$ into $1 - t$ which is the unique factor from the integrand which does not vanish anywhere on the contour.
In the second line, it is convenient to change variables $t \mapsto t^{-1}$ followed by taking $t_i = 1 - t$ and $t_j = 1 - z^{-1}$.
All in all,
\begin{align}
\mathcal{F}_{(1,1)}^{(1,2)(1,2)(1,2)(1,2)}(z) &= (1 - z)^{-\alpha_-^2/2} \frac{\Gamma(2 - 2\alpha_-^2)}{\Gamma(\alpha_-^2) \Gamma(1 - \alpha_-^2)} \int_{-i\infty}^{i\infty} \frac{\text{d}\gamma}{2\pi i} \left ( \frac{z}{1 - z} \right )^\gamma \frac{\Gamma(-\gamma) \Gamma(\alpha_-^2 + \gamma) \Gamma(1 - \alpha_-^2 + \gamma)}{\Gamma(2 - 2\alpha_-^2 + \gamma)} \nonumber \\
\mathcal{F}_{(1,3)}^{(1,2)(1,2)(1,2)(1,2)}(z) &= z^{2\alpha_-^2 - 1} (1 - z)^{-\alpha_-^2/2} \frac{\Gamma(2\alpha_-^2)}{\Gamma(1 - \alpha_-^2)} \int_{-i\infty}^{i\infty} \frac{\text{d}\gamma}{2\pi i} \left ( \frac{z}{1 - z} \right )^\gamma \frac{\Gamma(-\gamma) \Gamma(\alpha_-^2 + \gamma) \Gamma(1 - \alpha_-^2 + \gamma)}{\Gamma(2\alpha_-^2 + \gamma)}\,. \label{vir-blocks-mellin}
\end{align}
Notice that setting $\alpha_-^2 = 1$ before integration to extract the leading large-$m$ asymptotics is incorrect because this forces a pinching of the contour: the leading poles of $\Gamma(-\gamma) \Gamma(1 - \alpha_-^2 + \gamma)$ approach each other in this limit.
Starting from $\alpha_-^2<1$, the resolution is to move the contour past one of the poles so that once the $\alpha_-^2 \to 1$ limit is taken, the above pinching no longer occurs. This leads to a crucial residue as shown in Figure \ref{fig:deform}.
Taking this into account for the first line of \eqref{vir-blocks-mellin}, the result is\footnote{Notice that the residue at $\gamma=0$ of $\frac{\Gamma(-\gamma) \Gamma(\alpha_-^2 + \gamma) \Gamma(1 - \alpha_-^2 + \gamma)}{\Gamma(2 - 2\alpha_-^2 + \gamma)} $ is $\frac{ \Gamma(\alpha_-^2 ) \Gamma(1 - \alpha_-^2 )}{\Gamma(2 - 2\alpha_-^2 )} $, which goes to $2$ in the limit $\alpha_-^2 \to 1$. However, if we naively take the limit first in the integrand we find $\Gamma(-\gamma) \Gamma(1 + \gamma)$, whose residue at $\gamma=0$ is $1$. The mismatch is due to the pinching of the contour, and as explained the correct way of taking the limit in the integrand is to first move the contour past the pole at $\alpha_-^2 - 1$. As evident from \eqref{vir-expanded1}, this provides the missing half of the residue at $\gamma=0$.}
\ba
\mathcal{F}_{(1,1)}^{(1,2)(1,2)(1,2)(1,2)}(z) &= (1 - z)^{-\alpha_-^2/2} \frac{\Gamma(2 - 2\alpha_-^2)}{\Gamma(\alpha_-^2) \Gamma(1 - \alpha_-^2)} \left [ \underset{\alpha_-^2 - 1}{\text{Res}} + \int_{\Re \gamma \in (-1, 0)} \frac{\text{d}\gamma}{2\pi i} \right ] \label{vir-expanded1} \\
& \left ( \frac{z}{1 - z} \right )^\gamma \frac{\Gamma(-\gamma) \Gamma(\alpha_-^2 + \gamma) \Gamma(1 - \alpha_-^2 + \gamma)}{\Gamma(2 - 2\alpha_-^2 + \gamma)} \\
&= (1 - z)^{-\alpha_-^2/2} \frac{\Gamma(2 - 2\alpha_-^2)}{\Gamma(\alpha_-^2) \Gamma(1 - \alpha_-^2)} \left [ 1 - \frac{\log z - \log(1 - z) - 2\gamma_E}{m} \right . \\
&\left. + \int_{-i\infty}^{i\infty} \frac{\text{d}\gamma}{2\pi i} \left ( \frac{z}{1 - z} \right )^\gamma \Gamma(-\gamma) \Gamma(1 + \gamma) \left ( 1 - \frac{\psi(\gamma) + \psi(1 + \gamma)}{m} \right ) \right ] + O(m^{-2})\,.
\ea
Both the integral and the residue in the first line can be safely expanded at large $m$ and this yields a new integral whose contour is once again the standard Mellin-Barnes one. This does not occur in the remaining block but otherwise it is handled in the same way:
\ba
\mathcal{F}_{(1,3)}^{(1,2)(1,2)(1,2)(1,2)}(z) &= z^{2\alpha_-^2 - 1} (1 - z)^{-\alpha_-^2/2} \frac{\Gamma(2\alpha_-^2)}{\Gamma(1 - \alpha_-^2)} \left [ \underset{\gamma = \alpha_-^2 - 1}{\text{Res}} + \int_{\Re \gamma \in (-1, 0)} \frac{\text{d}\gamma}{2\pi i} \right ] \label{vir-expanded2} \\
& \left ( \frac{z}{1 - z} \right )^\gamma \frac{\Gamma(-\gamma) \Gamma(\alpha_-^2 + \gamma) \Gamma(1 - \alpha_-^2 + \gamma)}{\Gamma(2\alpha_-^2 + \gamma)} \\
&= z^{2\alpha_-^2 - 1} (1 - z)^{-\alpha_-^2/2} \frac{\Gamma(2\alpha_-^2)}{\Gamma(1 - \alpha_-^2)} \left [ m + 4 - \log \left ( \frac{z}{1 - z} \right ) - 2\gamma_E \right. \\
& \left. - \int_{\Re \gamma \in (-1, 0)} \frac{\text{d}\gamma}{2\pi i} \left ( \frac{z}{1 - z} \right )^\gamma \Gamma(-1 - \gamma) \Gamma(\gamma) \right ] + O(m^{-2})\,.
\ea
Evaluating the integrals in \eqref{vir-expanded1} and \eqref{vir-expanded2} produces
\ba
\mathcal{F}_{(1,1)}^{(1,2)(1,2)(1,2)(1,2)}(z) &= \frac{2 - z}{2\sqrt{1 - z}} + \frac{3z \log(1 - z)}{4m \sqrt{1 - z}} + O(m^{-2}) \label{vir-expanded3} \\
\mathcal{F}_{(1,3)}^{(1,2)(1,2)(1,2)(1,2)}(z) &= \frac{z}{\sqrt{1 - z}} + \frac{(z + 2)\log(1 - z) + 2z(3 - 4\log z)}{2m \sqrt{1 - z}} + O(m^{-2})\,.
\ea
The leading terms here are easy to derive using the hypergeometric functions in Appendix \ref{app:mmcorrelators} but the subleading terms are less trivial.\footnote{For a very different approach to this problem, see \cite{1303.3015} which performed a large-$m$ expansion of Virasoro blocks using the AGT relation \cite{0906.3219}.}

\subsection{The \texorpdfstring{$\phirs{1}{2}\chi$}{phi12chi} beta function}
\label{sec:CGmedium}

We now turn to the harder task of expanding integrated correlators of $\mathcal{O} = \phirs{1}{2}\chi$. The best approach here is to bypass the use of Virasoro blocks, which the chiral Coulomb gas formalism computes, and to instead build correlators directly from the non-chiral Coulomb gas. Once this is done, we can then proceed analogously by going to Mellin space and deforming the contours. It will be crucial to use Symanzik's formula \cite{s72}, which applies to star-like integrations that have $n>3$ legs with exponents satisfying $\sum_{i = 1}^n \D_i = 2$. It states that
\be
\int \frac{\text{d}^2 t}{\pi} \prod_{i = 1}^n \frac{\Gamma(\D_i)}{|t - z_i|^{2\D_i}} = \prod_{i < j} \int_{-i\infty}^{i\infty} \frac{d \gamma_{ij}}{2\pi i} \Gamma(\gamma_{ij}) |z_{ij}|^{-2\gamma_{ij}} \label{symanzik}\,,
\ee
where the integration variables $\gamma_{ij}$ satisfy
\be
\gamma_{ii} = 0, \quad \gamma_{ij} = \gamma_{ji}, \quad \sum_{j \neq i} \gamma_{ij} = \D_i \label{symanzik-rels}\,,
\ee
leaving $n(n-3)/2$ independent variables, and their contours separate sequences of poles which increase to the right from sequences of poles which increase to the left. The ability to do this with a straight contour (one which gives all gamma arguments in the measure a positive real part) is again not guaranteed. If it is possible however, the relations \eqref{symanzik-rels} imply that all exponents satisfy $0 < \D_i < 1$. This is precisely the condition for the original integral to converge without needing analytic continuation.

Before involving the four-point function for $\chi$, let us show that we can write down a four-point function for $\phirs{1}{2}$ in Mellin space which agrees with the result of \cite{psz19}. As should be familiar, we will use the charge $\alpha_{1,2}$ thrice and $2\alpha_0 - \alpha_{1,2}$ once in order to get a Coulomb gas representation with a single screening charge. Integrating its position and using \eqref{nonchiral-selberg} to normalize, we have
\begin{align}
\la \phirs{1}{2}(0) \phirs{1}{2}(z, \bar{z}) \phirs{1}{2}(1) \phirs{1}{2}(\infty) \ra &= \frac{\la V_{\alpha_{1,2}}(0) V_{\alpha_{1,2}}(z, \bar{z}) V_{\alpha_{1,2}}(1) V_{2\alpha_0 - \alpha_{1,2}}(\infty) \ra}{\la V_{\alpha_{1,2}}(0) V_{\alpha_{1,2}}(1) V_{2\alpha_0}(\infty) \ra} \\
&= [|z| |1 - z|]^{\alpha_-^2} \frac{\Gamma(\alpha_-^2)^2 \Gamma(2 - 2\alpha_-^2)}{\Gamma(1 - \alpha_-^2)^2 \Gamma(2\alpha_-^2 - 1)} \int \frac{\text{d}^2t}{\pi} [|t| |z - t| |1 - t|]^{-2\alpha_-^2}\,. \nonumber
\end{align}
If we were to move the point at infinity to a finite point, we would see that the exponents add up to $2$ which is a simple consequence of the screening charges being marginal. We can therefore use \eqref{symanzik} and solve for the $\gamma_{ij}$ as
\ba
& \gamma_{34} = -\frac{x}{2} \quad \gamma_{12} = 2\alpha_-^2 - 1 - \frac{x}{2} \quad \gamma_{13} = 1 - \alpha_-^2 + \frac{x + y}{2} \\
& \gamma_{14} = -\frac{y}{2} \quad \gamma_{23} = 2\alpha_-^2 - 1 - \frac{y}{2} \quad \gamma_{24} = 2 - 3\alpha_-^2 + \frac{x + y}{2}\,.
\ea
The resulting Mellin integral is therefore
\begin{align}
&\la \phirs{1}{2}(0) \phirs{1}{2}(z, \bar{z}) \phirs{1}{2}(1) \phirs{1}{2}(\infty) \ra = \frac{[|z| |1 - z|]^{2 - 3\alpha_-^2} \Gamma(2 - 2\alpha_-^2)}{\Gamma(\alpha_-^2) \Gamma(1 - \alpha_-^2)^2 \Gamma(2\alpha_-^2 - 1) \Gamma(2 - 3\alpha_-^2)} \label{nonchiral-mellin1} \\
& \int_{-i\infty}^{i\infty} \frac{\text{d}x \text{d}y}{(4\pi i)^2} |z|^x |1 - z|^y \Gamma(-\tfrac{x}{2}) \Gamma(-\tfrac{y}{2}) \Gamma(2\alpha_-^2 - 1 - \tfrac{x}{2}) \Gamma(2\alpha_-^2 - 1 - \tfrac{y}{2}) \Gamma(1 - \alpha_-^2 + \tfrac{x + y}{2}) \Gamma(2 - 3\alpha_-^2 + \tfrac{x + y}{2})\,. \nonumber
\end{align}
Since this has two Mandelstam-like variables instead of one, we should expect some complications as compared to the previous subsection.

It is now time to multiply by
\be
\la \chi(0) \chi(z, \bar{z}) \chi(1) \chi(\infty) \ra = 1 + |z|^{3\alpha_-^2 - 6} + |1 - z|^{3\alpha_-^2 - 6} \label{gff-part}\,,
\ee
but the subsequent integration reveals a problem. Although there exists a value of $\alpha_-^2$ which allows the pole sequences of \eqref{nonchiral-mellin1} to be separated by a straight contour, this is no longer true after we integrate. What we will need to do is shift the exponents of \eqref{gff-part} by additional regulator parameters so that we may keep track of them as they approach zero.\footnote{A less efficient but more conceptually pleasing approach is to analytically continue the charges of the external operators away from $\alpha_{1,2}$ such that the total charge remains $2\alpha_0 - \alpha_-$. This makes it possible to choose precise values such that no relevant operators are exchanged in any channel which is the condition for the integral to be finite.} The first term in \eqref{gff-part} for instance can be written as $[|z| |1 - z|]^{-2\delta}$ with $\delta \to 0$.\footnote{Note that this $\delta$ is a regulator parameter and should not be confused with the $\delta$ appearing in the classical term of the beta function.} Integrating just this piece shows that
\begin{align}
& I_{1,2} = \frac{\pi \Gamma(2 - 2\alpha_-^2)}{4 \Gamma(\alpha_-^2) \Gamma(1 - \alpha_-^2)^2 \Gamma(2\alpha_-^2 - 1) \Gamma(2 - 3\alpha_-^2)} \label{nonchiral-mellin2} \\
& \int_{-i\infty}^{i\infty} \frac{\text{d}x \text{d}y}{(2\pi i)^2} \Gamma(-\tfrac{x}{2}) \Gamma(-\tfrac{y}{2}) \Gamma(2\alpha_-^2 - 1 - \tfrac{x}{2}) \Gamma(2\alpha_-^2 - 1 - \tfrac{y}{2}) \Gamma(1 - \alpha_-^2 + \tfrac{x + y}{2}) \Gamma(2 - 3\alpha_-^2 + \tfrac{x + y}{2}) \nonumber \\
& \Gamma(\tfrac{4 - 3\alpha_-^2 - 2\delta + x}{2}) \Gamma(\tfrac{4 - 3\alpha_-^2 - 2\delta + y}{2}) \Gamma(\tfrac{6\alpha_-^2 - 6 + 4\delta - x - y}{2}) \Gamma(\tfrac{3\alpha_-^2 - 2 + 2\delta - x}{2})^{-1} \Gamma(\tfrac{3\alpha_-^2 - 2 + 2\delta - y}{2})^{-1} \Gamma(\tfrac{8 - 6\alpha_-^2 + 4\delta + x + y}{2})^{-1}\,, \nonumber
\end{align}
is the first of the three integrals we need to expand at large $m$. As it turns out, all three parts of \eqref{gff-part} contribute equally thus allowing us to define
\be
\beta_3 = -3 \frac{2\pi}{3!} I_{1,2}\,.
\ee
As we are integrating over all of space here, let us stress that the factor of $3$ in front does not refer to three copies of the region $\mathcal{R}$ but the three types of Wick contractions for generalized free fields.

The large-$m$ expansion we are seeking is now part of a double expansion with $\delta$ and $1 - \alpha_-^2$ both approaching zero from above. To begin the analysis of it, let us recall that the Virasoro block expressions \eqref{vir-expanded1} and \eqref{vir-expanded2} are both obtained from a certain operator acting on the original integrand --- a residue and an integral over a new contour which makes the $\alpha_-^2 \to 1$ limit safe to take. We can determine the analogous operator for $\delta \to 0$ here by identifying the gamma functions from \eqref{nonchiral-mellin2} which have colliding poles pinching the contour in this limit. Looking at the $y$ variable first, a simple check returns the sequences
\be
y = 6\alpha_-^2 - 4 - x - 2m, \quad y = 6\alpha_-^2 - 6 + 4\delta - x + 2n\,,
\ee
for non-negative integers $m$ and $n$. The parameter $\delta$ is what prevents these sequences from having overlap at $(m, n) \in \{(0, 1), (1, 0)\}$. We therefore need two residues to move the contour out of the region where poles can collide. As a result, the most convenient rewriting of $I_{1,2}$ is the one that has
\be
-\int_{-i\infty}^{i\infty} \frac{\text{d}x}{2\pi i} \left ( \underset{y = 6\alpha_-^2 - 6 + 4\delta - x}{\text{Res}} + \underset{y = 6\alpha_-^2 - 4 + 4\delta - x}{\text{Res}} \right ) + \int_{\Re (x + y - 6\alpha_-^2) \in (-8, -6)} \frac{\text{d}x \text{d}y}{(2\pi i)^2} \label{nonchiral-operator}\,,
\ee
acting on the integrand of \eqref{nonchiral-mellin2}. Another language for what we have done is that we have ``merged'' the sixth and ninth gamma functions by adding their arguments according to
\be
\Gamma_6 \left ( 2 - 3\alpha_-^2 + \frac{x + y}{2} \right ) \Gamma_9 \left ( 3\alpha_-^2 - 3 + 2\delta - \frac{x + y}{2} \right ) \to \Gamma_{\{6,9\}} \left ( 2\delta - 1 \right )\,. \label{first-merging}
\ee
The two residues are explained by the fact that the $\delta \to 0$ limit lands one unit away from the leading pole of the new ``effective gamma function''. To locate the singularities that remain in \eqref{nonchiral-operator}, it will be helpful to proceed in this way.

Fortunately, this approach was spelled out explicitly in \cite{y18} which developed an algorithm for extracting singularities from multi-dimensional Mellin-Barnes integrals. This algorithm will be especially useful for the second term of \eqref{nonchiral-operator} which is a double integral but we will warm up by applying it to the single integral first. Starting with this easier term, we can compute both residues and then take the constant term in the small $\delta$ expansion to arrive at
\ba
I'_{1,2} &= \frac{1}{2} \Gamma \left ( -\frac{x}{2} \right ) \Gamma \left ( 2\alpha_-^2 - 1 - \frac{x}{2} \right ) \Gamma \left ( 1 - \alpha_-^2 + \frac{x}{2} \right ) \Gamma \left ( 2 - 3\alpha_-^2 + \frac{x}{2} \right ) \Gamma(2\alpha_-^2 - 2) \\
& \left [ \left ( (3 - 2\alpha_-^2)x^2 + 14(\alpha_-^2 - 1)^2 x - 2\alpha_-^4 x - 2(\alpha_-^2 - 1)(3\alpha_-^2 - 2)(3\alpha_-^2 - 4) \right ) \right. \\
& \left. \times \left ( H_{\frac{x}{2} - \alpha_-^2} + H_{\frac{x}{2} - 3\alpha_-^2 + 1} - H_{2\alpha_-^2 - 3} \right ) + 2(3 - \alpha_-^2)x + 9\alpha_-^4 - 28\alpha_-^2 + 16 \right ]\,.
\ea
The important part of this long expression is simply the factor in front with four gamma functions. The other factor does not introduce new singularities. Following \cite{y18,psz19}, we will label the important gamma functions as $\Gamma_1$ through $\Gamma_4$ in the order shown and consider all possible sums of their arguments such that the dependence on $x$ disappears. We arrive at
\be
\Gamma_{\{1,3\}}(1 - \alpha_-^2) \Gamma_{\{1,4\}}(2 - 3\alpha_-^2) \Gamma_{\{2,4\}}(1 - \alpha_-^2) \Gamma_{\{2,3\}}(\alpha_-^2)\,,
\ee
where a subscript indicates the two gamma functions that were merged. We can go through the meaning of these four mergings in order.
\begin{enumerate}
\item Taking the $x = 0$ residue from $\Gamma_1$ will make $\Gamma_3$ encounter a leading pole at $\alpha_-^2 = 1$.
\item The $x = 0$ residue from $\Gamma_1$ will give $\Gamma_4$ a subleading pole while $x = 2$ will give it a leading pole.
\item The $x = 4\alpha_-^2 - 2$ residue from $\Gamma_2$ will give $\Gamma_3$ a leading pole.
\item No residues at poles resulting from $\Gamma_2$ will make $\Gamma_4$ encounter a pole at $\alpha_-^2 = 1$.
\end{enumerate}
The content of the above is that we need to take
\be
-\left [ \underset{x = 0}{\text{Res}} + \underset{x = 2}{\text{Res}} + \underset{x = 4\alpha_-^2 - 2}{\text{Res}} \right ] I'_{1,2} = -\frac{1}{3} m^3 + O(m^2)\,,
\ee
where we have used $(\alpha_-^2 - 1)^{-3} = -m^3 + O(m^2)$. Going back to the original integral,
\be
I_{1,2} = \frac{\pi \Gamma(2 - 2\alpha_-^2)}{4 \Gamma(\alpha_-^2) \Gamma(1 - \alpha_-^2)^2 \Gamma(2\alpha_-^2 - 1) \Gamma(2 - 3\alpha_-^2)} \left [ -\frac{1}{3} m^3 + \dots \right ] \label{final-before}\,,
\ee
where the ellipsis denotes not just $O(m^2)$ contributions but also $O(m^3)$ contributions from the second term in \eqref{nonchiral-operator}. These will be worked out next.

The power of this merging procedure will be demonstrated most convincingly for the second term in \eqref{nonchiral-operator}. This is the integral from \eqref{nonchiral-mellin2} at $\delta = 0$ except over a shifted contour. We will pair up gamma functions that have $x$ appearing with opposite signs and then do the same for $y$. Since merging now happens in two iterations, there is a rule from \cite{y18} which normally needs to be applied in order to filter out fake poles. It states that an effective gamma function should be deleted if its label contains the label of some other effective gamma function as a proper subset. This is part of what ensures that final results are independent of whether we start with $x$ or $y$. Given the functions $\Gamma_1$ through $\Gamma_9$ from the numerator of \eqref{nonchiral-mellin2} (in the order shown there), it is easy to verify that
\be
\Gamma_{\{1,5,7,9\}} ( \tfrac{\alpha_-^2}{2} ) \Gamma_{\{1,6,7,9\}} ( 1 - \tfrac{3\alpha_-^2}{2} ) \Gamma_{\{3,5,7,9\}} ( \tfrac{5\alpha_-^2}{2} - 1 ) \Gamma_{\{3,6,7,9\}} ( \tfrac{\alpha_-^2}{2} )\,,
\ee
need to be removed, leaving us with
\ba
& \Gamma_{\{1,7\}} ( 2 - \tfrac{3\alpha_-^2}{2} ) \Gamma_{\{2,8\}} ( 2 - \tfrac{3\alpha_-^2}{2} ) \Gamma_{\{3,7\}} ( 1 + \tfrac{\alpha_-^2}{2} ) \Gamma_{\{4,8\}} ( 1 + \tfrac{\alpha_-^2}{2} ) \label{kept-gammas} \\
& \times\Gamma_{\{1,4,5\}}(\alpha_-^2) \Gamma_{\{2,3,5\}}(\alpha_-^2) \Gamma_{\{3,4,6\}}(\alpha_-^2) \Gamma_{\{3,4,5\}}(3\alpha_-^2 - 1) \Gamma_{\{7,8,9\}}(1) \\
& \times\Gamma_{\{1,2,5\}}(1 - \alpha_-^2) \Gamma_{\{1,4,6\}}(1 - \alpha_-^2) \Gamma_{\{2,3,6\}}(1 - \alpha_-^2) \Gamma_{\{1,2,6\}}(3 - 2\alpha_-^2) \Gamma_{\{5,9\}}(2\alpha_-^2 - 2) \Gamma_{\{6,9\}}(-1)\,.
\ea
\begin{figure}[h]
\centering
\includegraphics[scale=0.56]{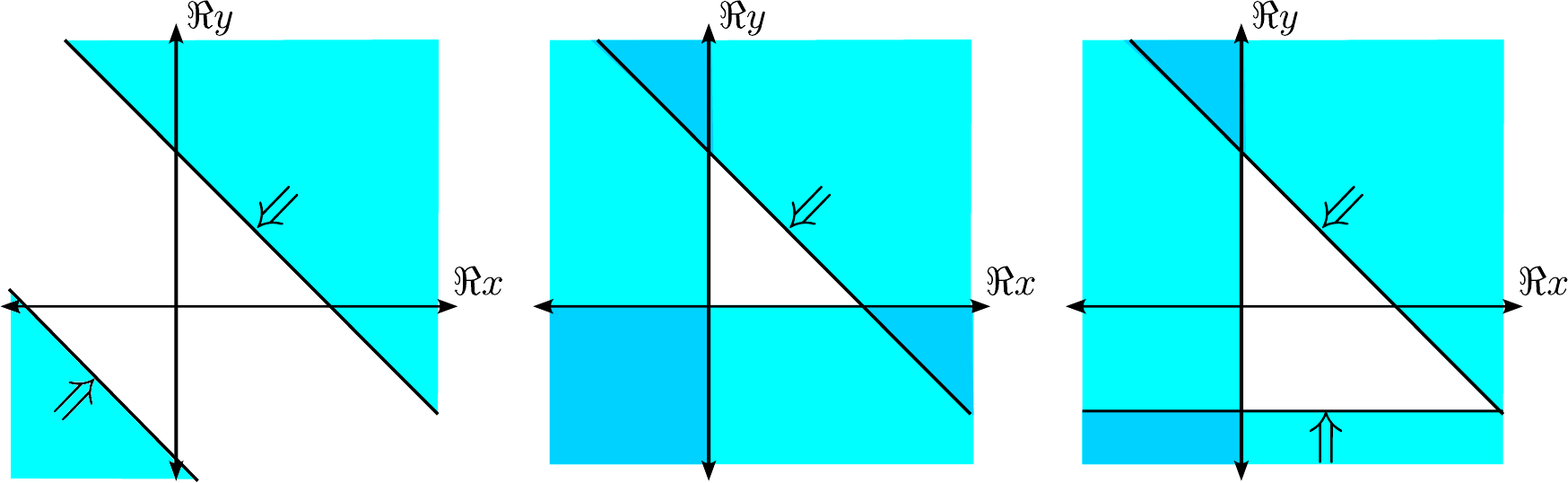}
\caption{Examples of how the contour can be pinched. Focusing on the real parts of $(x, y)$, each gamma function is analytic in a half-space. As $\alpha_-^2 \to 1$ (indicated by the double arrow), these half-spaces intersect in a locus called the pinching plane which is either a line or a point. The locus relevant to the strongest singularity is an intersection of several pinching planes and will therefore always be a point in this example. The diagram on the left corresponds to $P_{\{5,9\}}$ and one of the choices for $P_{\{6,9\}}$ --- the other choice is similar except the line where the half-spaces meet does not pass through the origin. The diagram in the middle corresponds to $P_{\{1,2,5\}}$ and one of the choices for $P_{\{1,2,6\}}$. The other choices are again translations of this. The diagram on the right for $P_{\{2,3,6\}}$ has two half-spaces move instead of one. We would simply change which ones these are for $P_{\{1,4,6\}}$.}
\label{fig:pinching}
\end{figure}
However, most of these effective gamma functions (including those removed by the filtering rule) only affect singularities away from $\alpha_-^2 = 1$. It is therefore sufficient to restrict attention to the last line of \eqref{kept-gammas}. One caveat is that the singular behaviour from \eqref{first-merging}, which exists for arbitrary $\alpha_-^2$, has already been avoided by the contour in \eqref{nonchiral-operator}. As such, $\Gamma_{\{6,9\}}$ plays a somewhat diminished role. The singularities as $\alpha_-^2 \to 1$, which indicate contour pinches, come from residues corresponding to the other five effective gamma functions but sometimes they are at poles overlapping with those of $\Gamma_{\{6,9\}}$. These overlaps enhance the strength of the singularity. To see how this works, define the pinching planes
\ba
& \hspace{1.7cm} P_{\{1,4,6\}} = (0, 4\alpha_-^2 - 2), \quad P_{\{1,2,5\}} = (0, 0), \quad P_{\{2,3,6\}} = (4\alpha_-^2 - 2, 0) \label{pinching-planes} \\
& \hspace{4.9cm} P_{\{1,2,6\}} \in \{ (0, 2), (0, 0), (2, 0) \} \,,\\
& P_{\{5,9\}} = V(x + y + 2 - 2\alpha_-^2), \quad P_{\{6,9\}} \in \{ V(x + y + 4 - 6\alpha_-^2), V(x + y + 6 - 6\alpha_-^2) \}\,,
\ea
where $V(L)$ is the set of points with $L = 0$. Examples of the pinchings are shown in Figure \ref{fig:pinching}. There are multiple choices for $P_{\{1,2,6\}}$ and $P_{\{6,9\}}$ because $\Gamma_{\{1,2,6\}}$ and $\Gamma_{\{6,9\}}$ allow one to take residues at subleading poles of the gamma functions involved and still achieve an $\alpha_-^2 \to 1$ singularity. Depending on the choice, different mutual intersections from
\ba
& P_{\{1,2,6\}} \cap P_{\{1,4,6\}} \cap P_{\{6,9\}}, \quad P_{\{1,2,6\}} \cap P_{\{2,3,6\}} \cap P_{\{6,9\}} \\
& \hspace{1.7cm} P_{\{1,2,6\}} \cap P_{\{1,2,5\}} \cap P_{\{5,9\}} \cap P_{\{6,9\}}\,,
\ea
can be non-empty. Looking at the first line, the maximal non-empty intersection involves three pinching planes and therefore produces an $O((\alpha_-^2 - 1)^{-3})$ term. In other words, an $O(m^3)$ which should be added to \eqref{final-before}. Even though the last line has four pinching planes, it does not produce an $O(m^4)$ because all of the same gamma functions still appear when either $P_{\{5,9\}}$ or $P_{\{6,9\}}$ is dropped.
Now that the pinchings contributing to the desired leading term are all known, it is straightforward to take double residues in the five ways indicated by \eqref{pinching-planes} to arrive at
\begin{align}
& \left [ \underset{x = 0, \; y = 0}{\text{Res}} + \underset{x = 0, \; y = 2}{\text{Res}} + \underset{x = 2, \; y = 0}{\text{Res}} + \underset{x = 4\alpha_-^2 - 2, \; y = 0}{\text{Res}} + \underset{x = 0, \; y = 4\alpha_-^2 - 2}{\text{Res}} \right ] \\
&\quad  \times \Gamma(-\tfrac{x}{2}) \Gamma(-\tfrac{y}{2}) \Gamma(2\alpha_-^2 - 1 - \tfrac{x}{2}) \Gamma(2\alpha_-^2 - 1 - \tfrac{y}{2}) \Gamma(1 - \alpha_-^2 + \tfrac{x + y}{2}) \Gamma(2 - 3\alpha_-^2 + \tfrac{x + y}{2}) \nonumber \\
& \qquad\times\Gamma(\tfrac{4 - 3\alpha_-^2 + x}{2}) \Gamma(\tfrac{4 - 3\alpha_-^2 + y}{2}) \Gamma(\tfrac{6\alpha_-^2 - 6 - x - y}{2}) \Gamma(\tfrac{3\alpha_-^2 - 2 - x}{2})^{-1} \Gamma(\tfrac{3\alpha_-^2 - 2 - y}{2})^{-1} \Gamma(\tfrac{8 - 6\alpha_-^2 + x + y}{2})^{-1} \nonumber\\
&\qquad\quad= \frac{4}{3} m^3 + O(m^2) \nonumber\,,
\end{align}
or
\be
I_{1,2} = \frac{\pi \Gamma(2 - 2\alpha_-^2)}{4 \Gamma(\alpha_-^2) \Gamma(1 - \alpha_-^2)^2 \Gamma(2\alpha_-^2 - 1) \Gamma(2 - 3\alpha_-^2)} \left [ m^3 + O(m^2) \right ]\,. \label{final-after}
\ee

Although the manipulations above are all correct, it is much better to carry them out in an automated way. This is what we explain how to do using the \texttt{Mathematica} package \texttt{MB} in Appendix \ref{app:coulombgas}. Running the code presented there for a few seconds leads to the result
\be
I_{1,2} = \frac{\pi \Gamma(2 - 2\alpha_-^2)}{4 \Gamma(\alpha_-^2) \Gamma(1 - \alpha_-^2)^2 \Gamma(2\alpha_-^2 - 1) \Gamma(2 - 3\alpha_-^2)} \left [ m^3 + (4 - 8 \log 2) m^2 + O(m) \right ]\,, \label{result-subleading1}
\ee
which has one extra term. As a prediction for the beta function, this is
\be
\beta_3 = \frac{3 \pi^2 m}{8} - \frac{3\pi^2}{8} (8 \log 2 + 1) + O(m^{-1}) \label{result-subleading2}\,.
\ee
as expected. The \texttt{MB} manipulations required for the next subsection will take closer to an hour.

\subsection{The \texorpdfstring{$\phirs{2}{2}\chi$}{phi22chi} beta function}
\label{sec:CGhard}

We finally come to the beta function for the $\phirs{2}{2} \chi$ deformation. While the \texttt{MB} package described in Appendix \ref{app:coulombgas} was convenient for the last subsection, it will be crucial here. This is because the four-point function which appears in $\beta_3$ requires one screening charge of each type. It is
\ba
& \la \phirs{2}{2}(0) \phirs{2}{2}(z, \bar{z}) \phirs{2}{2}(1) \phirs{2}{2}(\infty) \ra = \frac{\la V_{\alpha_{2,2}}(0) V_{\alpha_{2,2}}(z, \bar{z}) V_{\alpha_{2,2}}(1) V_{2\alpha_0 - \alpha_{2,2}}(\infty) \ra}{\la V_{\alpha_{2,2}}(0) V_{\alpha_{2,2}}(1) V_{2\alpha_0}(\infty) \ra} \\
&= [|z| |1 - z|]^{(\alpha_+ + \alpha_-)^2} \frac{\alpha_+^8 \Gamma(\alpha_-^2)^3 \Gamma(2 - \alpha_-^2) \Gamma(3 - 2\alpha_-^2) \Gamma(\alpha_+^2 - 1)^2 \Gamma(2 - 2\alpha_+^2)}{\Gamma(1 - \alpha_-^2)^3 \Gamma(\alpha_-^2 - 1) \Gamma(2\alpha_-^2 - 2) \Gamma(2 - \alpha_+^2)^2 \Gamma(2\alpha_+^2 - 1)} \\
& \int \frac{\text{d}^2t_1 \text{d}^2t_2}{\pi^2} [|t_1| |z - t_1| |1 - t_1|]^{-2 \alpha_+ (\alpha_+ + \alpha_-)} [|t_2| |z - t_2| |1 - t_2|]^{\alpha_- (\alpha_+ + \alpha_-)} |t_{12}|^{4\alpha_+ \alpha_-}\,,
\ea
where the normalization factor comes from \eqref{nonchiral-selberg}. In the next step, which is applying \eqref{symanzik} twice, we have ordered the marked points as $(0, z, 1, \infty, t_2)$ when eliminating $t_1$ and $(0, z, 1, \infty)$ when eliminating $t_2$. Using $\delta_{ij}$ variables in the first iteration and $\gamma_{ij}$ in the second, we can solve \eqref{symanzik-rels}, in terms of $\delta_{12}, \delta_{13}, \delta_{14}, \delta_{23}, \delta_{24}$ and $\gamma_{12}, \gamma_{13}$. This leads to the Mellin-Barnes representation
\begin{align}
& \la \phirs{2}{2}(0) \phirs{2}{2}(z, \bar{z}) \phirs{2}{2}(1) \phirs{2}{2}(\infty) \ra = \frac{\pi \alpha_+^8 (\alpha_-^2 - 1) \Gamma(\alpha_-^2)^2 \Gamma(2 - \alpha_-^2) \Gamma(3 - 2\alpha_-^2) \Gamma(2 - 2\alpha_+^2)}{\Gamma(1 - \alpha_-^2)^3 \Gamma(2\alpha_-^2 - 2) \Gamma(2 - \alpha_+^2)^2 \Gamma(2\alpha_+^2 - 1) \Gamma(\alpha_+^2 - 1) \Gamma(3 - 3\alpha_+^2)} \nonumber \\
& \int_{-i\infty}^{i\infty} \frac{\text{d}\delta_{12} \text{d}\delta_{13} \text{d}\delta_{14} \text{d}\delta_{23} \text{d}\delta_{24} \text{d}\gamma_{12} \text{d}\gamma_{13}}{(2\pi i)^7 |z|^{2\delta_{12} + 2\gamma_{12}} |1 - z|^{2\delta_{23} + 2\gamma_{23}}} [|z| |1 - z|]^{(\alpha_+ + \alpha_-)^2} \Gamma(\delta_{12}) \Gamma(\delta_{13}) \Gamma(\delta_{14}) \Gamma(\delta_{23}) \Gamma(\delta_{24}) \Gamma(\gamma_{12}) \Gamma(\gamma_{13}) \nonumber \\
& \Gamma(\alpha_+^2 + \alpha_+ \alpha_- - \delta_{12} - \delta_{13} - \delta_{14}) \Gamma(\alpha_+^2 + \alpha_+ \alpha_- - \delta_{12} - \delta_{23} - \delta_{24}) \Gamma(1 - 3\alpha_+^2 - 3\alpha_+ \alpha_- + \delta_{12} + \delta_{13} + \delta_{23}) \nonumber \\
& \Gamma(\alpha_+^2 - \alpha_+ \alpha_- - 1 + \delta_{12} + \delta_{14} + \delta_{24}) \Gamma(1 + 2\alpha_+ \alpha_- - \delta_{12} - \delta_{13} - \delta_{14} - \delta_{23} - \delta_{24}) \nonumber \\
& \Gamma((\alpha_+ + \alpha_-)^2 - \delta_{12} - \delta_{13} - \delta_{14} - \gamma_{12} - \gamma_{13}) \Gamma(1 - 2(\alpha_+ + \alpha_-)^2 + 2\delta_{12} + \delta_{13} + \delta_{14} + \delta_{23} + \delta_{24} + \gamma_{12}) \nonumber \\
& \Gamma(3\alpha_+^2 + 3\alpha_-^2 + 4\alpha_+ \alpha_- - 2 - \delta_{12} - \delta_{13} - \delta_{23} - \gamma_{12} - \gamma_{13}) \Gamma(2 - 2\alpha_+^2 - 2\alpha_-^2 - 2\alpha_+ \alpha_- + \delta_{13} - \delta_{24} + \gamma_{13}) \nonumber \\
& \Gamma(3 - 3\alpha_+^2 - 3\alpha_-^2 - 4\alpha_+ \alpha_- + \delta_{12} + \delta_{13} + \delta_{23})^{-1} \Gamma(\alpha_+^2 + \alpha_-^2 - 1 + \delta_{12} + \delta_{14} + \delta_{24})^{-1} \nonumber \\
& \Gamma((\alpha_+ + \alpha_-)^2 - \delta_{12} - \delta_{13} - \delta_{14})^{-1} \Gamma((\alpha_+ + \alpha_-)^2 - \delta_{12} - \delta_{23} - \delta_{24})^{-1}.
\end{align}

To separate the sequences of poles and allow the Mellin-Barnes contour to be straight, we will need to analytically continue in $\alpha_+^2$, $\alpha_-^2$ and $\alpha_+ \alpha_-$ as if they were independent variables. As before, there will also be a regulator parameter introduced by the integral over space. Looking at
\be
\la \chi(0) \chi(z, \bar{z}) \chi(1) \chi(\infty) \ra = 1 + |z|^{3(\alpha_+ + \alpha_-)^2 - 4} + |1 - z|^{3(\alpha_+ + \alpha_-)^2 - 4}\,,
\ee
specifically, we will recognize that all three terms contribute equally and change the $1$ to $[|z| |1 - z|]^\gamma$. The final integral to consider is then
\begin{align}
& I_{2,2} = \frac{\pi \alpha_+^8 (\alpha_-^2 - 1) \Gamma(\alpha_-^2)^2 \Gamma(2 - \alpha_-^2) \Gamma(3 - 2\alpha_-^2) \Gamma(2 - 2\alpha_+^2)}{\Gamma(1 - \alpha_-^2)^3 \Gamma(2\alpha_-^2 - 2) \Gamma(2 - \alpha_+^2)^2 \Gamma(2\alpha_+^2 - 1) \Gamma(\alpha_+^2 - 1) \Gamma(3 - 3\alpha_+^2)} \nonumber \\
& \int_{-i\infty}^{i\infty} \frac{\text{d}\delta_{12} \text{d}\delta_{13} \text{d}\delta_{14} \text{d}\delta_{23} \text{d}\delta_{24} \text{d}\gamma_{12} \text{d}\gamma_{13}}{(2\pi i)^7} \Gamma(\delta_{12}) \Gamma(\delta_{13}) \Gamma(\delta_{14}) \Gamma(\delta_{23}) \Gamma(\delta_{24}) \Gamma(\gamma_{12}) \Gamma(\gamma_{13}) \nonumber \\
& \Gamma(\alpha_+^2 + \alpha_+ \alpha_- - \delta_{12} - \delta_{13} - \delta_{14}) \Gamma(\alpha_+^2 + \alpha_+ \alpha_- - \delta_{12} - \delta_{23} - \delta_{24}) \Gamma(1 - 3\alpha_+^2 - 3\alpha_+ \alpha_- + \delta_{12} + \delta_{13} + \delta_{23}) \nonumber \\
& \Gamma(\alpha_+^2 - \alpha_+ \alpha_- - 1 + \delta_{12} + \delta_{14} + \delta_{24}) \Gamma(1 + 2\alpha_+ \alpha_- - \delta_{12} - \delta_{13} - \delta_{14} - \delta_{23} - \delta_{24}) \nonumber \\
& \Gamma((\alpha_+ + \alpha_-)^2 - \delta_{12} - \delta_{13} - \delta_{14} - \gamma_{12} - \gamma_{13}) \Gamma(1 - 2(\alpha_+ + \alpha_-)^2 + 2\delta_{12} + \delta_{13} + \delta_{14} + \delta_{23} + \delta_{24} + \gamma_{12}) \nonumber \\
& \Gamma(3\alpha_+^2 + 3\alpha_-^2 + 4\alpha_+ \alpha_- - 2 - \delta_{12} - \delta_{13} - \delta_{23} - \gamma_{12} - \gamma_{13}) \Gamma(2 - 2\alpha_+^2 - 2\alpha_-^2 - 2\alpha_+ \alpha_- + \delta_{13} - \delta_{24} + \gamma_{13}) \nonumber \\
& \Gamma(1 - \delta_{12} - \gamma_{12} + \tfrac{\gamma + (\alpha_+ + \alpha_-)^2}{2}) \Gamma(3 + \delta_{12} + \delta_{13} + \gamma_{12} + \gamma_{13} + \tfrac{\gamma - 5\alpha_+^2 - 5\alpha_-^2 - 6\alpha_+ \alpha_-}{2}) \nonumber \\
& \Gamma(2\alpha_+^2 + 2\alpha_-^2 + 2\alpha_+ \alpha_- - 3 - \delta_{13} - \gamma_{13} - \gamma) \Gamma(3 - 3\alpha_+^2 - 3\alpha_-^2 - 4\alpha_+ \alpha_- + \delta_{12} + \delta_{13} + \delta_{23})^{-1} \nonumber \\
& \Gamma(\alpha_+^2 + \alpha_-^2 - 1 + \delta_{12} + \delta_{14} + \delta_{24})^{-1} \Gamma((\alpha_+ + \alpha_-)^2 - \delta_{12} - \delta_{13} - \delta_{14})^{-1} \nonumber \\
& \Gamma((\alpha_+ + \alpha_-)^2 - \delta_{12} - \delta_{23} - \delta_{24})^{-1} \Gamma(4 - 2\alpha_+^2 - 2\alpha_-^2 - 2\alpha_+ \alpha_- + \delta_{13} + \gamma_{13} + \gamma)^{-1} \nonumber \\
& \Gamma(\delta_{12} + \gamma_{12} - \tfrac{\gamma - (\alpha_+ + \alpha_-)^2}{2})^{-1} \Gamma(-2 - \delta_{12} - \delta_{13} - \gamma_{12} - \gamma_{13} - \tfrac{\gamma - 5\alpha_+^2 - 5\alpha_-^2 - 6\alpha_+ \alpha_-}{2})^{-1}\,. \label{final-input1}
\end{align}

While it is clear that \texttt{MB} will have a longer runtime due to the sheer number of integrals, the setup of it is also more involved due to the two types of screening charges. This is explained in Appendix \ref{app:coulombgas} which shows the steps needed to produce the result
\be
I_{2,2} = \frac{\pi \alpha_+^8 (\alpha_-^2 - 1) \Gamma(\alpha_-^2)^2 \Gamma(2 - \alpha_-^2) \Gamma(3 - 2\alpha_-^2) \Gamma(2 - 2\alpha_+^2)}{\Gamma(1 - \alpha_-^2)^3 \Gamma(2\alpha_-^2 - 2) \Gamma(2 - \alpha_+^2)^2 \Gamma(2\alpha_+^2 - 1) \Gamma(\alpha_+^2 - 1) \Gamma(3 - 3\alpha_+^2)} \left [ -\frac{1}{6} m^4 + O(m^2) \right ]\,. \label{result-subleading3}
\ee
Expanding the prefactor, and recalling that $\beta_3 = -3 \frac{2\pi}{3!} I_{2,2}$, we see that
\be
\beta_3 = \frac{\pi^2}{2m^2} - \frac{\pi^2}{2m^3} + O(m^{-4})\,. \label{result-subleading4}
\ee
This proves the leading and first subleading asymptotic that we conjectured from numerical conformal perturbation theory.

\subsection{Further comments}

This section has explored the idea of performing conformal perturbation theory using the Mellin amplitudes of four-point functions in minimal models. Although the main application was the large-$m$ expansion, there are also some observations to be made for finite $m$.

One of these pertains to special anomalous dimensions in the $\phirs{1}{2}\chi$-flow. Starting with
\ba
& \la \phirs{r}{s}(0) \phirs{1}{2}(z, \bar{z}) \phirs{1}{2}(1) \phirs{r}{s}(\infty) \ra = \frac{\la V_{\alpha_{r,s}}(0) V_{\alpha_{1,2}}(z, \bar{z}) V_{\alpha_{1,2}}(1) V_{2\alpha_0 - \alpha_{r,s}}(\infty) \ra}{\la V_{\alpha_{1,2}}(0) V_{\alpha_{1,2}}(1) V_{2\alpha_0}(\infty) \ra} \label{anom12-coulomb} \\
&= |z|^{1 - r - (1 - s)\alpha_-^2} |1 - z|^{\alpha_-^2} \frac{\Gamma(\alpha_-^2)^2 \Gamma(2 - 2\alpha_-^2)}{\Gamma(1 - \alpha_-^2)^2 \Gamma(2\alpha_-^2 - 1)} \int \frac{\text{d}^2t}{\pi} |t|^{-2(1 - r) + 2(1 - s)\alpha_-^2} [|z - t| |1 - t|]^{-2\alpha_-^2}\,,
\ea
we can notice that one of the exponents of the integrand becomes a non-negative integer for $s = 1$. When $r$ is odd, \eqref{anom12-coulomb} is not sufficient for computing anomalous dimensions due to the mixing problem treated in section \ref{sec:mixing}. When $r$ is even however, Symanzik's formula shows that the anomalous dimension will vanish whether or not $m$ is taken to be large. This is the finding from \cite{psz19} that minimal model correlators of the form $\la \phirs{r}{1}(0) \phirs{1}{2}(z, \bar{z}) \phirs{1}{2}(1) \phirs{r}{1}(\infty) \ra$ have vanishing Mellin amplitudes. It greatly generalizes the calculation in \cite{Behan:2017emf} which showed that the $\varepsilon$ operator in the 2d long-range Ising model (which is $\phirs{2}{1}$) has a vanishing two-loop anomalous dimension. An obvious counterpart to this statement is that, in the $\phirs{2}{1}\chi$-flow, $\gamma^{(2)}_{1,2k}$ vanishes for all $m$ instead of $\gamma^{(2)}_{2k,1}$.

We can also ask how far the methods in this section take us for non-trivial perturbative data at finite $m$. The answer is that everything goes through exactly as above until the last step when an analytic continuation is done for $\alpha_-^2 \to 1$ or $\alpha_+^2 \to 1$. If one wants $\alpha_-^2 \to \frac{m}{m + 1}$ or $\alpha_+^2 \to \frac{m + 1}{m}$ instead, it is easy to ask for such a limit in \texttt{MBcontinue[]}. While this no longer produces the types of closed-form expressions seen in \eqref{result-subleading1} and \eqref{result-subleading3}, it produces Mellin-Barnes integrals which can be computed numerically to high precision. For the $\phirs{1}{2}\chi$-flow, which gives \texttt{MB} runtimes that are fairly reasonable, we have tested this method and found that it agrees with the rather different numerical method of section \ref{sec:phi12flow}.

The last point we would like to make is that the use of Mellin-Barnes integrals here allows this method to manifestly generate known special functions. As a simple example, consider the $\sigma$ four-point function in the 2d Ising model
\ba
\la \sigma(0) \sigma(z, \bar{z}) \sigma(1) \sigma(\infty) \ra &= -\frac{1}{8 \pi^2} [|z| |1 - z|]^{-\frac{1}{4}} \label{ising-mellin} \\
& \int_{-i\infty}^{i\infty} \frac{\text{d}x \text{d}y}{(4\pi i)^2} |z|^x |1 - z|^y \Gamma(-\tfrac{x}{2}) \Gamma(-\tfrac{y}{2}) \Gamma(\tfrac{1 - x}{2}) \Gamma(\tfrac{1 - y}{2}) \Gamma(\tfrac{x + y}{2} + \tfrac{1}{4}) \Gamma(\tfrac{x + y}{2} - \tfrac{1}{4}) \\
&= -\int_{-i\infty}^{i\infty} \frac{\text{d}x \text{d}y}{(2\pi i)^2} \Gamma(-x) \Gamma(-y) \Gamma(x + y - \tfrac{1}{2}) \frac{|z|^{x - \frac{1}{4}} |1 - z|^{y - \frac{1}{4}}}{2 \sqrt{2\pi}}\,.
\ea
The first line follows from Symanzik's formula and the second line uses the duplication formula to recover the expression which was first written in \cite{az15}. In this form, it is clear that the poles of $\Gamma(-x)$ and $\Gamma(-y)$ are kept while those of $\Gamma(x + y - \tfrac{1}{2})$ are not. If we now include the GFF and integrate,
\ba\label{eq:kdf-integral}
\beta_3 = -3\pi \int_{-i\infty}^{i\infty} \frac{\text{d}x \text{d}y}{2^{7/2} \pi^{5/2}} \Gamma(-x) \Gamma(-y) \Gamma \left ( x + y - \frac{1}{2} \right ) \frac{\Gamma(\tfrac{x}{2} + \tfrac{7}{8}) \Gamma(\tfrac{y}{2} + \tfrac{7}{8}) \Gamma(-\tfrac{2x + 2y + 3}{4})}{\Gamma(\tfrac{1}{8} - \tfrac{x}{2}) \Gamma(\tfrac{1}{8} - \tfrac{y}{2}) \Gamma(\tfrac{2x + 2y + 7}{4})}\,,
\ea
where we have included a factor of 3 instead of writing two extra terms. If we now make the change of variables $x' = \frac{7}{8} - \frac{x}{2}$ and $y' = \frac{7}{8} - \frac{y}{2}$, we can take the residue at $x' = m$ and $y' = n$ to find
\be\label{eq:kdf-residue}
-\frac{\Gamma(\tfrac{7}{8})^2 \Gamma(\tfrac{11}{8})^2}{2^6 \pi^{5/2}} \frac{(1)_{m + n}^2 (\tfrac{7}{8})_m (\tfrac{7}{8})_n (\tfrac{11}{8})_m (\tfrac{11}{8})_n}{m! n! (1)_m (1)_n (3)_{m + n} (\tfrac{5}{2})_{m + n}}\,,
\ee
after using both the duplication formula and the reflection formula several times. Summing \eqref{eq:kdf-residue} weighted by $u^m v^n$ would yield a Kamp\'{e} de F\'{e}riet hypergeometric function with arguments $(u, v)$. This sum diverges at $u = v = 1$ but we can take the convergent integral \eqref{eq:kdf-integral} to be the definition of the Kamp\'{e} de F\'{e}riet hypergeometric function in this case and write
\be
\beta_3 = -\frac{\Gamma(\tfrac{7}{8})^2 \Gamma(\tfrac{11}{8})^2}{2^6 \pi^{5/2}} F \left [ \begin{tabular}{c} $1,1; \frac{7}{8}, \frac{11}{8}; \frac{7}{8}, \frac{11}{8}$ \\ $\frac{5}{2}, 3; 1; 1$ \end{tabular}; 1, 1 \right ]\,.
\ee

%% file: sections/conclusions.tex

\section{Conclusions and future directions}
\label{sec:conclusions}

In this paper, we introduced and studied a class of two-dimensional QFTs, characterized by multi-critical universal behavior and long-range interactions, that arises when coupling a Virasoro minimal model to a generalized-free field (GFF) theory. 
More precisely, if $\phirs{r}{s}$ is a relevant primary in the $m$-th unitary Virasoro minimal model and $\chi$ is a generalized free field, we turn on the interaction
\be
g \int  \text{d}^2x \, \phirs{r}{s} \chi\,.
\ee

This is relevant when the operator $\cO \equiv \phirs{r}{s} \chi$ has scaling dimension $\D_\cO = 2 - \delta < 2$. In this case, the RG flow generically reaches an IR fixed point, defining a family of non-local CFTs labeled by $(m,r,s)$ and continuously parametrized by $\delta$. We call this family the long-range minimal models (LRMM) of type $(m,r,s)$.
For $m > 3$, these models naturally extend the 2d long-range Ising construction near the short-range end \cite{Behan:2017dwr,Behan:2017emf} to multicritical theories, i.e. with more than one $\mathbb{Z}_2$-invariant operator.

In the near short-range regime ($0<\delta \ll 1$), the fixed point is weakly coupled and CFT data admit a perturbative expansion in $\delta$. We analyzed LRMM of types $(m,1,2)$, $(m,2,1)$, and $(m,2,2)$, computing the beta function of the weakly-relevant coupling to leading order in conformal perturbation theory — numerically for finite $m$ and analytically in the $1/m$ expansion. We also determined anomalous dimensions of a wide set of low-lying UV operators, including Virasoro primaries $\phirs{r}{s}$ and higher-spin currents, at leading nontrivial order in $\d$.

In the opposite regime with large $\delta$, we have argued that the $(m,2,2)$ LRMM admits a dual, weakly-coupled, Ginzburg-Landau formulation in terms of an appropriate GFF $\varphi$, perturbed by a multi-critical potential (see Sec.~\ref{sec:duality}).
Determining the extent to which this is true for other LRMMs is an important open problem.

The constructions of long-range models via the coupling of a local CFT with a GFF is very general, and it can be extended to many other models.
Our reason for singling out the Virasoro minimal models is the high-degree of control that we have on them, but even in this case there are many open questions.

We list some future directions:

\begin{itemize}

\item It would be interesting to better understand the structure of multi-coupling renormalization that arises in the large-$m$ limit, as discussed in Sec.~\ref{sec:CGfirst}. A related point is that consecutive short-range minimal models $\mathcal{M}_{m+1,m}$ are related by the famous staircase RG flow \cite{Zamolodchikov:1987ti,Cardy:1989da}, that is:\footnote{See also \cite{Antunes:2024hrt} for a recent numerical bootstrap analysis of the staircase RG flow between minimal models, in the context of QFT in AdS$_2$.}
\be
\mathcal{M}_{m+1,m}+ h \int \text{d}^2x \, \phirs{1}{3} \to \mathcal{M}_{m,m-1}\,.
\ee
We can ask if there is also a similarly simple way to understand RG flows between the $(m,r,s)$ and the $(m-1,r,s)$ LRMMs.

\item Although the LRMMs are nonlocal CFTs, they are amenable to standard non-perturbative approaches, including the conformal bootstrap. The framework of \cite{Lauria:2020emq,Behan:2020nsf,Behan:2021tcn} in particular has been successful at finding kinks for the long-range Ising model \cite{Behan:2023ile}. Moreover, each such kink lives on a curve which admits a second kink and this second kink moves as a function of the spin-2 operator gap being imposed. A preliminary check indicated that it moves in a way which is consistent with the $(m,1,2)$ LRMM. Now that a large amount of perturbative data is available, the time is ripe to test this conjecture by taking more numerical data. Apart from the bootstrap, a functional renormalization group approach to the long-range Ising model \cite{Defenu:2014bea,Defenu:2020umv,Solfanelli:2024obb} is a also good candidate for studying long-range minimal models, at least in their Ginzburg-Landau description, when this is available.

\item Our focus so far has been on local operators in LRMMs but CFTs also admit defect operators. Defects in long-range $O(N)$ models have recently been studied in \cite{2412.08697}. In any CFT, there is a large class of defect operators called pinning field defects. It was recently shown in \cite{2504.06203}, subject to various assumptions, that pinning field defects of codimension one are generically factorizing. Nonlocal CFTs violate these assumptions and a counter-example in the long-range Ising model was subsequently constructed in \cite{2505.15018} --- namely a conformal interface which is perturbatively close to the trivial interface and therefore not factorizing. The long-range Ising example integrates $\vph^2$ along the interface and the natural analogue of this is $\vph^{m-1}$ in Ginzburg-Landau description of the $(m,2,2)$ LRMM.

\item Our construction can also be extended to other two-dimensional long-range systems, such as long-range $Q$-state Potts models. For $Q=3$, we expect the critical and tricritical long-range Potts models to be described by a subset of fields in the LRMM of type $(5,2,2)$ and $(6,2,2)$, respectively, as in the analog short-range case \cite{DiFrancesco:1997nk}. The status of other long-range Potts models is unclear to us. For $Q>4$, the short-range model has a first-order transition, but its ``walking" renormalization group flow can be understood in terms of complex fixed points \cite{Gorbenko:2018ncu,Gorbenko:2018dtm}. In analogy with the one-dimensional case \cite{Cardy_1981,Cannas:1995ja,Bayong:1999,Reynal_2004}, we expect that long-range interactions will affect the nature of the phase transition, at least in some range of the long-range exponent $s$, and up to some value of $Q$. It would be interesting to understand if a second order transition below some critical $s^\star$, and eventually how to describe the long-range to short-range crossover: for example, as a fixed point annihilation \cite{Kaplan:2009kr}?
The short-range $Q=4$ case is special, as it is described not by a minimal model, but by an orbifold construction \cite{Dijkgraaf:1987vp}. It would be interesting to explore its long-range deformation.

\item Another obvious generalization would be to apply the same type of long-range deformations to the non-unitary minimal models. 
It has been conjectured \cite{Amoruso,Zambelli:2016cbw,Lencses:2022ira,Lencses:2024wib} that the Ginzburg-Landau description for the nonunitary minimal models $\cM_{2,4m+1}$ is obtained similarly to the unitary case, but with the $\mathbb{Z}_2$-even potential replaced by the $\cP\cT$-symmetric perturbation $\im\vph^{2m+1}$. 
For other non-unitary minimal models, the Ginzburg-Landau description is even more open \cite{Klebanov:2022syt,Katsevich:2024jgq}.
Studying long-range deformations of such minimal models, and comparing them to long-range versions of their conjectured Ginzburg-Landau description, could perhaps help in corroborating such conjectures, as the long-range case provides a tunable parameter that allows to make the Ginzburg-Landau description weakly coupled, as we have seen in Sec.~\ref{sec:duality}.
	
\item Another interesting target is the LRMM of type $(m,3,3)$. 
The operator $\phi_{3,3}$ is the energy field, i.e.\ $\vph^2$ in the Ginzburg-Landau description. By such identification, we recognize our $\cO=\phirs{3}{3}\chi$ as the type of interaction that appears naturally in models with  weak disorder that results in an effective long-range correlated random-temperature disorder \cite{Weinrib:1983zz,Honkonen:1988fq,Prudnikov:2000,Chippari:2023vnx,Lecce:2024hnd}, where $\chi$ represents the disorder field.
The main difference is that our deformation corresponds to an annealed disorder, while in the quenched case one has to introduce replicas. 
For these type of long-range deformations, a first challenge is to compute the blocks, which are solutions of a ninth-order ODE.
However, simplifications occur for $m=3$ and $m=4$, as in the former case we have $\phi_{3,3}=\phi_{2,1}$, and in the latter case we have $\phi_{3,3}=\phi_{1,2}$.

\item Finally, it would be interesting to construct a theory for the long-range to short-range crossover for the one-dimensional version of the multi-critical models of Sec.\ref{sec:duality}. For the case of the one-dimensional long-range Ising model, which was understood only recently in \cite{Benedetti:2024wgx,Benedetti:2025nzp}, the crossover theory near $s=1$ is a generalized version of the bosonized spin-$1/2$ Kondo impurity model, with the scalar field having a negative dimension for $s<1$.  
It is natural to imagine the multi-critical case to be related to a spin-$j$ generalization of such model.

\end{itemize}

%% file: sections/app_CPT.tex

\section{Brief review of conformal perturbation theory}
\label{app:CPT}

In this appendix, we review the method of conformal perturbation theory, following closely  \cite{Komargodski:2016auf,Behan:2017emf}.\footnote{See also \cite{Zamolodchikov:1987ti,Dotsenko:1994sy,Cardy:1996xt,Gaberdiel:2008fn,1303.3015} for earlier references.}
The reason for repeating this here, besides the convenience of having an internal reference for formulas that have a crucial role in the bulk of the paper, is also to make some aspects of the subtraction process more explicit.

We start from \eqref{eq:deformCFT}, specialized to $d=2$, 
\be \label{eq:deformCFT-app}
\la\ldots\ra_{g_0} = \la\ldots e^{-g_0\int \rmd^2 z\, \cO(z,\zb) }\ra\,,
\ee
where $\cO=\Phi_i\chi$, and the composite operator is weakly relevant: $\D_{\cO}=2-\d$.
As usual, the computation of observables in the perturbed theory needs UV and IR regularizations and a renormalization prescription.

Following \cite{Komargodski:2016auf}, a convenient observable is $\la\cO(\infty) \ra_{g_0}$, where the operator at infinity is defined as the limit $\cO(\infty):= \lim_{z,\zb\to\infty}(z\zb)^{\D_{\cO}}\cO(z,\zb)$.
Expanding the observable in powers of $g_0$, we find
\ba \label{eq:Oinf-pert}
    \la\cO(\infty) \ra_{g_0} &= \la\cO(\infty)\ra - g_0 \int \rmd^2 z\ \la \cO(z,\zb)\cO(\infty)\ra +\f12 g_0^2 \int \rmd^2 z_1 \rmd^2 z_2\ \la \cO(z_1,\zb_1)\cO(z_2,\zb_2)\cO(\infty)\ra \\
    &- \frac{1}{3!} g_0^3  \int \rmd^2 z_1\rmd^2 z_2\rmd^2 z_3\ \la \cO(z_1,\zb_1)\cO(z_2,\zb_2)\cO(z_3,\zb_3)\cO(\infty)\ra +O(g^4) \,.
\ea
Evaluating the integrals, we encounter IR divergences, due to translation invariance of the integrands,\footnote{If rather than inserting the operator at infinity, we would place it at some point $w\in\cD$, these IR divergences would be replaced by UV divergences at $z_i\sim w$, and they would require a multiplicative renormalization  of $\cO(w)$, as well as a mixing with the identity operator.}
and UV divergences, originating from the integration of correlators of operators at coincident points.
For the IR regularization, we restrict the interaction to a finite domain $\cD=\{|z|\leq R\}$, with volume $V=\pi R^2$, thus replacing every integration as follows: $\int_{\mathbb{C}}\rmd^2 z \to \int_{\cD}\rmd^2 z$.
This also implies that the inserted operator is never at a coincident point with the integrated operators, and hence we do not need to renormalize it. 
UV divergences can instead be regularized either by a hard cutoff (i.e. $|z_{ij}|\geq a$) or by analytical continuation in $\d$.
Demanding that $\la\cO(\infty) \ra_g$ is independent of the UV cutoff or that it admits a regular limit for $\d\to 0$, requires that we implement a  coupling renormalization.

Equation \eqref{eq:Oinf-pert} is rather generic, but in our specific case it is notably simplified.
The first and third terms vanish identically, because they contain an odd number of fields $\chi$, and the latter follows a centered Gaussian distribution.
Moreover, a conformal two-point function with one operator at infinity reduces to a constant, that can be chosen to be equal to one.
Therefore, equation \eqref{eq:Oinf-pert} becomes
\be
  \la\cO(\infty) \ra_{g_0} = - V ( g_0 + g_0^3 I_{\cO\cO\cO\cO}  +O(g_0^5) \,,
\ee
with
\be \label{eq:I_OOOO}
I_{\cO\cO\cO\cO} =   \frac{1}{3! \, V}   \int_{\cD^3} \rmd^2 z_1\rmd^2 z_2\rmd^2 z_3\ \la \cO(z_1,\zb_1)\cO(z_2,\zb_2)\cO(z_3,\zb_3)\cO(\infty)\ra \,.
\ee

The UV divergences of the integral can result either separately from each of the pairwise coincident limits $|z_{12}|\sim 0$, $|z_{13}|\sim 0$, and $|z_{23}|\sim 0$,
or from the triple coincidence limit, when all three points collide.

The strongest (i.e.\ power-law) divergences originate from the pairwise coincident limits, which therefore we should deal with first.
Consider the region $|z_{12}|\sim 0$ at finite $|z_{13}|$ and $|z_{23}|$, and use the OPE
\be \label{eq:OO-OPE}
\cO(z_1,\zb_1)\cO(z_2,\zb_2) = \sum_k \f{C_{\cO\cO k}}{ |z_{12}|^{2\D_{\cO}-\D_k} } \cO_k(z_2,\zb_2) + \text{descendants}\,.
\ee
The integral over $z_1$ results in a singularity for each operator $\cO_k$ such that $2\D_{\cO}-\D_k\geq2$, schematically:
\ba \label{eq:power-div}
\int & \rmd^2 z_1\rmd^2 z_2\rmd^2 z_3\ \la \cO(z_1,\zb_1)\cO(z_2,\zb_2)\cO(z_3,\zb_3)\cO(\infty)\ra  \\
&\supset \int \rmd^2 z \f{C_{\cO\cO k}}{ |z|^{2\D_{\cO}-\D_k} } \int \rmd^2 z_2\rmd^2 z_3\  \la \cO_k(z_2,\zb_2)\cO(z_3,\zb_3)\cO(\infty)\ra  \,.
\ea
Since $\D_{\cO}=2-\d$ with $0<\d\ll 1$, this means that we will have power-law divergence for each relevant operator ($\D_k<2-2\d$), and at $\d=0$ a logarithmic one for $\D_k=2$.

The standard way to deal with the power-law divergences is to subtract them by adding a counterterm in \eqref{eq:deformCFT-app} for each relevant operator in the $\cO\times\cO$ OPE:\footnote{Notice that if we had divided \eqref{eq:deformCFT-app} by the partition function $\la e^{-g_0\int \rmd^2 z\, \cO(z,\zb) }\ra$ we would have automatically subtracted the divergences arising from the contribution of the identity operator. In other words, the counterterm associated to the identity operator is the free energy. For simplicity, here we choose to treat the identity as any other operator.}
\be \label{eq:deformCFT-ct}
\la\ldots\ra_{g_0} = \la\ldots \exp\Big\{-g_0\int \rmd^2 z\, \cO(z,\zb) -\sum_{k \mid \D_k<2} \l_k(g_0) \int \rmd^2 z\, \cO_k(z,\zb) \Big\} \ra\,,
\ee
where $\l_k$ is a series in $g_0$ with coefficients chosen to cancel the above divergences.
For example, in order to cancel the divergence in \eqref{eq:power-div} and the other two similar divergences at $|z_{13}|\sim 0$, and $|z_{23}|\sim 0$, we need
\be
\l_k(g_0) = \f{g_0^2}{a^{2-2\d-\D_k} } \f{\pi\, C_{\cO\cO k}}{ 2-2\d-\D_k} + O(g_0^3) \,,
\ee
where $a$ is a UV length cutoff.\footnote{Notice that we do not need to introduce an independent coupling for these counterterms, as the observable $\la\cO_k(\infty) \ra_{g_0}$ is automatically finite with these choices. Indeed, up to finite boundary contributions, we have
\ba \nn
\la\cO_k(\infty) \ra_{g_0} &= -\d\l_k  \int \rmd^2 z\ \la \cO_k(z,\zb)\cO_k(\infty)\ra +\f12 g_0^2 \int \rmd^2 z_1 \rmd^2 z_2\ \la \cO(z_1,\zb_1)\cO(z_2,\zb_2)\cO_k(\infty)\ra  + O(g_0^3) \\
& \sim -\d\l_k V  +\f12 g_0^2 V \int \rmd^2 z \f{C_{\cO\cO k}}{ |z|^{2\D_{\cO}-\D_k} }  + O(g_0^3) =   0 + O(g_0^3) \,.
\ea
}
At order $g_0^3$, this results in
\be
  \la\cO(\infty) \ra_{g_0} = - V ( g_0 + g_0^3 (I_{\cO\cO\cO\cO} - K)  +O(g_0^5) \,,
\ee
where
\be
K = \sum_{k \mid \D_k<2}  \f{\pi\, C_{\cO\cO k}}{ a^{2-2\d-\D_k}\, (2-2\d-\D_k) \, V} \int_{\cD^2} \rmd^2 z_2\rmd^2 z_3\  \la \cO_k(z_2,\zb_2)\cO(z_3,\zb_3)\cO(\infty)\ra   \,.
\ee

In practice, at this order of the perturbative expansion, the effect of the counterterms is equivalent to replacing the four-point function in \eqref{eq:I_OOOO} with
\ba \label{eq:F-subtraction}
F\big((z_1,\zb_1),& (z_2,\zb_2),(z_3,\zb_3)\big) = \la \cO(z_1,\zb_1) \cO(z_2,\zb_2)\cO(z_3,\zb_3)\cO(\infty)\ra \\
& - \sum_{k \mid \D_k<2} (C_{\cO\cO k})^2 \left(\f{1}{|z_{12}|^{2\D_{\cO}-\D_k}|z_{13}|^{\D_k}} +\f{1}{|z_{13}|^{2\D_{\cO}-\D_k}|z_{12}|^{\D_k}} +\f{1}{|z_{32}|^{2\D_{\cO}-\D_k}|z_{31}|^{\D_k}} \right) \,.
\ea
Equivalently, one can use analytic continuation in $\d$, i.e. compute the integrals at a large enough value of $\d$ such that $2-2\d-\D_k<0$ and then analytically continue to small $\d$; this way, the power divergences are automatically subtracted.

For the logarithmic divergences caused at $\d=0$ by other marginal operators $\cO_k$ with $\D_k=2$ things are a bit different.\footnote{Notice that the 3-point function in \eqref{eq:power-div} (proportional to $1/|z_{23}|^{\D_k}$) also leads to logarithmic divergence in this case, hence we have a double-log divergence. Since there is no $g^2$ term in the perturbative expansion of $\la\cO(\infty) \ra_g$, if in this type of computation we find such a double-log divergence, it is a clear signal that we have missed another marginal operator.}
First, these divergences are present even in analytic regularization, where they translate as usual into poles at $\d=0$.
Second, logarithms (or poles) necessarily carry another scale, for dimensional reasons. This other scale is a renormalization scale, or IR scale (such as $R$), and thus the counterterm  $\l_k(g_0)$ has a nontrivial RG flow.
Therefore, in this case in order to keep the bare theory fixed, we need to introduce an independent bare coupling $\l_{k,}$, to be treated on equal footing with $g_0$, and then re-express it in term of running renormalized couplings $\l_{k,0}=f(\l_k,g,R)$.
In most of our work we are not in this situation, hence in the rest of this appendix we assume that $\cO$ is the only (near-)marginal operator in the theory. See however the large-$m$ discussion in Sec.~\ref{sec:CGfirst} for an instance in which these remarks become relevant.

Lastly, we should consider what happens when all three integration points in \eqref{eq:I_OOOO} come close together, which we anticipate to lead to a logarithmic divergence in the integral.
This triple coincidence limit could be expressed in terms of a triple OPE \cite{Behan:2017emf}, but this results in a rather tautological statement.
Following \cite{Komargodski:2016auf,Behan:2017emf}, the divergence is extracted by the following manipulations:
\ba \label{eq:logdiv-steps}
\tilde{I}^{\,\rm div}_{\cO\cO\cO\cO} &=   \frac{1}{3! \, V}   \int \rmd^2 z_1\rmd^2 z_2\rmd^2 z_3\ F\big((z_1,\zb_1), (z_2,\zb_2),(z_3,\zb_3)\big) \\
&\sim  \frac{1}{3!}   \int \rmd^2 z_2\rmd^2 z_3\ F\big(0, (z_2,\zb_2),(z_3,\zb_3)\big) \\
&=  \frac{2\pi}{3!}   \int_0^R \f{\rmd |z_3|}{|z_3|^{2\D_{\cO}-3}} \int \rmd^2 z\ F\big(0, (z,\zb),\hat{e}\big) \\
&\sim  \frac{\pi}{6} \frac{R^{2\d}}{\d}    \int \rmd^2 z\ F\big(0, (z,\zb),\hat{e}\big) \,.
\ea 
Choosing to work with a hard cutoff $a$ at $\d=0$, we find as usual that the pole in $\d$ is replaced by a logarithm:
\be
\tilde{I}^{\,\rm div}_{\cO\cO\cO\cO} = \frac{\pi}{3} \log\left(R/a\right)    \int \rmd^2 z\ F\big(0, (z,\zb),\hat{e}\big) \,.
\ee

In \eqref{eq:logdiv-steps}, we have ignored boundary effects associated to changes of variables, because we are interested in the divergent part arising from the short distance limit.
In particular, in the second step, we have translated the integration variables in order to set $z_1=\zb_1=0$ in the integrand, and we have approximated the integral over $z_1$ with a simple volume factor. This step, is only true in the large-$V$ limit, hence the $\sim$ sign.
Next, we have shifted $(z_2,\, \zb_2) \to (|z_3| z, |z_3| \zb)$, so that the triple coincidence limit corresponds to $|z_3|\to 0$, and we used conformal invariance to pull out a $1/|z_3|^{2\D_{\cO}}$ from the four-point funciton. 
In the absence of other marginal operators, the final integral over the point $z$ is convergent,\footnote{However, it is not absolutely convergent, as pointed out in \cite{Behan:2017emf}, due to relevant or marginal operators with nonzero spin among the descendants in \eqref{eq:OO-OPE}, so it has to be evaluated with some care.} it is independent of the arbitrary unit vector $\hat{e}$, and in the large-$V$ limit also of $|z_3|$ and $R$.
The last statement is only true in the case that $C_{\cO\cO\cO}=0$, otherwise it results in a $\log\left(R/|z_3|\right)$ and a total double-log divergence for $\tilde{I}_{\cO\cO\cO\cO}$.\footnote{This can be shown by studying the large-$|z|$ behavior of $F\big(0, (z,\zb),\hat{e}\big)$ via an OPE in the $\cO(z,\zb)\times\cO(\infty)$ channel:
\[
\cO(z,\zb)\cO(\infty) \sim \sum_k \f{C_{\cO\cO k}}{|z|^{\D_k}} \cO_k(\infty) \, ,
\]
and remembering that relevant operators have been subtracted in \eqref{eq:F-subtraction}.
}

Therefore, for our $C_{\cO\cO\cO}=0$ case, the subtracted integral $\tilde{I}_{\cO\cO\cO\cO}$ has a simple pole in $\d$, or logarithmic divergence in $a$, which can be absorbed in the definition of a (dimensionless) renormalized coupling:
\be \label{eq:g_renorm}
g = R^{\d} \left( g_0 + g_0^3 \tilde{I}^{\,\rm div}_{\cO\cO\cO\cO}  + O(g_0^5) \right) \,.
\ee
The independence of the coupling $g_0$ on $R$ implies an RG flow of $g$:
\be \label{eq:beta-app}
-R \f{d g}{d R} = - \d g + \b_3 g^3 + O(g^5) \,,
\ee
where we defined
\be
\b_3  = - \frac{\pi}{3}   \int \rmd^2 z\ F\big(0, (z,\zb),\hat{e}\big) \,.
\ee
Alternatively, in the case of cutoff regularization, we can define an RG flow for the bare coupling with respect to $a$, using the inverse of \eqref{eq:g_renorm} and the independence of the coupling $g$ on $a$. The two schemes are related by a coupling redefinition (e.g.\ \cite{Montvay}), and $\b_3$ is scheme independent at leading-order in $\d$.

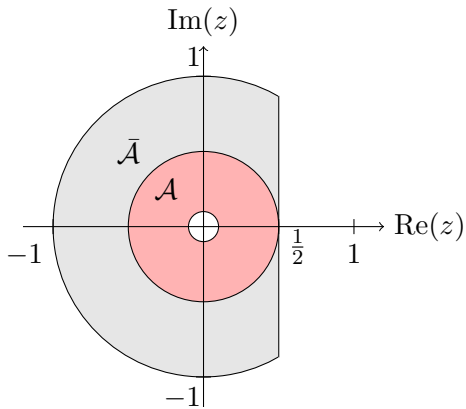
\begin{figure}[h]
\centering
\begin{tikzpicture}[scale=2]
    \tikzset{line/.style={thin}}
    \draw[fill=gray!20] (0,0) circle (1) node [black,yshift=1cm,xshift=-1cm] {$\bar{\mathcal{A}}$};
    \draw[fill=white,draw=none] (0.5,-1) rectangle ++(0.6,2);
    \draw (0.5,-0.87) -- (0.5,0.87);
    \draw[fill=red!30] (0,0) circle (0.5) node [black,yshift=0.5cm,xshift=-0.5cm] {$\mathcal{A}$};
    \draw[fill=white] (0,0) circle (0.1);
    \draw[->] (-1.2,0) -- (1.2,0) node [right] {$\text{Re}(z)$};
    \draw[->] (0,-1.2) -- (0,1.2) node [above] {$\text{Im}(z)$};
    \draw (-0.05,1) -- (0.05,1) node [above left] {$1$};
    \draw (-0.05,-1) -- (0.05,-1) node [below left] {$-1$};
    \draw (1,-0.05) -- (1,0.05) node [below, yshift=-2mm] {$1$};
    \draw (0.5,-0.05) -- (0.5,0.05) node [below right] {$\frac{1}{2}$};
    \draw (-1,-0.05) -- (-1,0.05) node [below left, yshift=-2mm] {$-1$};
\end{tikzpicture}
\caption{The integration region where a disk near the origin has been singled out for analytic integration of a truncated power series term-by-term. This allows power divergences to be subtracted systematically. In the remaining region, we can reliably perform the numerical integral all at once.}
\label{fig:regionR}
\end{figure}

The above derivation and resulting expression for $\b_3$ corresponds to what is called ``Method 1" in \cite{Behan:2017emf}.
For numerical computations, it is convenient to use ``Method 2" of \cite{Behan:2017emf}.
In the latter, the integration domain is divided in three regions $\cR_{ij} = \{z_1,z_2,z_3\,:\, |z_{ij}|<|z_{ik}|, |z_{ij}|<|z_{jk}|\}$, associated to each of the possible pairwise coincidence limits discussed above, and the subtraction of power divergences is done by hand in each region.
In the large-$V$ limit, the three regions give equal contributions, hence one defines 
\be
\b_3  = - \left( \pi   \int_{\cR} \rmd^2 z\  \la \cO(0) \cO(z,\zb)\cO(1)\cO(\infty)\ra \right)_{\text{finite}} \,,
\ee
where 
\be  \label{eq:integ-region}
\mathcal{R}=\{z\,:\, |z|<1, |z|<|z-1|\} \,.
\ee
As plotted in Figure \ref{fig:regionR}, the integration region is then split as $\mathcal{R}=\cA\,\cup\,\bar{\cA}$, where $\cA$ is the annulus
\begin{align}
\mathcal{A}= \{z\,:\, 0<a\leq |z|\leq 1/2\}\,,
\end{align}
and $\bar{\cA}=\cR \setminus \cA$ is its complement in $\cR$. In $\mathcal{A}$, we perform the integration analytically, while in $\bar{\mathcal{A}}$, we perform the integration numerically. 
The notation $(\ldots)_{\text{finite}}$ means that we keep only the terms that stay finite in the $a\to 0$ limit.

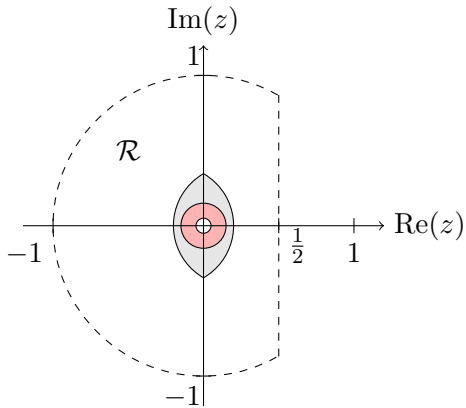
\begin{figure}[h]
\centering
\begin{tikzpicture}[scale=2]
    \tikzset{line/.style={thin}}
    \draw[fill=none, dashed] (0,0) circle (1) node [black,yshift=1cm,xshift=-1cm] {$\mathcal{R}$};
    \draw[fill=white,draw=none] (0.5,-1) rectangle ++(0.6,2);
    \draw[dashed] (0.5,-0.87) -- (0.5,0.87);
    \filldraw[fill=gray!20, draw=black] ([shift=(-60:0.4)]-0.2,0) arc (-60:60:0.4);
    \filldraw[fill=gray!20, draw=black] ([shift=(240:0.4)]0.2,0) arc (240:120:0.4);
    \draw[fill=red!30] (0,0) circle (0.15);
    \draw[fill=white] (0,0) circle (0.05);
    \draw[->] (-1.2,0) -- (1.2,0) node [right] {$\text{Re}(z)$};
    \draw[->] (0,-1.2) -- (0,1.2) node [above] {$\text{Im}(z)$};
    \draw (-0.05,1) -- (0.05,1) node [above left] {$1$};
    \draw (-0.05,-1) -- (0.05,-1) node [below left] {$-1$};
    \draw (1,-0.05) -- (1,0.05) node [below, yshift=-2mm] {$1$};
    \draw (0.5,-0.05) -- (0.5,0.05) node [below right] {$\frac{1}{2}$};
    \draw (-1,-0.05) -- (-1,0.05) node [below left, yshift=-2mm] {$-1$};
\end{tikzpicture}
\caption{The region $\cR$ (interior of the dashed line) together with its image under the $\r$ map (grey region). Notice that the annulus is not the image of the original annulus: we split the region into a new annulus and its complement after having applied the map.}
\label{fig:regionRho}
\end{figure}

One drawback of the region $\cR$ is that we use a series expansion in $z$ for the integrand, we are likely to find a divergent result because the integral partly extends to $|z|=1$, where such series cannot converge, due to the singularity of the four-point function at $z=1$.
It is then convenient to introduce radial coordinate $\r$ of \cite{Hogervorst:2013sma}
\be\label{eq:rhocoordapp}
\rho = \frac{z}{(1 + \sqrt{1 - z})^2}, \quad \bar{\rho} = \frac{\bar{z}}{(1 + \sqrt{1 - \bar{z}})^2},
\ee
that maps $\cR$ to a smaller region, where convergence of the series is ensured. Indeed the point $z=1$ is mapped to $\r=1$, while $z=\infty$ is mapped to $\r=-1$, hence the radius of convergence in the $\r$ plane is still $|\r|=1$, and the region $\cR$ is mapped to a region strictly inside the unit disk, see Fig.~\ref{fig:regionRho}.

\subsection{Anomalous dimensions}

Once we have computed the beta function as in \eqref{eq:beta-app}, we can compute the associated critical exponent, that for a one-dimensional flow is simply given by $d\b/dg$ evaluated at the fixed point $g_* = \pm \sqrt{\d/\b_3}$. 
From this, we can then obtain the scaling dimension of $\cO$ at the IR fixed point:
\be
\D_{\cO} = 2 + \f{d\b}{dg}\Big|_{g=g_*} = 2 + 2\d \, .
\ee

We are of course interested in computing also the scaling dimensions of other operators $\Phi_i$ at the fixed point.
In order to do that, we can compute the mixing matrix by renormalizing for example its two-point function. 
Alternatively, we can use the trick of introducing it as a perturbation that we will switch off at the fixed point, and obtain the anomalous dimension as a critical exponent, as for $\cO$. 
The two methods are of course completely equivalent.

Consider the two-point function of $\Phi_i$, expanded in powers of $g_0$:
\ba
\la \Phi_i(0)  \Phi_i(\infty) \ra_{g_0} = & \la \Phi_i(0)  \Phi_i(\infty) \ra - g_0 \int \rmd^2 z \, \la \cO(z,\zb) \Phi_i(0)  \Phi_i(\infty) \ra \\
& + \f12 g_0^2  \int \rmd^2 z_1 \rmd^2 z_2\ \la \cO(z_1,\zb_1)\cO(z_2,\zb_2) \Phi_i(0)  \Phi_i(\infty) \ra + O(g_0^3) \,.
\ea
As usual, the integrals lead to UV divergences, and we cure them by a multiplicative renormalization of the operator: $\Phi_i\to [\Phi_i]= Z_i{^{-1}}(g_0)\Phi_i$.
However, the point at infinity lies outside the integration region, and thus the operator $\Phi_i(\infty)$ does not need to be renormalized.

Assuming that $\Phi_i$ is just an operator in the original minimal model, i.e.\ it is independent of $\chi$, the three-point function $\la\cO\Phi_i\Phi_i\ra$ vanishes, and hence the result can be simplified to 
\be
Z_i{^{-1}}(g_0)\la\Phi_i(0)\Phi_i(\infty)\ra_{g_0}  = Z_i{^{-1}}(g_0) \left(1 + I_{\Phi  \cO\cO \Phi}\, g_0^2 +O(g_0^4) \right)\, ,
\ee
and therefore we need to set
\be \label{eq:Zi}
Z_i(g_0) = 1 {+} I^{\,\rm div}_{\Phi  \cO\cO \Phi}\, g_0^2  +O(g_0^4) \, .
\ee

The four-point function is again simplified by rescaling $z_1$ and effectively setting $z_2=1$ .
However, now the $s$, $t$ and $u$ channels do not contribute equally.
To compensate, we use translation invariance\footnote{Translation invariance is broken by the regularization, but we assume it only affect the finite part, not the diverging one.} to shift the operators around such that the integration regions are centered around each inserted operator and correspond to \eqref{eq:integ-region}
\ba\label{eq:Ipoop}
I^{\,\rm div}_{\Phi  \cO\cO \Phi} &= \f12\int_{\cD^2} \rmd^2 z_1\rmd^2z_2\,\la\Phi_i(0)\cO(z_1,\zb_1)\cO(z_2,\zb_2)\Phi_i(\infty)\ra\\
&\sim \f12 \int_0^R \f{\rmd |z_3|}{|z_3|^{2\D_{\cO}-3}} \int \rmd^2 z\, \la\Phi_i(0)\cO(z,\zb)\cO(1)\Phi_i(\infty)\ra\\
&=  R^{2\delta} \frac{\pi}{2\delta} \int_\mathcal{R} \rmd^2 z\, \left[2 \la\Phi_i(0)\cO(z,\zb)\cO(1)\Phi_i(\infty)\ra+ \la\cO(0)\cO(z,\zb)\Phi_i(1)\Phi_i(\infty)\ra\right]\, .
\ea
As in the discussion above, it should be understood that in general the integrand in this expression needs subtractions, corresponding to divergences caused by operators in the OPE $\Phi_i\times\cO$ with dimension smaller than that of $\Phi_i$. Moreover, if there exist other operators in the OPE $\Phi_i\times\cO$ with the same dimension as $\Phi_i$, these lead to new logarithmic divergences, and thus they require that $[\Phi_i]$ include a nontrivial mixing with them; in this case the renormalization factor $Z_i$ becomes a mixing matrix $Z_i{}^j$. 

In the method reviewed in \cite{Komargodski:2016auf}, one introduces instead $\l_{i,0}\Phi_i$ as a perturbation in the action. Considering the observable $\la\Phi_i(\infty)\ra_{g_0,\l_{i,0}}$, expanding to linear order in $\l_{i,0}$ and quadratic in $g_0$, and redefining 
$\l_{i,0}= R^{\D_i^0-2}\l_i Z_i{^{-1}}(g)$, one finds exactly the same formulas as above, times an overall factor $ R^{\D_i^0-2}\l_i $. The beta function of $\l_i$ then reads
\be
\b_i(g,\l_i) = R \f{d}{dR}\l_i = (\D_i^0-2)\, \l_i + \l_i \g_i(g) \, , \quad \g_i ={-} R \f{d}{dR}\ln Z_i   \,.
\ee

From a Callan-Symanzik argument (e.g.\ \cite{Benedetti:2020yvb}) we find that the scaling dimension $\D_i$ of $[\Phi_i]$ at the fixed point is
\be
\D_i = 2+ \f{d\b_\l}{d\l}\Big|_{g=g_*} = \D_i^0 + \g_i(g_*)  \,.
\ee
Using \eqref{eq:Zi}, and the general formula for $I^{\,\rm div}_{\Phi  \cO\cO \Phi} $, we have
\be
\g_i(g) = - \pi\, g^2 \int_\mathcal{R} \rmd^2 z\, \left[2 \la\Phi_i(0)\cO(z,\zb)\cO(1)\Phi_i(\infty)\ra+ \la\cO(0)\cO(z,\zb)\Phi_i(1)\Phi_i(\infty)\ra\right]\, .
\ee
%

%% file: sections/app_MFT.tex

\section{Near mean field theory computations}
\label{app:MFT}

While the Lagrangian \eqref{eq:LG} is suited to standard perturbation computation in momentum space (see for example \cite{Benedetti:2020rrq,Benedetti:2024mqx} for the case $m=3$), we will choose here an approach in direct space paralleling conformal perturbation theory, which allows us to simplify the computations for general $m$.

With respect to appendix~\ref{app:CPT}, the main differences in the near-MFT case are that $(i)$ the three-point function of the perturbing operator is non-vanishing, and $(ii)$ the unperturbed theory is Gaussian, hence we can construct the three and four-point functions by Wick contractions.

\subsection{Integrals toolkit}

We report here the integrals involved in the computation of a long-range Gaussian three and four-point functions, with dimension $2(m-1)\Delta = 2- \eps$.

\begin{itemize}
    \item The integral of a single power of $1/\abs{x}^{2 a\D}$ presents a pole in $1/\eps$ for $a=m-1$
\be
D_1(a)
= \int_V \rmd^2 x\, \frac{1}{\abs{x}^{2a\Delta}}=
\int_V \rmd^2 x\, \frac{1}{\abs{x}^{2-\eps}}\delta_{a,m-1}= \frac{2\pi}{\eps}R^{\eps}\,\delta_{a,m-1}+(1-\delta_{a,m-1})O(1).
\ee
    \item The double integral factorizes in the large $V$ limit and contributes only for $a$ or $b=m-1$ 
\be\label{eq:MFT-D2}
D_2(a,b)
= \int_V \rmd^2 x_1 \rmd^2 x_2\, \frac{1}{\abs{x_1}^{2a\Delta}}\frac{1}{\abs{x_2}^{2b\Delta}}\sim \int_V \rmd^2 x_1 \rmd^2 x_2\, \frac{1}{\abs{x_1}^{2a\Delta}}\frac{1}{\abs{x_1-x_2}^{2b\Delta}}=D_1(a)D_1(b).
\ee
\end{itemize}%
The $\sim$ stands for the large $V$ limit where boundary terms are neglected and translation invariance is assumed to hold, as explained in appendix \ref{app:CPT}.
\begin{itemize}
    \item As an intermediate result, the following integral is finite in the limit $R\rightarrow\infty$ and $\eps>0$
\be\label{eq:MFT-Dtilde}
    \tilde{D}(a,b) 
    = \int \rmd^2 x_2\,\frac{1}{\abs{x_2}^{2a\Delta}}\frac{1}{\abs{x_1-x_2}^{2b\Delta}}
    = \pi f\left(a\D,b\D,2-(a+b)\D\right) \abs{x_1}^{2-2(a+b)\Delta},
\ee
\end{itemize} 
with the function $f(a,b,c)$ given by:
\be\label{MFT-f}
f(a,b,c) = \frac{\Gamma\left(1-a\right)\Gamma\left(1-b\right)\Gamma\left(1-c\right)}{\Gamma\left(a\right)\Gamma\left(b\right)\Gamma\left(c\right)}.
\ee
\begin{itemize}
    \item Then, using the intermediate result \eqref{eq:MFT-Dtilde},
\be\label{eq:MFT-D3}
D_3(a,b,c)= \int_V \rmd^2 x_1\rmd^2 x_2\, \frac{1}{\abs{x_1}^{2c\Delta}}\frac{1}{\abs{x_2}^{2a\Delta}}\frac{1}{\abs{x_1-x_2}^{2b\Delta}} \sim \frac{\pi^2}{\eps} f\left(a\D,b\D,2-(a+b)\D\right) R^{2\eps}
\ee
\end{itemize}
Note that similarly to $D_2$ \eqref{eq:MFT-D2}, the permutation symmetry in $(a,b,c)$ inside the integral $D_3$ is broken by the regularization.
Indeed, the function $f(a,b,c)$ \eqref{MFT-f} is well defined only for $0< a,b,c<1$. Outside this domain, the function as to be analytically continued, which can generate extra terms corresponding to IR divergencies (see IR rearrangement in \cite{Kleinert:2001hn}).
As we are only interested in the UV divergencies, we will neglect those terms and freely use the assumed permutation symmetry of $(a,b,c)$ to only evaluate it on its domain of definition, that is for $(a+b)\D > 1$ or equivalently $a+b>m-1$.

For clarity, the function $f\left(a\D,b\D,2-(a+b)\D\right)$ can be rewritten under the additional condition $a+b+c=2(m-1)$ as $f\left(a\D,b\D,c\D+\eps\right)$. 
In particular, for $a$ or $b=m-1$,  it presents a second pole in $1/\eps$ in addition to the factorized pole \eqref{eq:MFT-D3}:
\be
f\left(a\D,b\D,c\D+\eps\right)\Big|_{\begin{subarray}{l}a+b+c = 2(m-1)\\ a+b > m-1\\a,b,c>0\end{subarray}} = \left\{ \begin{array}{l}
\frac{\Gamma\left(1-\frac{a}{m-1}\right)\Gamma\left(1-\frac{b}{m-1}\right)\Gamma\left(1-\frac{c}{m-1}\right)}{\Gamma\left(\frac{a}{m-1}\right)\Gamma\left(\frac{b}{m-1}\right)\Gamma\left(\frac{c}{m-1}\right)} + O(\eps),\quad a,b\neq m-1\\
\left[\frac{2}{\eps} - \left(H_{-\frac{a}{m-1}} + H_{-\frac{c}{m-1}}\right) + O(\eps)\right]\delta_{b,m-1} + (a\leftrightarrow b),
\end{array}\right.
\ee
with $H_{-n}$ denoting the Harmonic number. 

\subsection{Beta function}

From conformal perturbation theory the one and two-loops order of the beta function \eqref{eq:beta_MFT} are given by the poles of the integrals of the three and four-points functions of the interacting field $\phi^{2(m-1)}$ respectively:
\be\begin{split}
A &= \left.\frac{1}{2}\left(\frac{1}{(2m-2)!}\right)^2 \int_V \rmd^2 x\ \la\phi^{2(m-1)}(\infty)\phi^{2(m-1)}(x)\phi^{2(m-1)}(0)\ra_0\right|_{1/\eps \text{ pole}},\\
B_1 &= \left.\frac{1}{3!}\left(\frac{1}{(2m-2)!}\right)^3 \int_V \rmd^2 x_1\rmd x_2\ \la\phi^{2(m-1)}(\infty)\phi^{2(m-1)}(x_1)\phi^{2(m-1)}(x_2)\phi^{2(m-1)}(0)\ra_0\right|_{1/\eps \text{ pole}},\\
B_2 &= \left.\frac{1}{3!}\left(\frac{1}{(2m-2)!}\right)^3 \int_V \rmd^2 x_1\rmd x_2\ \la\phi^{2(m-1)}(\infty)\phi^{2(m-1)}(x_1)\phi^{2(m-1)}(x_2)\phi^{2(m-1)}(0)\ra_0\right|_{1/\eps^2 \text{ pole}}.\\
\end{split}
\ee
The notation $|_{1/\eps}$ represents the coefficient of the $1/\eps$ pole.
By consistency of the RG flow, the double poles contribution arising at two-loops cancel as 
\be\label{eq:MFTconststency}
A^2=B_2.
\ee%
The monomial $\phi^{2(m-1)}$ is a primary operator of the generalized free field CFT.
The convention is choosen such that its two-point functions in normalized by the coupling's prefactor.
In addition, the field $\phi$ is a protected operator of long-range theory and do not get renormalized.
Therefore, no other integrals are expected to contribute to the beta function at this order.

The correlators are evaluated in the unperturbed theory -- a Gaussian theory -- and can be evaluated through wick contractions.
The integrals are represented diagrammatically by drawing a line between two vertices for each possible Wick contraction. 
For simplicity, we denote by a single line all the contractions between fields placed at the same point with an associated label representing its multiplicity. 
Then, as for Feynman diagrams, each line of multiplicity $a$ contribute to the integral as $1/\abs{x}^{2a\D}$.
The point at infinity plays the role of the "external legs" because it is not integrated over.
In table \ref{tab:MFT-betadiag} the diagrams contributing to the beta function are listed along side their associated combinatorial number (the number of equivalent diagrams). By dashed lines, we mean lines with zero multiplicity, i.e removed lines.
As in dimensional regularization, no contribution of tadpoles and disconnected diagrams are taken into account.\medskip

As an example, we compute the combinatorial factor of the one-loop diagram \ref{tab:MFT-betadiag} explicitly:
\ba
&S_{\text{1-loop}}=\\
&{\scriptstyle\underbrace{\frac{1}{2} \left(\frac{1}{(2m-2)!}\right)^2\vphantom{\binom{2(m-1)}{m-1}}}_{\text{Prefactor}}\,  \times\,  \underbrace{\binom{2(m-1)}{m-1}}_{\substack{\text{Choose $(m-1)$}\\ \text{legs from $\infty$}}}\,\underbrace{\binom{2(m-1)}{m-1}}_{\substack{\text{Choose $(m-1)$}\\ \text{legs from $0$}}}\underbrace{(m-1)!\vphantom{\binom{2(m-1)}{m-1}}}_{\substack{\text{Ways to connect}\\ \text{0 to $\infty$}}}\times\,  \underbrace{\binom{2(m-1)}{m-1}}_{\substack{\text{Choose $(m-1)$}\\ \text{legs from $x$}}}\underbrace{(m-1)!\vphantom{\binom{2(m-1)}{m-1}}}_{\substack{\text{Connect left over}\\ \text{from 0 to $x$}}} \times\,  \underbrace{(m-1)!\vphantom{\binom{2(m-1)}{m-1}}}_{\substack{\text{Connect}\\ \text{$x$ to $\infty$}}}}.
\ea
All other combinatorial factors are computed in a similar way.
\vspace{-0.28cm}
\begin{table}[H]
\centering
\begin{tabular}{|c|c|c|c|c|}
    \hline
     & Diagram & Conditions & Combinatorial factor& Integral\\
    \hline\hline
$\lambda_0^1$
& \begin{tikzpicture}[anchor=base, baseline,scale=0.7]
    \tikzset{line/.style={thin}}

    \node[circle,draw=black,fill=black,inner sep=0pt,minimum size=4pt,label=right:{$\infty$}] (x) at (3,0) {};
    \node[circle,draw=black,fill=black,inner sep=0pt,minimum size=4pt,label=left:{$0$}] (0) at (0,0) {};
    \draw (x) -- (0) node [midway, above] {$2(m-1)$};
\end{tikzpicture} & & $1$ & \\
\hline
$\lambda_0^2$ & \begin{tikzpicture}[anchor=base, baseline=(current bounding box.center),scale=0.5]
    \tikzset{line/.style={thin}}

    \node[circle,draw=black,fill=black,inner sep=0pt,minimum size=4pt,label=right:{$x$}] (x) at (4,0) {};
    \node[circle,draw=black,fill=black,inner sep=0pt,minimum size=4pt,label=left:{$0$}] (0) at (0,0) {};
    \node[circle,draw=black,fill=black,inner sep=0pt,minimum size=4pt,label=above:{$\infty$}] (inf) at (2, 2*sqrt 3){}; 
    \draw (0) -- (inf) node [midway, left] {$a$};
    \draw (inf) -- (x) node [midway, right] {$b$};
    \draw (x) -- (0) node [midway, below] {$c$};
\end{tikzpicture}
& $a=b=c=m-1$ & $S_{\text{1-loop}}(m)= \frac{1}{2} \frac{(2m-2)!}{(m-1)!^3}$ & $D_1(c)$\\
    \hline
$\lambda_0^3$ & \begin{tikzpicture}[anchor=base, baseline=(current bounding box.center),scale=0.55]
    \tikzset{line/.style={thin}}

    \node[circle,draw=black,fill=black,inner sep=0pt,minimum size=4pt,label=right:{$x_1$}] (x1) at (4,0) {};
    \node[circle,draw=black,fill=black,inner sep=0pt,minimum size=4pt,label=left:{$0$}] (0) at (0,0) {};
    \node[circle,draw=black,fill=black,inner sep=0pt,minimum size=4pt,label=above:{$x_2$}] (x2) at (2, 2 * sqrt 3) {}; 
    \node[circle,draw=black,fill=black,inner sep=0pt,minimum size=4pt,label=below:{$\infty$}] (inf) at (2,2/3 * sqrt 3) {}; 
    \draw (0) -- (x2) node [midway, left] {$a$};
    \draw (x2) -- (x1) node [midway, right] {$b$};
    \draw (x1) -- (0) node [midway, below] {$c$};
    \draw (0) -- (inf) node [midway, above] {$d$};
    \draw (x2) -- (inf) node [midway, right] {$e$};
    \draw (x1) -- (inf) node [midway, above] {$f$};
\end{tikzpicture}&
$\begin{aligned}
        &a=f,\,b=d,\,c=e,\\
        &a+b+c=2(m-1),\\
        &a,b,c\neq0
\end{aligned}$ &
$\begin{aligned}
        &\sum_{\substack{a+b+c=2(m-1)\\ a,\,b,\,c\,\neq 0}}S_{a,b,c}(m)\\
        &=\sum_{\substack{a+b+c=2(m-1)\\ a,\,b,\,c\,\neq 0}}\frac{1}{3!}\frac{(2m-2)!}{(a!b!c!)^2}
\end{aligned}$&$D_3(a,b,c)$\\
\cline{2-5}
& \begin{tikzpicture}[anchor=base, baseline=(current bounding box.center),scale=0.55]
    \tikzset{line/.style={thin}}

    \node[circle,draw=black,fill=black,inner sep=0pt,minimum size=4pt,label=right:{$x_1$}] (x1) at (4,0) {};
    \node[circle,draw=black,fill=black,inner sep=0pt,minimum size=4pt,label=left:{$0$}] (0) at (0,0) {};
    \node[circle,draw=black,fill=black,inner sep=0pt,minimum size=4pt,label=above:{$x_2$}] (x2) at (2,2 *sqrt 3) {}; 
    \node[circle,draw=black,fill=black,inner sep=0pt,minimum size=4pt,label=below:{$\infty$}] (inf) at (2,2/3 * sqrt 3) {}; 
    \draw (0) -- (x2) node [midway, left] {$a$};
    \draw (x2) -- (x1) node [midway, right] {$b$};
    \draw[dashed] (x1) -- (0);
    \draw (0) -- (inf) node [midway, above] {$d$};
    \draw[dashed] (x2) -- (inf);
    \draw (x1) -- (inf) node [midway, above] {$f$};
\end{tikzpicture}&
$\begin{aligned}
        &a=b=d=f,\\
        &a=m-1\\
\end{aligned}$&
$\begin{aligned}
        &3\, S_{m-1,m-1,0}(m)\\
        &=3\,\frac{1}{3!}\frac{(2m-2)!}{((m-1)!)^4}
\end{aligned}$&$D_2(a,b)$\\
\hline
\end{tabular}
\caption{Diagrams contributing to the beta function for generic $m$.}
\label{tab:MFT-betadiag}
\end{table}


Combining all contributions, the factors $A$ \eqref{eq:1loop_MFT} and $B_1$ \eqref{eq:hard_sums} of the beta function are given by:
\be\begin{split}
A &= \left. \vphantom{\sum}
S_{\text{1-loop}}(m) D_1(m-1)
\right|_{1/\eps \text{ pole}}\\
B_1 &=
\sum_{\substack{a+b+c=2(m-1)\\ a,\,b,\,c\,\neq 0\\a,\,b,\,c\,\neq m-1}} S_{a,b,c}(m) D_3(a,b,c)
+ 3\sum_{\substack{a+b=(m-1)\\ a,\,b\,\neq 0}}S_{a,b,m-1}(m) D_3(a,m-1,b)
\left.\vphantom{\sum}\right|_{1/\eps \text{ pole}}\\
B_2 &=
 3\sum_{\substack{a+b=(m-1)\\ a,\,b\,\neq 0}} S_{a,b,m-1}(m) D_3(a,m-1,b)\left.\vphantom{\sum}+ 3\, S_{(m-1),(m-1),0}(m) D_2(m-1,m-1)
\right|_{1/\eps^2 \text{ pole}}\\
\end{split}
\ee
On one hand, the sums in $B_1$ do not admit an easily reachable closed form.
Nonetheless, they have a finite number of terms for a given value of $m$ and can be straightforwardly implemented in Mathematica.
On the other hand, it can be shown that the contributions in $B_2$ satisfy the consistency condition \eqref{eq:MFTconststency}. 

\subsection{Anomalous dimension of monomial operators}

The monomial operator $\phi^\alpha$, $\alpha\in \N$ is the simplest primary operator of the GFF theory.
As for the beta function, its anomalous dimension \eqref{eq:gamma} is controlled at one and two-loop order by the poles of the integrals of the three and four-point functions of the interacting field $\phi^{2(m-1)}$ with the operator considered $\phi^\alpha$:
\be\begin{split}
    \tilde{A} &= \left.\frac{1}{(2m-2)!}\int_V \rmd^2 x\, \la\phi^\alpha(\infty)\phi^\alpha(0)\phi^{2(m-1)}(x)\ra_0\right|_{1/\eps \text{ pole}},\\
    \tilde{B}_1 &= \left.\left(\frac{1}{(2m-1)!}\right)^2 \int_V \rmd^2 x_1\rmd x_2\, \la\phi^{\alpha}(\infty)\phi^{\alpha}(0)\phi^{2(m-1)}(x_1)\phi^{2(m-1)}(x_2)\ra_0\right|_{1/\eps \text{ pole}},\\
    \tilde{B}_2 &= \left.\left(\frac{1}{(2m-1)!}\right)^2 \int_V \rmd^2 x_1\rmd x_2\, \la\phi^{\alpha}(\infty)\phi^{\alpha}(0)\phi^{2(m-1)}(x_1)\phi^{2(m-1)}(x_2)\ra_0\right|_{1/\eps^2 \text{ pole}}.
\end{split}\ee
Once again, there is a consistency condition
\be
\tilde{B}_2 = \tilde{A}^2 + A\tilde{A}
\ee
which makes the two-loop coefficient in $\overline{\text{MS}}$ scheme regular in $\eps$.
The list of diagrams contributing are given in table \ref{tab:MFT-anodiag}. Tadpoles and disconnected diagrams do not contribute to the UV divergence.
The combinatorial factor involved is defined as follow:
\be
\tilde{S}^{a,b,c}_{d}(m,\alpha)  =\left(\frac{\alpha!}{a!c!}\right)^2\frac{1}{b!d!}.
\ee
Noteworthy, the diagrams are heavily constrained by the value of $\alpha$ taken.
Indeed, for $\alpha=1$, no connected diagram can be drawn at this order in perturbation. This is consistent with the knowledge that $\phi$ does not get renormalized in long-range theories.
For $\alpha \le (m-1)$, the one-loop order also vanishes under the Wick contractions' constrains.\bigskip

Putting all contributions together gives the one and two-loop results:
\be
\tilde{A} = \left.\tilde{S}_{\text{1-loop}}(m,\alpha) D_1(m-1)\right|_{1/\eps \text{ pole}},
\ee
and
\be\begin{split}
    \tilde{B}_1 =& \sum_{a=1}^{\min(\alpha,2(m-1))-2}\sum_{b=\max(0,2(m-1)-\alpha)+1}^{2(m-1)-a-1}\tilde{S}^{a,b,2(m-1)-a-b}_{b+\alpha-2(m-1)}D_3(a,b,2m-2-a-b)\\
    &+ \Theta(2m-2-\alpha)\sum_{a=1}^{\alpha-1}\tilde{S}^{a,2(m-1)-\alpha,\alpha-a}_{0}D_3(a,2(m-1)-\alpha,\alpha-a) \left.\vphantom{\sum}\right|_{1/\eps \text{ pole}},\\
    \tilde{B}_2 =& \sum_{a=1}^{\min(\alpha,2(m-1))-2}\sum_{b=\max(0,2(m-1)-\alpha)+1}^{2(m-1)-a-1}\tilde{S}^{a,b,2(m-1)-a-b}_{b+\alpha-2(m-1)}D_3(a,b,2m-2-a-b)\\
    &+ \Theta(2m-2-\alpha)\sum_{a=1}^{\alpha-1}\tilde{S}^{a,2(m-1)-\alpha,\alpha-a}_{0}D_3(a,2(m-1)-\alpha,\alpha-a)\\
    &+\left[\Theta(\alpha-2m+2)\tilde{S}^{m-1,0,m-1}_{m-1}+2\Theta(\alpha-m+1)\tilde{S}^{m-1,m-1,0}_{m-1}\right.\\
    &\left.\vphantom{\sum}\left.+\delta_{2m-2-\alpha}\tilde{S}^{m-1,0,m-1}_{0}\right]D_2(m-1,m-1)\right|_{1/\eps^2 \text{ pole}}
\end{split}\ee
where $\Theta$ is the Heaviside function with $\Theta(0) = 0$.
Note that the computation above are consistent with respect to the RG flow and all double poles cancel each other.
In fact, $D_2(m-1,m-1)$ \eqref{eq:MFT-D2} is proportional to $1/\eps^2$ and only the first two diagrams at two-loops in table \ref{tab:MFT-anodiag} contribute to the simple pole.

\begin{table}[H]
\begin{tabular}{|c|c|l|c|c|}
    \hline
     & Diagram & Conditions & Combinatorial factor & Integral\\
    \hline\hline
$\lambda_0^0$
& \begin{tikzpicture}[anchor=base, baseline,scale=0.7]
    \tikzset{line/.style={thin}}

    \node[circle,draw=black,fill=black,inner sep=0pt,minimum size=4pt,label=right:{$\infty$}] (x) at (3,0) {};
    \node[circle,draw=black,fill=black,inner sep=0pt,minimum size=4pt,label=left:{$0$}] (0) at (0,0) {};
    \draw (x) -- (0) node [midway, above] {$\alpha$};
\end{tikzpicture} & & $\alpha!$ & \\
\hline
$\lambda_0^1$ & \begin{tikzpicture}[anchor=base, baseline=(current bounding box.center),scale=0.6]
    \tikzset{line/.style={thin}}

    \node[circle,draw=black,fill=black,inner sep=0pt,minimum size=4pt,label=right:{$x$}] (x) at (3,0) {};
    \node[circle,draw=black,fill=black,inner sep=0pt,minimum size=4pt,label=left:{$0$}] (0) at (0,0) {};
    \node[circle,draw=black,fill=black,inner sep=0pt,minimum size=4pt,label=above:{$\infty$}] (inf) at (1.5, 3/2*sqrt 3){}; 
    \draw (0) -- (inf) node [midway, left] {$a$};
    \draw (inf) -- (x) node [midway, right] {$b$};
    \draw (x) -- (0) node [midway, below] {$c$};
\end{tikzpicture}
& $\begin{aligned}
        &a=\alpha-(m-1),\\
        &b=c=2(m-1)\\
        &a,b,c>0
\end{aligned}$
&$\begin{aligned}&\tilde{S}_{\text{1-loop}}(m,\alpha)  =\\ 
    &\left\{\begin{array}{l}
\frac{\alpha!^2}{(m-1)!^2(\alpha-m+1)!}, \quad \alpha> m-1\\
0, \quad \alpha\le m-1
\end{array}\right.\end{aligned}$ & $D_1(m - 1)$ \\
\hline
$\lambda_0^2$ & \begin{tikzpicture}[anchor=base, baseline=(current bounding box.center),scale=0.55]
    \tikzset{line/.style={thin}}

    \node[circle,draw=black,fill=black,inner sep=0pt,minimum size=4pt,label=right:{$x_1$}] (x1) at (4,0) {};
    \node[circle,draw=black,fill=black,inner sep=0pt,minimum size=4pt,label=left:{$0$}] (0) at (0,0) {};
    \node[circle,draw=black,fill=black,inner sep=0pt,minimum size=4pt,label=above:{$x_2$}] (x2) at (2, 2 * sqrt 3) {}; 
    \node[circle,draw=black,fill=black,inner sep=0pt,minimum size=4pt,label=below:{$\infty$}] (inf) at (2,2/3 * sqrt 3) {}; 
    \draw (0) -- (x2) node [midway, left] {$a$};
    \draw (x2) -- (x1) node [midway, right] {$b$};
    \draw (x1) -- (0) node [midway, below] {$c$};
    \draw (0) -- (inf) node [midway, above] {$d$};
    \draw (x2) -- (inf) node [midway, right] {$c$};
    \draw (x1) -- (inf) node [midway, above] {$a$};
\end{tikzpicture}
& \small$\begin{aligned}
        &a<\min\left(\alpha,2m-2\right)-1\\
        &\max\left(0,2(m-1)-\alpha\right)<b\\
        &b<2(m-1)-a\\
        &c = 2(m-1)-a-b\\
        &d = b+\alpha-2(m-1)\\
        &a,c,d>0
    \end{aligned}$
&\makecell{$\begin{aligned}\displaystyle\sum_{a=1}^{\min(\alpha,2(m-1))-2}\hspace{-12pt}&\sum_{b=\max(0,2(m-1)-\alpha)+1}^{2(m-1)-a-1}\\&\vphantom{\int^{a^{a^a}}}\tilde{S}^{a,b,2(m-1)-a-b}_{b+\alpha-2(m-1)}\end{aligned}$
} & $D_3(a,b,c)$ \\
\cline{2-5}
 & \begin{tikzpicture}[anchor=base, baseline=(current bounding box.center),scale=0.55]
    \tikzset{line/.style={thin}}

    \node[circle,draw=black,fill=black,inner sep=0pt,minimum size=4pt,label=right:{$x_1$}] (x1) at (4,0) {};
    \node[circle,draw=black,fill=black,inner sep=0pt,minimum size=4pt,label=left:{$0$}] (0) at (0,0) {};
    \node[circle,draw=black,fill=black,inner sep=0pt,minimum size=4pt,label=above:{$x_2$}] (x2) at (2, 2 * sqrt 3) {}; 
    \node[circle,draw=black,fill=black,inner sep=0pt,minimum size=4pt,label=below:{$\infty$}] (inf) at (2,2/3 * sqrt 3) {}; 
    \draw (0) -- (x2) node [midway, left] {$a$};
    \draw (x2) -- (x1) node [midway, right] {$b$};
    \draw (x1) -- (0) node [midway, below] {$c$};
    \draw[dashed] (0) -- (inf);
    \draw (x2) -- (inf) node [midway, right] {$c$};
    \draw (x1) -- (inf) node [midway, above] {$a$};
\end{tikzpicture}
& \small$\begin{aligned}
        &0<a<\alpha<2(m-1)\\
        &b=2(m-1)-\alpha\\
        &c = \alpha-a\\
        &b,c>0
    \end{aligned}$
&$\begin{aligned}\Theta(2m-2-\alpha)\times\\\displaystyle\sum_{a=1}^{\alpha-1}\tilde{S}^{a,2(m-1)-\alpha,\alpha-a}_{0}\end{aligned}$ & $D_3(a,b,c)$ \\
\cline{2-5}
 & \begin{tikzpicture}[anchor=base, baseline=(current bounding box.center),scale=0.55]
    \tikzset{line/.style={thin}}

    \node[circle,draw=black,fill=black,inner sep=0pt,minimum size=4pt,label=right:{$x_1$}] (x1) at (4,0) {};
    \node[circle,draw=black,fill=black,inner sep=0pt,minimum size=4pt,label=left:{$0$}] (0) at (0,0) {};
    \node[circle,draw=black,fill=black,inner sep=0pt,minimum size=4pt,label=above:{$x_2$}] (x2) at (2, 2 * sqrt 3) {}; 
    \node[circle,draw=black,fill=black,inner sep=0pt,minimum size=4pt,label=below:{$\infty$}] (inf) at (2,2/3 * sqrt 3) {}; 
    \draw (0) -- (x2) node [midway, left] {$a$};
    \draw[dashed] (x2) -- (x1);
    \draw (x1) -- (0) node [midway, below] {$c$};
    \draw (0) -- (inf) node [midway, above] {$d$};
    \draw (x2) -- (inf) node [midway, right] {$c$};
    \draw (x1) -- (inf) node [midway, above] {$a$};
\end{tikzpicture}
& \small$\begin{aligned}
        &0<a<2(m-1)<\alpha\\
        &c = 2(m-1)-a\\
        &d = \alpha-2(m-1)\\
        &c,d>0
    \end{aligned}$
&$\Theta(\alpha-2m+2)\tilde{S}^{m-1,0,m-1}_{m-1}$ & $D_2(a,c)$ \\
\cline{2-5}
 & \begin{tikzpicture}[anchor=base, baseline=(current bounding box.center),scale=0.55]
    \tikzset{line/.style={thin}}

    \node[circle,draw=black,fill=black,inner sep=0pt,minimum size=4pt,label=right:{$x_1$}] (x1) at (4,0) {};
    \node[circle,draw=black,fill=black,inner sep=0pt,minimum size=4pt,label=left:{$0$}] (0) at (0,0) {};
    \node[circle,draw=black,fill=black,inner sep=0pt,minimum size=4pt,label=above:{$x_2$}] (x2) at (2, 2 * sqrt 3) {}; 
    \node[circle,draw=black,fill=black,inner sep=0pt,minimum size=4pt,label=below:{$\infty$}] (inf) at (2,2/3 * sqrt 3) {}; 
    \draw (0) -- (x2) node [midway, left] {$a$};
    \draw (x2) -- (x1) node [midway, right] {$b$};
    \draw[dashed] (x1) -- (0);
    \draw (0) -- (inf) node [midway, above] {$d$};
    \draw[dashed] (x2) -- (inf);
    \draw (x1) -- (inf) node [midway, above] {$a$};
\end{tikzpicture}
& \small$\begin{aligned}
        &0<a<\min\left(\alpha,2(m-1)\right)\\
        &b = 2(m-1)-a\\
        &d = \alpha-a\\
        &b,d>0
    \end{aligned}$
&$2\,\Theta(\alpha-m+1)\tilde{S}^{m-1,m-1,0}_{m-1}$ & $D_2(a,b)$ \\
\cline{2-5}
 & \begin{tikzpicture}[anchor=base, baseline=(current bounding box.center),scale=0.55]
    \tikzset{line/.style={thin}}

    \node[circle,draw=black,fill=black,inner sep=0pt,minimum size=4pt,label=right:{$x_1$}] (x1) at (4,0) {};
    \node[circle,draw=black,fill=black,inner sep=0pt,minimum size=4pt,label=left:{$0$}] (0) at (0,0) {};
    \node[circle,draw=black,fill=black,inner sep=0pt,minimum size=4pt,label=above:{$x_2$}] (x2) at (2, 2 * sqrt 3) {}; 
    \node[circle,draw=black,fill=black,inner sep=0pt,minimum size=4pt,label=below:{$\infty$}] (inf) at (2,2/3 * sqrt 3) {}; 
    \draw (0) -- (x2) node [midway, left] {$a$};
    \draw[dashed] (x2) -- (x1);
    \draw (x1) -- (0) node [midway, below] {$c$};
    \draw[dashed] (0) -- (inf);
    \draw (x2) -- (inf) node [midway, right] {$c$};
    \draw (x1) -- (inf) node [midway, above] {$a$};
\end{tikzpicture}
& \small$\begin{aligned}
        &\alpha=2(m-1)\\
        &0<a<2(m-1)\\
        &c = 2(m-1)-a\\
        &c>0
    \end{aligned}$
&$\delta_{2m-2-\alpha}\tilde{S}^{m-1,0,m-1}_{0}$ & $D_2(a,c)$ \\
\hline
\end{tabular}
\caption{Diagrams contributing to the anomalous dimension of $\phi^\alpha$, $\alpha\in \N$ for generic $m$.}
\label{tab:MFT-anodiag}
\end{table}

%% file: sections/mmcorrelators.tex

\section{Correlators in minimal models}
\label{app:mmcorrelators}

\subsection{Conventions}
We focus on unitary and diagonal minimal models $\mathcal{M}_{m+1,m}$ with central charge
\begin{equation}\label{central}
	c=1-\frac{6}{m(m+1)}\,,\qquad {m=3,\,4,\,5,\ldots\,}
\end{equation}
Holomorphic Virasoro primaries are labeled by positive-integer pairs $(r,s)$,
\begin{align}\label{eq:kactable}
	\phirs{r}{s}(z)\,,\quad 1\leq r\leq m-1\,,\quad 1\leq s\leq m\,,\quad (r,s) \cong (m-r,m+1-s)\,,
\end{align}
where $\phirs{1}{1}\equiv \id$ is the identity, and $\phirs{r}{s}$ has the following (holomorphic) scaling dimensions\footnote{Notice that $(r,s,m) \leftrightarrow (s,r,-1-m)$ happens to be a symmetry of this formula. This has been used in the main text to relate large-$m$ expansions of CFT data in the $\phi_{1,2}\chi$ flow to those in the $\phi_{2,1}\chi$ flow. In the case of $\phi_{2,2}\chi$, it has been used as a check.}
\begin{equation}\label{VirasoroPrimaries}
	h_{r,s}= \frac{\bigl( (m+1)r-ms \bigr)^2-1}{4m (m+1)}\,.
\end{equation}
We will always take the Virasoro primaries to be unit-normalized.

The (holomorphic) fusion rules are
\begin{align}\label{fusionrulesholo}
	\phirs{r}{s} \times \phirs{r'}{s'} = 
	\sum_{\substack{r''=|r-r'|+1 \\ r+r'+r'' {\rm odd}}}^{r_{\rm max}} \,\,
	\sum_{\substack{s''=|s-s'|+1 \\ s+s'+s'' {\rm odd}}}^{s_{\rm max}} \
	\phirs{r''}{s''}\,,
\end{align}
where 	$r_{\rm max}={\rm min}(r+r'-1,\;2m-r-r'-1)$ and $s_{\rm max}={\rm min}(s+s'-1,\;2m-s-s'+1)$.

The diagonal minimal models are obtained by gluing the holomorphic and anti-holomorphic sectors, such that the resulting physical spectrum contains only scalar Virasoro primaries, with scaling dimensions
\begin{align}
	\D_{r,s} = 2 h_{r,s}\,.
\end{align}

 These theories enjoy a $\mathbb{Z}_2$ symmetry, under which a Virasoro primary with labels $(r,s)$ in $\mathcal{M}_{m+1,m}$ can be assigned a definite charge \cite{Cardy:1986gw,Cappelli:1986hf,Ruelle:1998zu}
\begin{align}
	\epsilon^{(m)}_{(r,s)}=(-1)^{(m+1) r+m s+1}\,.
\end{align}

\subsection{\texorpdfstring{The four-point function with $\phirs{1}{2}$ and $\phirs{r}{s}$}{The four-point function with phirs and phi12}}
Consider the following correlator
\begin{align}\label{rs1212rs}
	\la \phirs{r}{s}(z_1,\bar{z}_1)\phirs{1}{2}(z_2,\bar{z}_2)\phirs{1}{2}(z_3,\bar{z}_3)\phirs{r}{s}(z_4,\bar{z}_4)\ra\,.
\end{align}

To compute such a correlator, we first solve an ODE for the holomorphic correlator
\begin{align}
	\mathcal{D}_2^{(z_2)}\la \phirs{r}{s}(z_1)\phirs{1}{2}(z_2)\phirs{1}{2}(z_3)\phirs{r}{s}(z_4)\ra=0\,,
\end{align}
where $\mathcal{D}_2^{(z_2)}$,  acting on $z_2$,  is the differential operator:
\begin{align}\label{secondD}
	\mathcal{D}_2^{(\cdot)}&= 	{\mathcal L}_{-2}^{(\cdot)}-\frac{3}{4 h_{1,2}+2}({\mathcal L}_{-1}^{(\cdot)})^2\,,
\end{align}
and ${\cal L}_{-n}^{(\cdot)}$ are defined as:
\begin{align}\label{Lmkdef}
	\mathcal{L}_{-k}^{(w)}&\equiv\sum_{i=1}^{3}\left(\frac{ (k-1)h_i}{(z_i-w)^k}-\frac{1}{(z_i-w)^{k-1}}\partial_i\right)\,,\qquad {\cal L}_{-1}^{(w)}=\partial_w\,.
\end{align}

This holomorphic correlator takes the following form ($z_{ij}\equiv z_i-z_j$):
\begin{align}\label{eq:rs1212rs}
	\la \phirs{r}{s}(z_1)\phirs{1}{2}(z_2)&\phirs{1}{2}(z_3)\phirs{r}{s}(z_4)\ra\nonumber\\
	&=\left(\frac{z_{14}}{z_{13}}\right)^{h_{1,2}-h_{r,s}}\left(\frac{z_{24}}{z_{14}}\right)^{h_{r,s}-h_{1,2}} \frac{\mathcal{F}_{(r,s)(1,2)(1,2)(r,s)}({z})}{(z_{12}z_{34})^{h_{r,s}+h_{1,2}}} \,,
\end{align}
with
\begin{align}\label{eta_def}
	{z}&= \frac{z_{12}z_{34}}{z_{13}z_{24}}\,, \quad \bar{z} = \frac{ \bar{z}_{12} \bar{z}_{34}}{ \bar{z}_{13} \bar{z}_{24}}\,.
\end{align}

The differential equation for $\mathcal{F}({z})\equiv\mathcal{F}_{(r,s)(1,2)(1,2)(r,s)}({z})$ then reads
\begin{equation}
	\begin{split}
		0&=-16 ({z} -1)^2 {z} ^2 (m+1)^2  \mathcal{F}''({z} )\nonumber\\
		&+8 ({z} -1) {z}  (m+1) [-2 {z} +4 ({z} -1) h_{r,s} (m+1)-3 {z}  m+m+2] \mathcal{F}'({z} )\nonumber\\
		&-[16 ({z} -1)^2 h_{r,s}^2 (m+1)^2+8 \left(3 {z} ^2-4 {z} +1\right) h_{r,s} m (m+1)\nonumber\\
		&\qquad+(m-2) \left(2 ({z} -1)^2+\left({z} ^2-6 {z} +1\right) m\right)] \mathcal{F}({z} )\,.
	\end{split}
\end{equation}
The two independent solutions to this differential equation, the Virasoro blocks, are found to be:
\begin{align}
	\mathcal{F}^{(r,s)(1,2)(1,2)(r,s)}_{(r,s+1)}(z)&=(1-{z} )^{\frac{m}{2 m+2}} {z} ^{h_{r,s+1}} \, _2F_1\left(\frac{m}{m+1},\frac{-r m+s m+m-r}{m+1};-r+\frac{m s}{m+1}+1;{z} \right)\,,\nonumber\\
	\mathcal{F}^{(r,s)(1,2)(1,2)(r,s)}_{(r,s-1)}(z)&=(1-{z} )^{\frac{m}{2 m+2}} {z} ^{h_{r,s-1}} \, _2F_1\left(\frac{m}{m+1},\frac{r m-s m+m+r}{m+1};r-\frac{m s}{m+1}+1;{z} \right)\,.
\end{align}

The correlator \eqref{rs1212rs} is a sesquilinear combination of holomorphic and anti-holomorphic blocks, with coefficients determined by crossing symmetry. The final result is:
\begin{align}
	\la \phirs{r}{s}(z_1,\bar{z}_1)&\phirs{1}{2}(z_2,\bar{z}_2)\phirs{1}{2}(z_3,\bar{z}_3)\phirs{r}{s}(z_4,\bar{z}_4)\ra\nonumber\\
	&=\left(\frac{|z_{14}|}{|z_{13}|}\right)^{\D_{1,2}-\D_{r,s}}\left(\frac{|z_{24}|}{|z_{14}|}\right)^{\D_{r,s}-\D_{1,2}} \frac{\mathcal{F}_{(r,s)(1,2)(1,2)(r,s)}({z},\bar{{z}})}{(|z_{12}||z_{34}|)^{\D_{r,s}+\D_{1,2}}}\,.
\end{align}
where
\begin{equation}
	\begin{split}
		\mathcal{F}_{(r,s)(1,2)(1,2)(r,s)}({z},\bar{{z}})= &(C_{(r,s)(1,2)(r,s+1)})^2\mathcal{F}^{(r,s)(1,2)(1,2)(r,s)}_{(r,s+1)}({z})\bar{\mathcal{F}}^{(r,s)(1,2)(1,2)(r,s)}_{(r,s+1)}(\bar{{z}})\nonumber\\
		&+(C_{(r,s)(1,2)(r,s-1)})^2\mathcal{F}^{(r,s)(1,2)(1,2)(r,s)}_{(r,s-1)}({z})\bar{\mathcal{F}}^{(r,s)(1,2)(1,2)(r,s)}_{(r,s-1)}(\bar{{z}})\,.
	\end{split}
\end{equation}

The anti-holomorphic blocks are obtained by evaluating the holomorphic blocks at ${z}\to\bar{z}$, i.e.  $\bar{\mathcal{F}}_{\phi}(\bar{{z}})\equiv \mathcal{F}_{\phi}(\bar{{z}})$. The coefficients of the sesquilinear combination are the OPE coefficients, and the latter can be computed via the Coulomb gas formalism \cite{Dotsenko:1984nm,Dotsenko:1984ad,Dotsenko:1985hi} reviewed in section \ref{sec:CGreview}.\footnote{See also the Mathematica notebook attached to the submission of \cite{Esterlis:2016psv}, for an implementation of Coulomb gas formulae.}

Next, we compute
\begin{align}
	\la \phirs{1}{2}(z_1,\bar{z}_1)\phirs{1}{2}(z_2,\bar{z}_2)\phirs{r}{s}(z_3,\bar{z}_3)\phirs{r}{s}(z_4,\bar{z}_4)\ra\,.
\end{align}

Again, we first solve the following second order ODE for the holomorphic correlator
\begin{align}
	\mathcal{D}_2^{(z_2)}\la \phirs{1}{2}(z_1)\phirs{1}{2}(z_2)\phirs{r}{s}(z_3)\phirs{r}{s}(z_4)\ra=0\,,
\end{align}
where $\mathcal{D}_2^{(z_2)}$ is the differential operator of eq.~\eqref{secondD} and we have (${z}$ is defined as in \eqref{eta_def})
\begin{align}
	\la \phirs{1}{2}(z_1)\phirs{1}{2}(z_2)\phirs{r}{s}(z_3)\phirs{r}{s}(z_4)\ra= \frac{\mathcal{F}_{(1,2)(1,2)(r,s)(r,s)}({z})}{(z_{12})^{2h_{1,2}}(z_{34})^{2h_{r,s}}} \,.
\end{align}

The differential equation for $\mathcal{F}({z})\equiv\mathcal{F}_{(1,2)(1,2)(r,s)(r,s)}({z})$
\begin{equation}
	0=	3 ({z} -1) \left(({z} -1) {z}  (m+1) \mathcal{F}''({z})+({z}  (m+2)-2)  \mathcal{F}'({z})\right)-3 {z}  h_{r,s} m  \mathcal{F}({z})\,,
\end{equation}
leads to the following Virasoro blocks
\begin{align}
	\mathcal{F}^{(1,2)(1,2)(r,s)(r,s)}_{(1,1)}(z)&=(1-{z} )^{\frac{m r-m s+r+1}{2 m+2}} \, _2F_1\left(\frac{1}{m+1},\frac{m r+r-m s+1}{m+1};\frac{2}{m+1};{z} \right)\,,\nonumber\\
	\mathcal{F}^{(1,2)(1,2)(r,s)(r,s)}_{(1,3)}(z)&={z} ^{h_{1,3}} (1-{z} )^{\frac{m r-m s+r+1}{2 m+2}} \, _2F_1\left(\frac{m}{m+1},\frac{r m-s m+m+r}{m+1};\frac{2 m}{m+1};{z} \right)\,.
\end{align}

Finally, we have
\begin{align}
	\la \phirs{1}{2}(z_1,\bar{z}_1)\phirs{1}{2}(z_2,\bar{z}_2)\phirs{r}{s}(z_3,\bar{z}_3)\phirs{r}{s}(z_4,\bar{z}_4)\ra= \frac{\mathcal{F}_{(1,2)(1,2)(r,s)(r,s)}({z},\bar{{z}})}{(|z_{12}|)^{2\D_{1,2}}(|z_{34}|)^{2\D_{r,s}}} \,,
\end{align}
where
\begin{equation}
	\begin{split}
		\mathcal{F}_{(1,2)(1,2)(r,s)(r,s)}({z},\bar{{z}})&=\mathcal{F}^{(1,2)(1,2)(r,s)(r,s)}_{(1,1)}({z})\bar{\mathcal{F}}^{(1,2)(1,2)(r,s)(r,s)}_{(1,1)}(\bar{{z}})\nonumber\\
		&+C_{(1,2)(1,2)(1,3)}C_{(r,s)(r,s)(1,3)}\mathcal{F}^{(1,2)(1,2)(r,s)(r,s)}_{(1,3)}({z})\bar{\mathcal{F}}^{(1,2)(1,2)(r,s)(r,s)}_{(1,3)}(\bar{{z}})\,,
	\end{split}
\end{equation}
with OPE coefficient
\begin{align}\label{bulkOPEcoeffrsrs13}
			C_{(r,s)(r,s)(1,3)}&=\frac{\Gamma \left(\frac{m}{m+1}\right) \Gamma \left(\frac{2}{m+1}-1\right) \sqrt{\frac{(m-1) \Gamma \left(\frac{m}{m+1}\right) \Gamma \left(\frac{2 m}{m+1}\right) \Gamma \left(\frac{3}{m+1}-1\right)}{m \Gamma \left(\frac{m-1}{m+1}\right) \Gamma \left(2-\frac{3}{m+1}\right) \Gamma \left(\frac{1}{m+1}-1\right)}} \Gamma \left(\frac{m (s+1)}{m+1}-r\right) \Gamma \left(r+\frac{m-m s}{m+1}\right)}{\Gamma \left(\frac{1}{m+1}\right) \Gamma \left(\frac{2 m}{m+1}\right) \Gamma \left(-r+\frac{m (s-1)}{m+1}+1\right) \Gamma \left(r-\frac{m (s+1)}{m+1}+1\right)}\,.
\end{align}

\subsection{\texorpdfstring{The four-point function with $\phirs{2}{2}$ and $\phirs{r}{s}$}{The four-point function with phi22 and phirs}}\label{sec:2222rsrsfourptsol}
Consider the following correlator
\begin{align}
	\la \phirs{2}{2}(z_1,\bar{z}_1)\phirs{2}{2}(z_2,\bar{z}_2)\phirs{r}{s}(z_3,\bar{z}_3)\phirs{r}{s}(z_4,\bar{z}_4)\ra\,.
\end{align}
To compute such a correlator, we first solve an ODE for the holomorphic correlator
\begin{align}
	\mathcal{D}_4^{(z_2)}\la \phirs{2}{2}(z_1)\phirs{2}{2}(z_2)\phirs{r}{s}(z_3)\phirs{r}{s}(z_4)\ra=0\,,
\end{align}
where $\mathcal{D}_4^{(z_2)}$ (acting on point $z_2$)  is the differential operator:
\begin{align}\label{fourthorderD}
	\mathcal{D}_4^{(\cdot)}&= {\mathcal L}_{-4}^{(\cdot)}+\frac{4 h_{2,2}-3}{6 h_{2,2}}{\mathcal L}_{-3}^{(\cdot)}{\mathcal L}_{-1}^{(\cdot)}-\frac{4}{9} (h_{2,2}+3){\mathcal L}_{-2}^{(\cdot)}{\mathcal L}_{-2}^{(\cdot)}+\frac{4 h_{2,2}+6}{6 h_{2,2}}{\mathcal L}_{-2}^{(\cdot)}{\mathcal L}_{-1}^{(\cdot)}{\mathcal L}_{-1}^{(\cdot)}-\frac{1}{4h_{2,2}}({\mathcal L}_{-1}^{(\cdot)})^4\,.
\end{align}

The holomorphic correlator takes the following form
\begin{align}
	\la \phirs{2}{2}(z_1)\phirs{2}{2}(z_2)\phirs{r}{s}(z_3)\phirs{r}{s}(z_4)\ra=\frac{\mathcal{F}_{(2,2)(2,2)(r,s)(r,s)}({z})}{(z_{12})^{2h_{2,2}}(z_{34})^{2h_{r,s}}}\,,
\end{align}
 (${z}$ is defined in \eqref{eta_def}) and $\mathcal{F}({z})\equiv\mathcal{F}_{(2,2)(2,2)(r,s)(r,s)}({z})$ satisfies:
\begin{equation}\label{diff2222rsrs}
	\begin{split}
		0&=A\,\mathcal{F}^{''''}({z} )+B\,\mathcal{F}'''({z} ) + C\,\mathcal{F}''({z} )+D\,\mathcal{F}'({z})+E\,\mathcal{F}({z}) \,,
	\end{split}
\end{equation}
with
\begin{equation}
	\begin{split}
		A&=-\frac{9}{2}  ({z} -1)^4 {z} ^3\,,\quad B=-\frac{9 ({z} -1)^3 {z} ^2 (-{z} +2 (2 {z} -1) m (m+1)+2)}{m (m+1)}\,,\\
		C&=\frac{9 ({z} -1)^2 {z}  }{2 m^2 (m+1)^2}\left[{z} ^2 (2 m (m+1) (2 h_{r,s} m (m+1)+h_{r,s}-7 m (m+1)+5)-1)\right.\\
		&\qquad\qquad\qquad\quad\left.-2 \left(m (m+1) \left(m^2+m-5\right)+2\right)+2 {z}  (m (m+1) (7 m (m+1)-11)+2)\right]\,,\\
		D&=\frac{9 ({z} -1)}{2 m^2 (m+1)^2} \left[2 {z} ^2 \left(2 h_{r,s} \left(m^2+m+1\right)+3 (m-1) (m+2) \left(m^2+m-1\right)\right)+\right.\\
		&\qquad\qquad\qquad\qquad\left.{z} ^3 \left(h_{r,s} \left(4 m^2 (m+1)^2-2\right)-4 (m+2) m^3+(3 m+7) m-3\right)\right.\\
		&\left.\qquad\qquad\qquad\quad-2 {z}  \left(m (m+1) \left(m^2+m-8\right)+8\right)-4 \left(m^2+m-2\right)\right]\,,\\
		E&=\frac{9 {z}  h_{r,s}}{2 m^2 (m+1)^2}\left({z} ^2 \left(-h_{r,s} (2 m+1)^2+m^2+m-2\right)-6 {z}  \left(m^2+m-1\right)+6 \left(m^2+m-1\right)\right)\,.
	\end{split}
\end{equation}

The four independent solutions to this differential equation are the Virasoro blocks, $$\mathcal{F}_\phi({z})\equiv\mathcal{F}^{(2,2)(2,2)(r,s)(r,s)}_\phi({z})\,,$$ where $\phi\in \{\id, \phirs{1}{3}, \phirs{3}{1}, \phirs{3}{3}\}$. To compute them, we go to the radial coordinate frame \eqref{eq:rhocoord}, in terms of which we can write
\begin{align}
	\mathcal{F}_\phi(\rho)=\sum_{n=0}^\infty a^{(\phi)}_n \rho^{h_{\phi}+n}\,,
\end{align}
and we fix the coefficients $a^{(\phi)}$ recursively using the differential equation \eqref{diff2222rsrs}. The first non-zero coefficients are found to be:
\begin{equation}
	\begin{split}
		a^{(1,1)}_2 &=\frac{24 h_{r,s}}{(m-2) (m+3)}\,, \quad
		a^{(1,1)}_4 = \frac{16 h_{r,s} \left(8 h_{r,s} m^2+8 h_{r,s} m+30 h_{r,s}+7 m^2+7 m\right)}{(m-2) (m+3) (3 m-2) (3 m+5)}\,,\\
		&\\
		a^{(1,3)}_1 &=\frac{2 (m-1)}{m+1}\,,\quad	a^{(1,3)}_2 =\frac{4 \left(-8 h_{r,s} m^3-14 h_{r,s} m^2-4 h_{r,s} m+2 h_{r,s}+3 m^4-9 m^3+11 m^2-m-4\right)}{(m+1)^2 (3 m-2) (3 m+1)}\,,\\
		&\\
		a^{(3,1)}_1 &=\frac{2 (m+2)}{m}\,,\quad 
		a^{(3,1)}_2=\frac{4 \left(8 h_{r,s} m^3+10 h_{r,s} m^2+3 m^4+21 m^3+56 m^2+62 m+20\right)}{m^2 (3 m+2) (3 m+5)}\,,\\
		&\\
		a^{(3,3)}_1 &=\frac{4}{m (m+1)}\,,\quad 
		a^{(3,3)}_2=-\frac{8 \left(h_{r,s} m^4+2 h_{r,s} m^3+h_{r,s} m^2-m^2-m+2\right)}{(m-1) m^2 (m+1)^2 (m+2)}\,.\\
	\end{split}
\end{equation}

Finally, we have
\begin{align}
	\la \phirs{2}{2}(z_1,\bar{z}_1)\phirs{2}{2}(z_2,\bar{z}_2)\phirs{r}{s}(z_3,\bar{z}_3)\phirs{r}{s}(z_4,\bar{z}_4)\ra= \frac{\mathcal{F}_{(2,2)(2,2)(r,s)(r,s)}(\rho,\bar{\rho})}{(|z_{12}|)^{2\D_{2,2}}(|z_{34}|)^{2\D_{r,s}}}\,,
\end{align}
where 
\begin{equation}\label{FinalRes22rsrs}
	\begin{split}
		\mathcal{F}_{(2,2)(2,2)(r,s)(r,s)}(\rho,\bar{\rho})&= \mathcal{F}^{(2,2)(2,2)(r,s)(r,s)}_{(1,1)}(\rho)\bar{\mathcal{F}}^{(2,2)(2,2)(r,s)(r,s)}_{(1,1)}(\bar{\rho})\\
		&+C_{(2,2)(2,2)(1,3)}C_{(r,s)(r,s)(1,3)} \mathcal{F}^{(2,2)(2,2)(r,s)(r,s)}_{(1,3)}(\rho)\bar{\mathcal{F}}^{(2,2)(2,2)(r,s)(r,s)}_{(1,3)}(\bar{\rho})\\
		&+C_{(2,2)(2,2)(3,1)}C_{(r,s)(r,s)(3,1)} \mathcal{F}^{(2,2)(2,2)(r,s)(r,s)}_{(3,1)}(\rho)\bar{\mathcal{F}}^{(2,2)(2,2)(r,s)(r,s)}_{(3,1)}(\bar{\rho})\\
		&+C_{(2,2)(2,2)(3,3)}C_{(r,s)(r,s)(3,3)} \mathcal{F}^{(2,2)(2,2)(r,s)(r,s)}_{(3,3)}(\rho)\bar{\mathcal{F}}^{(2,2)(2,2)(r,s)(r,s)}_{(3,3)}(\bar{\rho})\,,
	\end{split}
\end{equation}
with OPE coefficients \cite{Dotsenko:1984nm,Dotsenko:1984ad,Dotsenko:1985hi}
\begin{align}
		C_{(r,s)(r,s)(1,3)}&=\frac{\Gamma \left(\frac{m}{m+1}\right) \Gamma \left(\frac{2}{m+1}-1\right) \sqrt{\frac{(m-1) \Gamma \left(\frac{m}{m+1}\right) \Gamma \left(\frac{2 m}{m+1}\right) \Gamma \left(\frac{3}{m+1}-1\right)}{m \Gamma \left(\frac{m-1}{m+1}\right) \Gamma \left(2-\frac{3}{m+1}\right) \Gamma \left(\frac{1}{m+1}-1\right)}} \Gamma \left(\frac{m (s+1)}{m+1}-r\right) \Gamma \left(r+\frac{m-m s}{m+1}\right)}{\Gamma \left(\frac{1}{m+1}\right) \Gamma \left(\frac{2 m}{m+1}\right) \Gamma \left(-r+\frac{m (s-1)}{m+1}+1\right) \Gamma \left(r-\frac{m (s+1)}{m+1}+1\right)}\,,\nonumber \\
		C_{(r,s)(r,s)(3,1)}&=\frac{\Gamma \left(1+\frac{1}{m}\right) \Gamma \left(-\frac{m+2}{m}\right) \sqrt{-\frac{(m+2)^2 \Gamma \left(\frac{1}{m}\right) \Gamma \left(-\frac{m+3}{m}\right)}{m^3 \Gamma \left(2+\frac{3}{m}\right) \Gamma \left(-\frac{1}{m}\right)}} \Gamma \left(r+\frac{r+1}{m}-s+1\right) \Gamma \left(s-\frac{(m+1) (r-1)}{m}\right)}{\Gamma \left(2+\frac{2}{m}\right) \Gamma \left(-\frac{1}{m}\right) \Gamma \left(s-\frac{m r+r+1}{m}\right) \Gamma \left(\frac{m r+r-m s-1}{m}\right)}\,,\nonumber \\
		C_{(r,s)(r,s)(3,3)}&=\frac{m^2 (m+1)^2 \Gamma \left(1+\frac{1}{m}\right) \Gamma \left(\frac{m-2}{m}\right) \Gamma \left(\frac{m}{m+1}\right) \Gamma \left(\frac{m+3}{m+1}\right) \Gamma \left(r+\frac{r+1}{m}-s+1\right) \Gamma \left(s-\frac{(m+1) (r-1)}{m}\right)}{8 \pi ^{3/4} \Gamma \left(-\frac{1}{m}\right) \Gamma \left(\frac{2}{m}\right) \Gamma \left(-\frac{2}{m+1}\right) \Gamma \left(\frac{1}{m+1}\right) \Gamma \left(\frac{r+m (r-s+1)-1}{m}\right)} \nonumber \\
		&\times \frac{\Gamma \left(\frac{r m-s m+m+r}{m+1}\right) \Gamma \left(\frac{-r m+s m+m-r}{m+1}\right)}{\Gamma\left(-r+s-\frac{r+1}{m}+1\right) \Gamma \left(\frac{r m-s m+m+r+2}{m+1}\right) \Gamma \left(\frac{-r m+s m+m-r+2}{m+1}\right)}\\
		&\times \sqrt{-\frac{2^{\frac{m+3}{m+1}} \sin \left(\frac{2 \pi }{m+1}\right) \Gamma \left(\frac{1}{m}-1\right) \Gamma \left(-\frac{3}{m}\right) \Gamma \left(-\frac{1}{m+1}\right) \Gamma \left(\frac{2 m}{m+1}\right) \Gamma \left(\frac{m+4}{m+1}\right) \Gamma \left(\frac{1}{2}+\frac{1}{m+1}\right)}{\Gamma \left(\frac{3}{m}\right) \Gamma \left(\frac{m-2}{m+1}\right) \Gamma \left(-\frac{m+1}{m}\right)}}\,. \nonumber
\end{align}

Next, we consider:
\begin{align}
	\la \phirs{r}{s}(z_1,\bar{z}_1)\phirs{2}{2}(z_2,\bar{z}_2)\phirs{2}{2}(z_3,\bar{z}_3)\phirs{r}{s}(z_4,\bar{z}_4)\ra\,.
\end{align}
We start from the holomorphic correlator
\begin{align}
	\la \phirs{r}{s}(z_1)\phirs{2}{2}(z_2)\phirs{2}{2}(z_3)\phirs{r}{s}(z_4)\ra=\left(\frac{z_{14}}{z_{13}}\right)^{h_{2,2}-h_{r,s}}\left(\frac{z_{24}}{z_{14}}\right)^{h_{r,s}-h_{2,2}} \frac{\mathcal{F}_{(r,s)(2,2)(2,2)(r,s)}({z})}{(z_{12}z_{34})^{h_{r,s}+h_{2,2}}} \,,
\end{align}
(with cross-ration \eqref{eta_def}) which satisfies the following ODE
\begin{align}
	\mathcal{D}_4^{(\cdot)}\la\phirs{r}{s}(z_1,\bar{z}_1)\phirs{2}{2}(z_2,\bar{z}_2)\phirs{2}{2}(z_3,\bar{z}_3)\phirs{r}{s}(z_4,\bar{z}_4)\ra=0\,,
\end{align}
where $\mathcal{D}_4^{(z_2)}$ is the differential operator of \eqref{fourthorderD}. The differential equation for $$\mathcal{F}({z})\equiv\mathcal{F}_{(r,s)(2,2)(2,2)(r,s)}({z})\,,$$ reads
\begin{equation}\label{diffrs2222rs}
	\begin{split}
		0&=A\,\mathcal{F}^{''''}({z} )+B\,\mathcal{F}'''({z} ) + C\,\mathcal{F}''({z} )+D\,\mathcal{F}'({z})+E\,\mathcal{F}({z}) \,,
	\end{split}
\end{equation}
with
\begin{equation}
	\begin{split}
		A&=-9 ({z} -1)^4 {z} ^4\,,\quad B=-12 ({z} -1)^3 {z} ^3 [{z}  (h_{2,2}-3 h_{r,s}+6)+h_{2,2}+3 h_{r,s}-3]\,,\\
		C&=2 ({z} -1)^2 {z} ^2 \left[({z} +1)^2 h_{2,2}^2+3 h_{2,2} ({z}  (4-7 {z} )+2 ({z} -1) (5 {z} +1) h_{r,s}+5)\right.\\
		&\qquad\qquad\qquad\qquad\left.+9 \left(-7 {z}  ({z} -1)-3 ({z} -1)^2 h_{r,s}^2+(11 {z} -5) ({z} -1) h_{r,s}-1\right)\right]\,,\\
		D&=2 ({z} -1) {z}  \left[2 ({z} +1)^3 h_{2,2}^3+2 ({z} +1) h_{2,2}^2 (3 {z}  ({z} +1)+({z} -1) (3 {z} -5) h_{r,s}+2)\right.\\
		&\qquad\qquad\qquad+6 h_{2,2} \left(-2 ({z} -3) {z} ^2+{z} -(7 {z} -1) ({z} -1)^2 h_{r,s}^2+2 ({z}  ({z}  (5 {z} -9)+2)+2) h_{r,s}-1\right)\\
		&\left.\qquad\qquad\qquad+9 ({z} -1) \left({z} +2 {z} ^2 ((h_{r,s}-4) (h_{r,s}-1) h_{r,s}-1)\right.\right.\\
		&\left.\left.\qquad\qquad\qquad\qquad -2 {z}  h_{r,s} (h_{r,s} (2 h_{r,s}-7)+4)+2 h_{r,s} (h_{r,s}-1)^2\right)\right]\,,\\
		E&=-({z} +1)^4 h_{2,2}^4-2 ({z} +1)^2 h_{2,2}^3 \left({z}  ({z} +8)+\left(6 {z} ^2-8 {z} +2\right) h_{r,s}+1\right)\\
		&\quad	+h_{2,2}^2 \left[3 {z}  ({z}  (({z} -2) {z} +6)-2)-2 (3 {z}  (5 {z} -6)-1) ({z} -1)^2 h_{r,s}^2\right.\\
		&\qquad\left.-2 (3 {z} +1) ({z}  (5 {z} +8)-5) ({z} -1) h_{r,s}+3\right]\\
		&\qquad\qquad+6 ({z} -1) h_{2,2} h_{r,s} \left[{z} ^2 (3 {z} -11)+{z} +2 (3 {z} -1) ({z} -1)^2 h_{r,s}^2-({z}  (17 {z} -20)+1) ({z} -1) h_{r,s}-1\right] \\
		&\qquad\qquad-9 ({z} -1)^2 h_{r,s}^2 \left[{z} ^2 ((h_{r,s}-6) h_{r,s}+1)-2 {z}  (h_{r,s}-4) h_{r,s}+(h_{r,s}-1)^2\right]\,.
	\end{split}
\end{equation}

There are four Virasoro blocks, $\mathcal{F}_\phi\equiv\mathcal{F}^{(r,s)(2,2)(2,2)(r,s)}_\phi\,$, corresponding to $$ \phi\in \{ \phirs{r+1,s+1}, \phirs{r+1,s-1}, \phirs{r-1,s+1}, \phirs{r-1,s-1}\}\,.$$
 Going to radial coordinates \eqref{eq:rhocoord}, we write:
\begin{align}
	\mathcal{F}_\phi(\rho)=\sum_{n=0}^\infty a^{(\phi)}_n \rho^{h_{\phi}+n}\,,
\end{align}
and fix the coefficients $a^{(\phi)}$ recursively using \eqref{diffrs2222rs}. The first non-zero coefficients are found to be:
\begin{equation}
	\begin{split}
		a^{(r+1,s+1)}_1 &=\frac{2 (m r-m s+r+2)}{m (m+1) (m r-m s+r)}\,,
                \\
		&\\
		a^{(r+1,s-1)}_1 &\frac{2 \left(2 m^2 (r-s+1)+m (3 r-s+2)+r+2\right)^2}{m (m+1) (m (r-s+2)+r) (m (r-s+2)+r+2)}\,,
                \\
		&\\
		a^{(r-1,s+1)}_1 &=\frac{2 \left(2 m^2 (r-s-1)+m (3 r-s-2)+r-2\right)^2}{m (m+1) (m (r-s-2)+r-2) (m (r-s-2)+r)}\,,
                \\
		&\\
		a^{(r-1,s-1)}_1 &=\frac{2 (m r-m s+r-2)}{m (m+1) (m r-m s+r)}\,.
	\end{split}
\end{equation}

The final result reads
\begin{align}
	\la \phirs{r}{s}(z_1,\bar{z}_1)&\phirs{2}{2}(z_2,\bar{z}_2)\phirs{2}{2}(z_3,\bar{z}_3)\phirs{r}{s}(z_4,\bar{z}_4)\ra\nonumber\\
	&=\left(\frac{|z_{14}|}{|z_{13}|}\right)^{\D_{2,2}-\D_{r,s}}\left(\frac{|z_{24}|}{|z_{14}|}\right)^{\D_{r,s}-\D_{2,2}} \frac{\mathcal{F}_{(r,s)(2,2)(2,2)(r,s)}(\rho,\bar{\rho})}{(|z_{12}||z_{34}|)^{\D_{r,s}+\D_{2,2}}}\,.
\end{align}
where
\begin{equation}
	\begin{split}
		\mathcal{F}_{(r,s)(2,2)(2,2)(r,s)}(\rho,\bar{\rho})= &(C_{(r,s)(2,2)(r+1,s+1)})^2\mathcal{F}^{(r,s)(2,2)(2,2)(r,s)}_{(r+1,s+1)}(\rho)\bar{\mathcal{F}}^{(r,s)(2,2)(2,2)(r,s)}_{(r+1,s+1)}(\bar{\rho})\\
		&+(C_{(r,s)(2,2)(r+1,s-1)})^2\mathcal{F}^{(r,s)(2,2)(2,2)(r,s)}_{(r+1,s-1)}(\rho)\bar{\mathcal{F}}^{(r,s)(2,2)(2,2)(r,s)}_{(r+1,s-1)}(\bar{\rho})\\
		&+(C_{(r,s)(2,2)(r-1,s+1)})^2\mathcal{F}^{(r,s)(2,2)(2,2)(r,s)}_{(r-1,s+1)}(\rho)\bar{\mathcal{F}}^{(r,s)(2,2)(2,2)(r,s)}_{(r-1,s+1)}(\bar{\rho})\\
		&+(C_{(r,s)(2,2)(r-1,s-1)})^2\mathcal{F}^{(r,s)(2,2)(2,2)(r,s)}_{(r-1,s-1)}(\rho)\bar{\mathcal{F}}^{(r,s)(2,2)(2,2)(r,s)}_{(r-1,s-1)}(\bar{\rho})\,.
	\end{split}
\end{equation}
The OPE coefficients can be again computed via the Coulomb gas formalism discussed in section \ref{sec:large-m} and implemented in \cite{Esterlis:2016psv}.

%% file: sections/coulomb.tex

\section{Contour deformations with \texorpdfstring{$\texttt{MB}$}{MB}}
\label{app:coulombgas}

Section \ref{sec:large-m} explored the idea of writing perturbative data in long-range deformations of $\mathcal{M}_{m+1,m}$ as multi-dimensional Mellin-Barnes integrals. Computing the large-$m$ expansion of such integrals analytically was possible but only after a long sequence of contour deformations designed to avoid pinchings. The warm-up calculations in subsections \ref{sec:CGeasy} and \ref{sec:CGmedium} are some of the few that are practical enough to do by hand. For the $\phi_{2,2}\chi$ beta function in subsection \ref{sec:CGhard} (which is also hard to study using a multi-coupling RG flow), we made essential use of an automated solution. This is the \texttt{Mathematica} package \texttt{MB} \cite{c05} which we ran for about an hour. In this appendix, we will review the basics of this package and then give explicit code which performs the calculations in the main text.

\subsection{Overview}

The main property which makes Coulomb gas integrals different from the most familiar massless Feynman integrals is that they involve non-integer propagator powers. Any package which is able to analyze their singularities needs to be written with this in mind.\footnote{See \cite{cgm25} for a recent state-of-the-art package assuming integer powers.} This requirement is met by \texttt{MB} \cite{c05} which runs very quickly when there is one screening charge and slows down dramatically when this number is increased. Let us now delve into some of the details.

The first input for \texttt{MB} is a Mellin-Barnes integrand which distinguishes between two types of symbols --- integration variables and constant parameters. The second input is a set of substitution rules which make all gamma (and polygamma) arguments positive. The output is a set of Laurent expansion coefficients which are generated as we let the constant parameters approach some final values.

Starting with the inputs, the Mellin-Barnes integral is found by using the Coulomb gas and applying \eqref{chiral-mellin} and \eqref{symanzik}. With $N = N_+ + N_-$ non-chiral screening charges and $n$ other operators, these generate an integral of dimension
\be
D = \sum_{k = n}^{N + n - 1} \frac{k(k - 3)}{2} = N \frac{N^2 + 3Nn + 3n^2 - 6N - 12n + 5}{6}. \label{num-vars}
\ee
Coming to the substitution rules, there are a few possibilities.
\begin{enumerate}
\item Sometimes it is possible to make the gamma arguments positive by setting constant parameters to their final values right away. In this case, one simply has to Taylor expand and all non-trivial functionality of \texttt{MB} is skipped.
\item Sometimes it is impossible to make the gamma arguments positive. In this case, extra regulator parameters analogous to $\delta$ in \eqref{nonchiral-mellin2} need to be introduced.
\item In all other cases, one chooses suitable initial values for all constant (including regulator) parameters. This choice is usually inspired by the physics of the problem although \texttt{MB} provides a helper function which can be used as a last resort.
\end{enumerate}
The heart of \texttt{MB} records all pole crossings that take place as a constant parameter is moved between its initial and final values. By taking residues every time, it generates a list of new integrals with dimensions between $0$ and $D$ (recall that we saw dimensions $1$ and $2$ in \eqref{nonchiral-operator}). Each of these integrals in turn produces its own list of integrals once the next constant parameter is varied. When none are left, the last layer in the tree of integrals needs to be expanded. When the first few terms in the large-$m$ expansion have a simple analytic form, it is usually becasue non-trivial integrals in this last layer drop out at leading order.

\subsection{Example code}

We will now explain how to rederive \eqref{vir-expanded3} and \eqref{final-after} with \texttt{MB}. As expected, the code snippets below will constitute analytic proofs of these results. Terms that \texttt{MB} only knows how to evaluate numerically happen to be absent until higher orders in the large-$m$ expansion. 

Starting with the Virasoro block $\mathcal{F}_{(1,1)}^{(1,2)(1,2)(1,2)(1,2)}$, the large-$m$ expansion is handled with the following four lines.
\begin{align}
& \texttt{int = (z/(1 - z))\^{}g*Gamma[-g]*Gamma[g+am2]*Gamma[1-am2+g]/Gamma[2-2*am2+g];} \nonumber \\
& \texttt{rules = MBoptimizedRules[int, am2->1, \{\}, \{am2,z\}];} \nonumber \\
& \texttt{cont = MBcontinue[int, am2->1, rules];} \nonumber \\
& \texttt{list = MBmerge[MBexpand[cont, 1, \{am2,1,1\}]];} \nonumber
\end{align}
Clearly, the first line defines the Mellin-Barnes integral from \eqref{vir-blocks-mellin}. The second line then uses the code \texttt{MBoptimizedRules[]} to find values of $\gamma$ and $\alpha_-^2$ which make all gamma arguments positive.\footnote{Note that this function succeeds whenever it returns a non-empty set of rules. Error messages which are sometimes displayed before this happens, saying for example that no rules could be found, should be ignored \cite{c05}.} The non-trivial step of performing the analytic continuation in $\alpha_-^2$ is done by \texttt{MBcontinue[]} in the third line. Finally, the fourth line computes a list of terms which may be expanded in $1/m$ to yield the second equality in \eqref{vir-expanded1}. 

The exact same procedure works for $\mathcal{F}_{(1,3)}^{(1,2)(1,2)(1,2)(1,2)}$ if we edit the first line to read
\begin{align}
& \texttt{int = (z/(1 - z))\^{}g*Gamma[-g]*Gamma[g+am2]*Gamma[1-am2+g]/Gamma[2+g];} \nonumber
\end{align}
instead.

Moving onto the multi-variable example $I_{1,2}$, we should run
\begin{align}
& \texttt{int = Gamma[-x/2]*Gamma[-y/2]*Gamma[2*am2-1-x/2]*Gamma[2*am2-1-y/2];} \nonumber \\
& \texttt{int *= Gamma[1-am2+(x+y)/2]*Gamma[2-3*am2+(x+y)/2];} \nonumber \\
& \texttt{int *= Gamma[2-d-(3*am2-x)/2]*Gamma[2-d-(3*am2-y)/2]*Gamma[3*am2-3+2*d-(x+y)/2];} \nonumber \\
& \texttt{int /= Gamma[d-1+(3*am2-x)/2]*Gamma[d-1+(3*am2-y)/2]*Gamma[2*d+4-3*am2+(x+y)/2];} \nonumber \\
& \texttt{rules = \{\{am2->89/192, d->105/128\}, \{x->-1/2, y->-3/8\}\};} \nonumber \\
& \texttt{rules = MBcorrectContours[rules, 10000];} \nonumber \\
& \texttt{cont1 = MBcontinue[int, d->0, rules];} \nonumber \\
& \texttt{list1 = MBmerge[MBexpand[cont1, 1, \{d,0,0\}]];} \nonumber \\
& \texttt{cont2 = MBcontinue[\#[[1]], am2->1, \#[[2]]] \&/@ list1;} \nonumber \\
& \texttt{list2 = MBmerge[MBexpand[cont2, 1, \{am2,1,-2\}]];} \nonumber
\end{align}
which takes a few seconds. The first difference is that we have supplied a set of rules manually which is faster than calling \texttt{MBoptimizedRules[]}. The second difference is that the continuation now has two steps. The result of taking $\delta \to 0$ (even after taking the constant term and dropping $O(\delta)$) is a list of integrals. This entire layer must be fed into \texttt{MBcontinue[]} so that $\alpha_-^2 \to 1$ can be applied to each entry. Fortunately, the final result, once expanded in $1/m$ and integrated, yields the closed form result \eqref{result-subleading1}.

The most time consuming contour deformation in this paper (and the main reason to use \texttt{MB}) is the one for $I_{2,2}$, namely \eqref{final-input1}. To arrive at a set of substitution rules which make the contour straight, demand that all exponents are between $0$ and $1$ whenever \eqref{symanzik} is used. When using it on $t_1$, this produces values of $\alpha_+^2$ and $\alpha_+ \alpha_-$ which make the $\Re \delta_{ij} > 0$ linear programming problem easy to solve. Next for $t_2$, these partial rules plus a suitable value of $\alpha_-^2$ allows $\Re \gamma_{ij} > 0$ to be solved quickly as well. It is then not difficult to tune $\gamma$ until all arguments are positive. Using this method, a set of rules which does the job is found to be
\ba
& \alpha_+^2 = \frac{41}{72}, \quad \alpha_-^2 = \frac{361}{576}, \quad \alpha_+ \alpha_- = -\frac{11}{48}, \quad \gamma = -\frac{989}{576}, \quad \Re \gamma_{12} = \frac{1}{8}, \quad \Re \gamma_{13} = \frac{11}{96} \\
& \Re \delta_{12} = \frac{1}{24}, \quad \Re \delta_{13} = \frac{1}{12}, \quad \Re \delta_{14} = \frac{1}{6}, \quad \Re \delta_{23} = \frac{1}{24}, \quad \Re \delta_{24} = \frac{1}{8}. \label{final-input2}
\ea
Let us assume that \texttt{int} has been set to \eqref{final-input1} and \texttt{rules} has been set to \eqref{final-input2}. Before finishing, we must contend with the fact that the large $m$ limit sends $\alpha_+^2 \to 1$ and $\alpha_-^2 \to 1$ at the same time. This is a problem which only appears when both types of screening charges are present.\footnote{A naive way to account for this would be to take $\alpha_+^2 \to \alpha_-^{-2}$ followed by $\alpha_-^2 \to 1$ but this would give the wrong answer. Gamma functions containing the sum of a parameter and its inverse lead to very expensive analytic continuations which \texttt{MB} does not implement. As stated in \cite{c05}, parameters known to \texttt{MB} must enter linearly in the arguments of gamma functions.} If we define $\epsilon \equiv m^{-1}$, a viable strategy is to cut off the infinite sum in
\be
\alpha_+^2 = 1 + \epsilon, \quad \alpha_-^2 = \sum_{k = 0}^\infty (-\epsilon)^k
\ee
and then treat different functions of $\epsilon$ as separate parameters which go to zero. The number of terms to keep is set by the number of powers of $m$ which are desired in the final result. To extract the leading and first subleading singularities, it is enough to use
\be
\alpha_+^2 = 1 + \epsilon_1, \quad \alpha_-^2 = 1 - \epsilon_1 + \epsilon_2 \label{two-epsilons}
\ee
where we later set $\epsilon_1 = \epsilon$ and $\epsilon_2 = \epsilon^2 - \epsilon^3$. The code for expanding in these two parameters is now straightforward. Using \texttt{g, apam, e1, e2} to denote $(\gamma, \alpha_+ \alpha_-, \epsilon_1, \epsilon_2)$ respectively, we should run
\begin{align}
& \texttt{rules = MBcorrectContours[rules, 10000];} \nonumber \\
& \texttt{cont1 = MBcontinue[int, g->0, rules];} \nonumber \\
& \texttt{list1 = MBmerge[MBexpand[MBpreselect[cont1, \{g,0,0\}], 1, \{g,0,0\}]];} \nonumber \\
& \texttt{cont2 = MBcontinue[\#[[1]], apam->-1, \#[[2]]] \&/@ list1;} \nonumber \\
& \texttt{list2 = MBmerge[MBexpand[MBpreselect[cont2, \{apam,-1,0\}], 1, \{apam,-1,0\}]];} \nonumber \\
& \texttt{cont3 = MBcontinue[\#[[1]], e2->0, \#[[2]]] \&/@ list2;} \nonumber \\
& \texttt{list3 = MBmerge[MBexpand[MBpreselect[cont3, \{e2,0,0\}], 1, \{e2,0,0\}]];} \nonumber \\
& \texttt{cont4 = MBcontinue[\#[[1]], e1->0, \#[[2]]] \&/@ list3;} \nonumber \\
& \texttt{list4 = MBmerge[MBexpand[MBpreselect[cont3, \{e1,0,-1\}], 1, \{e1,0,-1\}]];} \nonumber
\end{align}
which finishes in about an hour.\footnote{To prevent it from taking much longer, we have modified \texttt{MB} so that the \texttt{MBmerge[]} command no longer calls \texttt{Simplify[]}.} A few optimizations are present but the important feature of this code is that only the $\gamma \to 0$ and $\alpha_+ \alpha_- \to -1$ continuations allow us to pick out the constant term. For $\epsilon_1$ and $\epsilon_2$, we must consider a genuine double expansion containing inverse powers of both. The output \texttt{list4} ultimately has all of its $O(m^4)$ dependence come from $O(\epsilon_1^{-2} \epsilon_2^{-2})$, specifically
\be
-\frac{1}{6} \epsilon_1^{-2} \epsilon_2^{-1} = -\frac{1}{6} \epsilon^{-4} (1 + \epsilon)^{-1} = -\frac{1}{6} (m^4 + m^3) + O(m^2). \label{m34}
\ee
The $-\frac{1}{6}$ is actually expressed in terms of a few unevaluated single integrals but they can all be computed with the first Barnes lemma. The remaining pieces at $O(m^3)$ now come from different types of terms as follows.\footnote{By keeping enough powers of $\epsilon_1$ and $\epsilon_2$ to learn about $O(m^3)$, \texttt{MB} also produces spurious singularities like $O((\alpha_+ \alpha_- + 1)^{-1})$ and it must be checked that they ultimately cancel.}
\be
O(\epsilon_1^{-1} \epsilon_2^{-1}): 0, \quad O(\epsilon_1^{-3}): \frac{1}{6} m^3 \label{just-m3}
\ee
The $0$ comes from taking a result with unevaluated double integrals and using the first Barnes lemma iteratively. Arranging \eqref{m34} and \eqref{just-m3} into a final result, \eqref{result-subleading3} follows.

%% file: LRmm.bbl
\providecommand{\href}[2]{#2}\begingroup\raggedright\begin{thebibliography}{100}

\bibitem{Belavin:1984vu}
A.A.~Belavin, A.M.~Polyakov and A.B.~Zamolodchikov, \emph{{Infinite Conformal
  Symmetry in Two-Dimensional Quantum Field Theory}},
  \href{https://doi.org/10.1016/0550-3213(84)90052-X}{\emph{Nucl. Phys. B}
  {\bfseries 241} (1984) 333}.

\bibitem{Polchinski:1987}
J.~Polchinski, \emph{{Scale and Conformal Invariance in Quantum Field Theory}},
  \href{https://doi.org/10.1016/0550-3213(88)90179-4}{\emph{Nucl. Phys. B}
  {\bfseries 303} (1988) 226}.

\bibitem{Ginsparg:1988ui}
P.H.~Ginsparg, \emph{Applied conformal field theory},  in \emph{{Les Houches
  Summer School in Theoretical Physics: Fields, Strings, Critical Phenomena}},
  9, 1988 [\href{https://arxiv.org/abs/hep-th/9108028}{{\ttfamily
  hep-th/9108028}}].

\bibitem{Mussardo:2010mgq}
G.~Mussardo, \emph{{Statistical field theory}: {an introduction to exactly
  solved models in statistical physics}}, Oxford Univ. Press, New York, NY
  (2010).

\bibitem{DiFrancesco:1997nk}
P.~Di~Francesco, P.~Mathieu and D.~Senechal, \emph{{Conformal Field Theory}},
  Graduate Texts in Contemporary Physics, Springer-Verlag, New York (1997),
  \href{https://doi.org/10.1007/978-1-4612-2256-9}{10.1007/978-1-4612-2256-9}.

\bibitem{d69a}
F.J.~Dyson, \emph{{Existence of a phase transition in a one-dimensional Ising
  ferromagnet}}, \href{https://doi.org/10.1007/BF01645907}{\emph{Commun. Math.
  Phys.} {\bfseries !2} (1969) 91}.

\bibitem{d69b}
F.J.~Dyson, \emph{{Non-existence of spontaneous magnetization in a
  one-dimensional Ising ferromagnet}},
  \href{https://doi.org/10.1007/BF01661575}{\emph{Commun. Math. Phys.}
  {\bfseries 12} (1969) 212}.

\bibitem{t69}
D.J.~Thouless, \emph{{Long-range order in one-dimensional Ising systems}},
  \href{https://doi.org/10.1103/PhysRev.187.732}{\emph{Phys. Rev.} {\bfseries
  187} (1969) 732}.

\bibitem{Fisher:1972zz}
M.E.~Fisher, S.-k.~Ma and B.~Nickel, \emph{Critical exponents for long-range
  interactions},
  \href{https://doi.org/10.1103/PhysRevLett.29.917}{\emph{Phys.Rev.Lett.}
  {\bfseries 29} (1972) 917}.

\bibitem{Sak:1973}
J.~Sak, \emph{Recursion relations and fixed points for ferromagnets with
  long-range interactions},
  \href{https://doi.org/10.1103/PhysRevB.8.281}{\emph{Phys. Rev. B} {\bfseries
  8} (1973) 281}.

\bibitem{Paulos:2015jfa}
M.F.~Paulos, S.~Rychkov, B.C.~van Rees and B.~Zan, \emph{{Conformal Invariance
  in the Long-Range Ising Model}},
  \href{https://doi.org/10.1016/j.nuclphysb.2015.10.018}{\emph{Nucl.\ Phys.\ B}
  {\bfseries 902} (2016) 246}
  [\href{https://arxiv.org/abs/1509.00008}{{\ttfamily 1509.00008}}].

\bibitem{Sak:1977}
J.~Sak, \emph{Low-temperature renormalization group for ferromagnets with
  long-range interactions},
  \href{https://doi.org/10.1103/PhysRevB.15.4344}{\emph{Phys. Rev. B}
  {\bfseries 15} (1977) 4344}.

\bibitem{Behan:2017dwr}
C.~Behan, L.~Rastelli, S.~Rychkov and B.~Zan, \emph{{Long-range critical
  exponents near the short-range crossover}},
  \href{https://doi.org/10.1103/PhysRevLett.118.241601}{\emph{Phys.\ Rev.\
  Lett.} {\bfseries 118} (2017) 241601}
  [\href{https://arxiv.org/abs/1703.03430}{{\ttfamily 1703.03430}}].

\bibitem{Behan:2017emf}
C.~Behan, L.~Rastelli, S.~Rychkov and B.~Zan, \emph{{A scaling theory for the
  long-range to short-range crossover and an infrared duality}},
  \href{https://doi.org/10.1088/1751-8121/aa8099}{\emph{J. Phys.} {\bfseries
  A50} (2017) 354002} [\href{https://arxiv.org/abs/1703.05325}{{\ttfamily
  1703.05325}}].

\bibitem{Benedetti:2024wgx}
D.~Benedetti, E.~Lauria, D.~Maz\'a\v{c} and P.~van Vliet, \emph{{1d Ising model
  with $1/r^{1.99}$ interaction}},
  \href{https://doi.org/10.1103/PhysRevLett.134.201602}{\emph{Phys. Rev. Lett.}
  {\bfseries 134} (2025) 201602}
  [\href{https://arxiv.org/abs/2412.12243}{{\ttfamily 2412.12243}}].

\bibitem{Benedetti:2025nzp}
D.~Benedetti, E.~Lauria, D.~Mazac and P.~van Vliet, \emph{{A strong-weak
  duality for the 1d long-range Ising model}},
  \href{https://arxiv.org/abs/2509.05250}{{\ttfamily 2509.05250}}.

\bibitem{Zamolodchikov:1987ti}
A.B.~Zamolodchikov, \emph{{Renormalization Group and Perturbation Theory Near
  Fixed Points in Two-Dimensional Field Theory}}, {\emph{Sov. J. Nucl. Phys.}
  {\bfseries 46} (1987) 1090}.

\bibitem{Komargodski:2016auf}
Z.~Komargodski and D.~Simmons-Duffin, \emph{{The Random-Bond Ising Model in
  2.01 and 3 Dimensions}},
  \href{https://doi.org/10.1088/1751-8121/aa6087}{\emph{J. Phys.} {\bfseries
  A50} (2017) 154001} [\href{https://arxiv.org/abs/1603.04444}{{\ttfamily
  1603.04444}}].

\bibitem{Zamolodchikov:1986db}
A.B.~Zamolodchikov, \emph{{Conformal Symmetry and Multicritical Points in
  Two-Dimensional Quantum Field Theory. (In Russian)}}, {\emph{Sov. J. Nucl.
  Phys. 44 (1986) 529-533, Yad. Fiz. 44 (1986) 821-827} (Also in: Conformal
  invariance and applications to statistical mechanics, Ed. by C. Itzykson et
  al., 258-264 (1998). World Scientific) }.

\bibitem{Behan:2023ile}
C.~Behan, E.~Lauria, M.~Nocchi and P.~van Vliet, \emph{{Analytic and numerical
  bootstrap for the long-range Ising model}},
  \href{https://doi.org/10.1007/JHEP03(2024)136}{\emph{JHEP} {\bfseries 03}
  (2024) 136} [\href{https://arxiv.org/abs/2311.02742}{{\ttfamily
  2311.02742}}].

\bibitem{Behan:2021tcn}
C.~Behan, L.~Di~Pietro, E.~Lauria and B.C.~van Rees, \emph{{Bootstrapping
  boundary-localized interactions II. Minimal models at the boundary}},
  \href{https://doi.org/10.1007/JHEP03(2022)146}{\emph{JHEP} {\bfseries 03}
  (2022) 146} [\href{https://arxiv.org/abs/2111.04747}{{\ttfamily
  2111.04747}}].

\bibitem{Cardy:1989da}
J.L.~Cardy, \emph{Conformal invariance and statistical mechanics},  in
  \emph{{Les Houches Summer School in Theoretical Physics: Fields, Strings,
  Critical Phenomena}}, 1, 1989.

\bibitem{Recknagel:2000ri}
A.~Recknagel, D.~Roggenkamp and V.~Schomerus, \emph{{On relevant boundary
  perturbations of unitary minimal models}},
  \href{https://doi.org/10.1016/S0550-3213(00)00519-8}{\emph{Nucl. Phys. B}
  {\bfseries 588} (2000) 552}
  [\href{https://arxiv.org/abs/hep-th/0003110}{{\ttfamily hep-th/0003110}}].

\bibitem{Graham:2001pp}
K.~Graham, \emph{{On perturbations of unitary minimal models by boundary
  condition changing operators}},
  \href{https://doi.org/10.1088/1126-6708/2002/03/028}{\emph{JHEP} {\bfseries
  03} (2002) 028} [\href{https://arxiv.org/abs/hep-th/0111205}{{\ttfamily
  hep-th/0111205}}].

\bibitem{dns01}
V.S.~Dotsenko, X.S.~Nguyen and R.~Santachiara, \emph{{Models $WD_{n}$ in the
  presence of disorder and the coupled models}},
  \href{https://doi.org/10.1016/S0550-3213(01)00392-3}{\emph{Nuclear Physics B}
  {\bfseries 613} (2001) 445}
  [\href{https://arxiv.org/abs/hep-th/0104197}{{\ttfamily hep-th/0104197}}].

\bibitem{Fredenhagen:2009tn}
S.~Fredenhagen, M.R.~Gaberdiel and C.~Schmidt-Colinet, \emph{{Bulk flows in
  Virasoro minimal models with boundaries}},
  \href{https://doi.org/10.1088/1751-8113/42/49/495403}{\emph{J. Phys. A}
  {\bfseries 42} (2009) 495403}
  [\href{https://arxiv.org/abs/0907.2560}{{\ttfamily 0907.2560}}].

\bibitem{1303.3015}
R.~Poghossian, \emph{{Two Dimensional Renormalization Group Flows in Next to
  Leading Order}}, \href{https://doi.org/10.1007/JHEP01(2014)167}{\emph{JHEP}
  {\bfseries 01} (2014) 167} [\href{https://arxiv.org/abs/1303.3015}{{\ttfamily
  1303.3015}}].

\bibitem{2205.05091}
H.~Poghosyan and R.~Poghossian, \emph{{RG flow between $W_3$ minimal models by
  perturbation and domain wa}},
  \href{https://doi.org/10.1007/JHEP08(2022)307}{\emph{JHEP} {\bfseries 08}
  (2022) 307} [\href{https://arxiv.org/abs/2205.05091}{{\ttfamily
  2205.05091}}].

\bibitem{2211.16503}
A.~Antunes and C.~Behan, \emph{{Coupled minimal models revisited}},
  \href{https://doi.org/10.1103/PhysRevLett.130.071602}{\emph{Phys. Rev. Lett.}
  {\bfseries 130} (2023) 071602}
  [\href{https://arxiv.org/abs/2211.16503}{{\ttfamily 2211.16503}}].

\bibitem{Lauria:2023uca}
E.~Lauria, M.N.~Milam and B.C.~van Rees, \emph{{Perturbative RG flows in AdS.
  An {\'e}tude}}, \href{https://doi.org/10.1007/JHEP03(2024)005}{\emph{JHEP}
  {\bfseries 03} (2024) 005}
  [\href{https://arxiv.org/abs/2309.10031}{{\ttfamily 2309.10031}}].

\bibitem{2412.21107}
A.~Antunes and C.~Behan, \emph{{Coupled minimal models revisited II:
  Constraints from permutation symmetry}},
  \href{https://doi.org/10.21468/SciPostPhys.18.4.132}{\emph{SciPost Phys.}
  {\bfseries 18} (2025) 132}
  [\href{https://arxiv.org/abs/2412.21107}{{\ttfamily 2412.21107}}].

\bibitem{Suzuki:1973a}
M.~Suzuki, \emph{Critical exponents for long-range interactions. i:
  Dimensionality, symmetry and potential-range},
  \href{https://doi.org/10.1143/PTP.49.424}{\emph{Progress of Theoretical
  Physics} {\bfseries 49} (1973) 424}
  [\href{https://arxiv.org/abs/https://academic.oup.com/ptp/article-pdf/49/2/424/5253807/49-2-424.pdf}{{\ttfamily
  https://academic.oup.com/ptp/article-pdf/49/2/424/5253807/49-2-424.pdf}}].

\bibitem{Suzuki:1973b}
M.~Suzuki, \emph{Critical exponents for long-range interactions. ii:
  Universality and scaling relations},
  \href{https://doi.org/10.1143/PTP.49.1106}{\emph{Progress of Theoretical
  Physics} {\bfseries 49} (1973) 1106}
  [\href{https://arxiv.org/abs/https://academic.oup.com/ptp/article-pdf/49/4/1106/5443731/49-4-1106.pdf}{{\ttfamily
  https://academic.oup.com/ptp/article-pdf/49/4/1106/5443731/49-4-1106.pdf}}].

\bibitem{Brezin:2014rkt}
E.~Brezin, G.~Parisi and F.~Ricci-Tersenghi, \emph{{The Crossover Region
  Between Long-Range and Short-Range Interactions for the Critical Exponents}},
  \href{https://doi.org/10.1007/s10955-014-1081-0}{\emph{J. Statist. Phys.}
  {\bfseries 157} (2014) 855}
  [\href{https://arxiv.org/abs/1407.3358}{{\ttfamily 1407.3358}}].

\bibitem{Benedetti:2020rrq}
D.~Benedetti, R.~Gurau, S.~Harribey and K.~Suzuki, \emph{{Long-range
  multi-scalar models at three loops}},
  \href{https://doi.org/10.1088/1751-8121/abb6ae}{\emph{J. Phys. A} {\bfseries
  53} (2020) 445008} [\href{https://arxiv.org/abs/2007.04603}{{\ttfamily
  2007.04603}}].

\bibitem{Benedetti:2024mqx}
D.~Benedetti, R.~Gurau and S.~Harribey, \emph{{Addendum: Long-range
  multi-scalar models at three loops}},
  \href{https://doi.org/10.1088/1751-8121/adbfe4}{\emph{J. Phys. A} {\bfseries
  58} (2025) 129401} [\href{https://arxiv.org/abs/2411.00805}{{\ttfamily
  2411.00805}}].

\bibitem{Chai:2021arp}
N.~Chai, M.~Goykhman and R.~Sinha, \emph{{Long-range vector models at large
  N}}, \href{https://doi.org/10.1007/JHEP09(2021)194}{\emph{JHEP} {\bfseries
  09} (2021) 194} [\href{https://arxiv.org/abs/2107.08052}{{\ttfamily
  2107.08052}}].

\bibitem{Tarnopolsky:2016vvd}
G.~Tarnopolsky, \emph{{Large $N$ expansion of the sphere free energy}},
  \href{https://doi.org/10.1103/PhysRevD.96.025017}{\emph{Phys. Rev. D}
  {\bfseries 96} (2017) 025017}
  [\href{https://arxiv.org/abs/1609.09113}{{\ttfamily 1609.09113}}].

\bibitem{Fraser-Taliente:2025udk}
L.~Fraser-Taliente, \emph{{The sphere free energy of the vector models to order
  $1/N$}},  \href{https://arxiv.org/abs/2507.16896}{{\ttfamily 2507.16896}}.

\bibitem{rt15}
S.~Rychkov and Z.M.~Tan, \emph{{The epsilon expansion from conformal field
  theory}}, \href{https://doi.org/10.1088/1751-8113/48/29/29FT01}{\emph{J.
  Phys. A: Math. Theor.} {\bfseries 48} (2015) 29FT01}
  [\href{https://arxiv.org/abs/1505.00963}{{\ttfamily 1505.00963}}].

\bibitem{1601.01310}
S.~Giombi and V.~Kirilin, \emph{{Anomalous dimensions in CFT with weakly broken
  higher spin symmetry}},
  \href{https://doi.org/10.1007/JHEP11(2016)068}{\emph{JHEP} {\bfseries 11}
  (2016) 068} [\href{https://arxiv.org/abs/1601.01310}{{\ttfamily
  1601.01310}}].

\bibitem{Dotsenko:1984nm}
V.S.~Dotsenko and V.A.~Fateev, \emph{{Conformal Algebra and Multipoint
  Correlation Functions in Two-Dimensional Statistical Models}},
  \href{https://doi.org/10.1016/0550-3213(84)90269-4}{\emph{Nucl. Phys. B}
  {\bfseries 240} (1984) 312}.

\bibitem{Dotsenko:1984ad}
V.S.~Dotsenko and V.A.~Fateev, \emph{{Four Point Correlation Functions and the
  Operator Algebra in the Two-Dimensional Conformal Invariant Theories with the
  Central Charge c \ensuremath{<} 1}},
  \href{https://doi.org/10.1016/S0550-3213(85)80004-3}{\emph{Nucl. Phys. B}
  {\bfseries 251} (1985) 691}.

\bibitem{Dotsenko:1985hi}
V.S.~Dotsenko and V.A.~Fateev, \emph{{Operator Algebra of Two-Dimensional
  Conformal Theories with Central Charge C \ensuremath{<}= 1}},
  \href{https://doi.org/10.1016/0370-2693(85)90366-1}{\emph{Phys. Lett. B}
  {\bfseries 154} (1985) 291}.

\bibitem{c05}
M.~Czakon, \emph{{Automatized analytic continuation of Mellin-Barnes
  integrals}}, \href{https://doi.org/10.1016/j.cpc.2006.07.002}{\emph{Comput.
  Phys. Common.} {\bfseries 175} (2006) 559}
  [\href{https://arxiv.org/abs/hep-ph/0511200}{{\ttfamily hep-ph/0511200}}].

\bibitem{Lohmann:2017}
M.~Lohmann, G.~Slade and B.C.~Wallace, \emph{Critical two-point function for
  long-range o(n) models below the upper critical dimension},
  \href{https://doi.org/10.1007/s10955-017-1904-x}{\emph{Journal of Statistical
  Physics} {\bfseries 169} (2017) 1132}
  [\href{https://arxiv.org/abs/1705.08540}{{\ttfamily 1705.08540}}].

\bibitem{Amit:1982az}
D.J.~Amit and L.~Peliti, \emph{On dangerous irrelevant operators},
  \href{https://doi.org/10.1016/0003-4916(82)90159-2}{\emph{Annals Phys.}
  {\bfseries 140} (1982) 207}.

\bibitem{Lencses:2022ira}
M.~Lencs\'es, A.~Miscioscia, G.~Mussardo and G.~Tak\'acs,
  \emph{{Multicriticality in Yang-Lee edge singularity}},
  \href{https://doi.org/10.1007/JHEP02(2023)046}{\emph{JHEP} {\bfseries 02}
  (2023) 046} [\href{https://arxiv.org/abs/2211.01123}{{\ttfamily
  2211.01123}}].

\bibitem{Hogervorst:2013sma}
M.~Hogervorst and S.~Rychkov, \emph{{Radial Coordinates for Conformal Blocks}},
  \href{https://doi.org/10.1103/PhysRevD.87.106004}{\emph{Phys. Rev.}
  {\bfseries D87} (2013) 106004}
  [\href{https://arxiv.org/abs/1303.1111}{{\ttfamily 1303.1111}}].

\bibitem{Behan:2018hfx}
C.~Behan, \emph{{Bootstrapping the long-range Ising model in three
  dimensions}}, \href{https://doi.org/10.1088/1751-8121/aafd1b}{\emph{J. Phys.}
  {\bfseries A52} (2019) 075401}
  [\href{https://arxiv.org/abs/1810.07199}{{\ttfamily 1810.07199}}].

\bibitem{m09}
G.~Mack, \emph{{D-independent representation of conformal field theories in D
  dimensions via transformation to auxiliary dual resonance models. Scalar
  amplitudes}},  \href{https://arxiv.org/abs/0907.2407}{{\ttfamily 0907.2407}}.

\bibitem{p10}
J.~Penedones, \emph{{Writing CFT correlation functions as AdS scattering
  amplitudes}}, \href{https://doi.org/10.1007/JHEP03(2011)025}{\emph{JHEP}
  {\bfseries 03} (2011) 025} [\href{https://arxiv.org/abs/1011.1485}{{\ttfamily
  1011.1485}}].

\bibitem{fkprv11}
A.L.~Fitzpatrick, J.~Kaplan, J.~Penedones, S.~Raju and B.C.~van Rees, \emph{{A
  natural language for AdS/CFT correlators}},
  \href{https://doi.org/10.1007/JHEP11(2011)095}{\emph{JHEP} {\bfseries 11}
  (2011) 095} [\href{https://arxiv.org/abs/1107.1499}{{\ttfamily 1107.1499}}].

\bibitem{psz19}
J.~Penedones, J.A.~Silva and A.~Zhiboedov, \emph{{Nonperturbative Mellin
  amplitudes: Existence, properties and applications}},
  \href{https://doi.org/10.1007/JHEP08(2020)031}{\emph{JHEP} {\bfseries 08}
  (2020) 031} [\href{https://arxiv.org/abs/1912.11100}{{\ttfamily
  1912.11100}}].

\bibitem{Esterlis:2016psv}
I.~Esterlis, A.L.~Fitzpatrick and D.~Ramirez, \emph{{Closure of the Operator
  Product Expansion in the Non-Unitary Bootstrap}},
  \href{https://doi.org/10.1007/JHEP11(2016)030}{\emph{JHEP} {\bfseries 11}
  (2016) 030} [\href{https://arxiv.org/abs/1606.07458}{{\ttfamily
  1606.07458}}].

\bibitem{km20}
D.~Kapec and R.~Mahajan, \emph{{Comments on the quantum field theory of the
  Coulomb gas}}, \href{https://doi.org/10.1007/JHEP04(2021)136}{\emph{JHEP}
  {\bfseries 04} (2021) 136}
  [\href{https://arxiv.org/abs/2010.10428}{{\ttfamily 2010.10428}}].

\bibitem{fgp90}
P.~Furlan, A.C.~Ganchev and V.B.~Petkova, \emph{{Fusion matrices and $c < 1$
  (quasi) local conformal theories}},
  \href{https://doi.org/10.1142/S0217751X90001252}{\emph{Int. J. Mod. Phys.}
  {\bfseries A5} (1990) 2721}.

\bibitem{p09}
V.B.~Petkova, \emph{{Lecture notes on conformal field theory}},  in
  \emph{https://smallperturbation.com/sites/default/files/Petkova2009.pdf}
  (2009).

\bibitem{nrj23}
R.~Nivesvivat, S.~Ribault and J.L.~Jacobsen, \emph{{Critical loop models are
  exactly solvable}},
  \href{https://doi.org/10.21468/SciPostPhys.17.2.029}{\emph{SciPost Phys.}
  {\bfseries 17} (2024) 029}
  [\href{https://arxiv.org/abs/2311.17558}{{\ttfamily 2311.17558}}].

\bibitem{nr25}
R.~Nivesvivat and S.~Ribault, \emph{{Fusion rules and structure constants of
  E-series minimal models}},
  \href{https://doi.org/10.21468/SciPostPhys.18.5.163}{\emph{SciPost Phys.}
  {\bfseries 18} (2025) 163}
  [\href{https://arxiv.org/abs/2502.14295}{{\ttfamily 2502.14295}}].

\bibitem{dsz87}
P.~Di~Francesco, H.~Saleur and J.B.~Zuber, \emph{{Relations between the Coulomb
  gas picture and conformal invariance of two-dimensional critical models}},
  \href{https://doi.org/10.1007/BF01009954}{\emph{J. Stat. Phys.} {\bfseries
  48} (1987) 57}.

\bibitem{ggqzz24}
B.~Gabai, V.~Gorbenko, J.~Qiao, B.~Zan and A.~Zhabin, \emph{{Quantum groups as
  global symmetries II: Coulomb gas construction}},
  \href{https://arxiv.org/abs/2410.24143}{{\ttfamily 2410.24143}}.

\bibitem{0906.3219}
L.F.~Alday, D.~Gaiotto and Y.~Tachikawa, \emph{{Liouville Correlation Functions
  from Four-dimensional Gauge Theories}},
  \href{https://doi.org/10.1007/s11005-010-0369-5}{\emph{Lett. Math. Phys.}
  {\bfseries 91} (2010) 167} [\href{https://arxiv.org/abs/0906.3219}{{\ttfamily
  0906.3219}}].

\bibitem{s72}
K.~Symanzik, \emph{{On calculations in conformal invariant field theories}},
  \href{https://doi.org/10.1007/BF02824349}{\emph{Lett. Nuovo Cim.} {\bfseries
  3} (1972) 734}.

\bibitem{y18}
E.Y.~Yuan, \emph{{Simplicity in AdS perturbative dynamics}},
  \href{https://arxiv.org/abs/1801.07283}{{\ttfamily 1801.07283}}.

\bibitem{az15}
L.F.~Alday and A.~Zhiboedov, \emph{{Conformal bootstrap with slightly broken
  higher spin symmetry}},
  \href{https://doi.org/10.1007/JHEP06(2016)091}{\emph{JHEP} {\bfseries 06}
  (2016) 091} [\href{https://arxiv.org/abs/1506.04659}{{\ttfamily
  1506.04659}}].

\bibitem{Antunes:2024hrt}
A.~Antunes, E.~Lauria and B.C.~van Rees, \emph{{A bootstrap study of minimal
  model deformations}},  \href{https://arxiv.org/abs/2401.06818}{{\ttfamily
  2401.06818}}.

\bibitem{Lauria:2020emq}
E.~Lauria, P.~Liendo, B.C.~Van~Rees and X.~Zhao, \emph{{Line and surface
  defects for the free scalar field}},
  \href{https://doi.org/10.1007/JHEP01(2021)060}{\emph{JHEP} {\bfseries 01}
  (2021) 060} [\href{https://arxiv.org/abs/2005.02413}{{\ttfamily
  2005.02413}}].

\bibitem{Behan:2020nsf}
C.~Behan, L.~Di~Pietro, E.~Lauria and B.C.~Van~Rees, \emph{{Bootstrapping
  boundary-localized interactions}},
  \href{https://doi.org/10.1007/JHEP12(2020)182}{\emph{JHEP} {\bfseries 12}
  (2020) 182} [\href{https://arxiv.org/abs/2009.03336}{{\ttfamily
  2009.03336}}].

\bibitem{Defenu:2014bea}
N.~Defenu, A.~Trombettoni and A.~Codello, \emph{{Fixed-point structure and
  effective fractional dimensionality for O$(N)$ models with long-range
  interactions}}, \href{https://doi.org/10.1103/PhysRevE.92.052113}{\emph{Phys.
  Rev. E} {\bfseries 92} (2015) 052113}
  [\href{https://arxiv.org/abs/1409.8322}{{\ttfamily 1409.8322}}].

\bibitem{Defenu:2020umv}
N.~Defenu, A.~Codello, S.~Ruffo and A.~Trombettoni, \emph{{Criticality of spin
  systems with weak long-range interactions}},
  \href{https://doi.org/10.1088/1751-8121/ab6a6c}{\emph{J. Phys. A} {\bfseries
  53} (2020) 143001} [\href{https://arxiv.org/abs/1908.05158}{{\ttfamily
  1908.05158}}].

\bibitem{Solfanelli:2024obb}
A.~Solfanelli and N.~Defenu, \emph{{Universality in long-range interacting
  systems: The effective dimension approach}},
  \href{https://doi.org/10.1103/PhysRevE.110.044121}{\emph{Phys. Rev. E}
  {\bfseries 110} (2024) 044121}
  [\href{https://arxiv.org/abs/2406.14651}{{\ttfamily 2406.14651}}].

\bibitem{2412.08697}
L.~Bianchi, L.S.~Cardinale and E.~de~Sabbata, \emph{{Defects in the long-range
  O(N) model}}, \href{https://doi.org/10.1088/1751-8121/adf788}{\emph{J. Phys.}
  {\bfseries A58} (2025) 335401}
  [\href{https://arxiv.org/abs/2412.08697}{{\ttfamily 2412.08697}}].

\bibitem{2504.06203}
F.K.~Popov and Y.~Wang, \emph{{Factorizing Defects from Generalized Pinning
  Fields}},  \href{https://arxiv.org/abs/2504.06203}{{\ttfamily 2504.06203}}.

\bibitem{2505.15018}
D.~Ge and Y.~Nakayama, \emph{{Non-Factorizing Interface in the Two-Dimensional
  Long-Range Ising Model}},  \href{https://arxiv.org/abs/2505.15018}{{\ttfamily
  2505.15018}}.

\bibitem{Gorbenko:2018ncu}
V.~Gorbenko, S.~Rychkov and B.~Zan, \emph{{Walking, Weak first-order
  transitions, and Complex CFTs}},
  \href{https://doi.org/10.1007/JHEP10(2018)108}{\emph{JHEP} {\bfseries 10}
  (2018) 108} [\href{https://arxiv.org/abs/1807.11512}{{\ttfamily
  1807.11512}}].

\bibitem{Gorbenko:2018dtm}
V.~Gorbenko, S.~Rychkov and B.~Zan, \emph{{Walking, Weak first-order
  transitions, and Complex CFTs II. Two-dimensional Potts model at $Q>4$}},
  \href{https://doi.org/10.21468/SciPostPhys.5.5.050}{\emph{SciPost Phys.}
  {\bfseries 5} (2018) 050} [\href{https://arxiv.org/abs/1808.04380}{{\ttfamily
  1808.04380}}].

\bibitem{Cardy_1981}
J.L.~Cardy, \emph{One-dimensional models with $1/r^2$ interactions},
  \href{https://doi.org/10.1088/0305-4470/14/6/017}{\emph{J. Phys. A}
  {\bfseries 14} (1981) 1407}.

\bibitem{Cannas:1995ja}
S.A.~Cannas and A.C.N.~de~Magalhaes, \emph{{One-dimensional Potts model with
  long range interactions: A Renormalization group approach}},
  \href{https://doi.org/10.1088/0305-4470/30/10/014}{\emph{J. Phys. A}
  {\bfseries 30} (1997) 3345}.

\bibitem{Bayong:1999}
E.~Bayong, H.T.~Diep and V.~Dotsenko, \emph{Potts model with long-range
  interactions in one dimension},
  \href{https://doi.org/10.1103/PhysRevLett.83.14}{\emph{Phys. Rev. Lett.}
  {\bfseries 83} (1999) 14}.

\bibitem{Reynal_2004}
S.~Reynal and H.T.~Diep, \emph{Reexamination of the long-range potts model: A
  multicanonical approach},
  \href{https://doi.org/10.1103/physreve.69.026109}{\emph{Physical Review E}
  {\bfseries 69} (2004) }
  [\href{https://arxiv.org/abs/cond-mat/0306493}{{\ttfamily
  cond-mat/0306493}}].

\bibitem{Kaplan:2009kr}
D.B.~Kaplan, J.-W.~Lee, D.T.~Son and M.A.~Stephanov, \emph{{Conformality
  Lost}}, \href{https://doi.org/10.1103/PhysRevD.80.125005}{\emph{Phys. Rev. D}
  {\bfseries 80} (2009) 125005}
  [\href{https://arxiv.org/abs/0905.4752}{{\ttfamily 0905.4752}}].

\bibitem{Dijkgraaf:1987vp}
R.~Dijkgraaf, E.P.~Verlinde and H.L.~Verlinde, \emph{{C = 1 Conformal Field
  Theories on Riemann Surfaces}},
  \href{https://doi.org/10.1007/BF01224132}{\emph{Commun. Math. Phys.}
  {\bfseries 115} (1988) 649}.

\bibitem{Amoruso}
N.~Amoruso, \emph{Renormalization group flows between non-unitary conformal
  models},  Master's thesis, Universit\`a di Bologna,
  http://amslaurea.unibo.it/11308/, 2015.

\bibitem{Zambelli:2016cbw}
L.~Zambelli and O.~Zanusso, \emph{{Lee-Yang model from the functional
  renormalization group}},
  \href{https://doi.org/10.1103/PhysRevD.95.085001}{\emph{Phys. Rev. D}
  {\bfseries 95} (2017) 085001}
  [\href{https://arxiv.org/abs/1612.08739}{{\ttfamily 1612.08739}}].

\bibitem{Lencses:2024wib}
M.~Lencs\'es, A.~Miscioscia, G.~Mussardo and G.~Tak\'acs,
  \emph{{Ginzburg-Landau description for multicritical Yang-Lee models}},
  \href{https://doi.org/10.1007/JHEP08(2024)224}{\emph{JHEP} {\bfseries 08}
  (2024) 224} [\href{https://arxiv.org/abs/2404.06100}{{\ttfamily
  2404.06100}}].

\bibitem{Klebanov:2022syt}
I.R.~Klebanov, V.~Narovlansky, Z.~Sun and G.~Tarnopolsky,
  \emph{{Ginzburg-Landau description and emergent supersymmetry of the (3, 8)
  minimal model}}, \href{https://doi.org/10.1007/JHEP02(2023)066}{\emph{JHEP}
  {\bfseries 02} (2023) 066}
  [\href{https://arxiv.org/abs/2211.07029}{{\ttfamily 2211.07029}}].

\bibitem{Katsevich:2024jgq}
A.~Katsevich, I.R.~Klebanov and Z.~Sun, \emph{{Ginzburg-Landau description of a
  class of non-unitary minimal models}},
  \href{https://doi.org/10.1007/JHEP03(2025)170}{\emph{JHEP} {\bfseries 03}
  (2025) 170} [\href{https://arxiv.org/abs/2410.11714}{{\ttfamily
  2410.11714}}].

\bibitem{Weinrib:1983zz}
A.~Weinrib and B.I.~Halperin, \emph{{Critical phenomena in systems with
  long-range-correlated quenched disorder}},
  \href{https://doi.org/10.1103/PhysRevB.27.413}{\emph{Phys. Rev. B} {\bfseries
  27} (1983) 413}.

\bibitem{Honkonen:1988fq}
J.~Honkonen and M.Y.~Nalimov, \emph{Crossover between field theories with short
  range and long range exchange or correlations},
  \href{https://doi.org/10.1088/0305-4470/22/6/024}{\emph{J. Phys. A}
  {\bfseries 22} (1989) 751}.

\bibitem{Prudnikov:2000}
V.V.~Prudnikov, P.V.~Prudnikov and A.A.~Fedorenko, \emph{Field-theory approach
  to critical behavior of systems with long-range correlated defects},
  \href{https://doi.org/10.1103/PhysRevB.62.8777}{\emph{Phys. Rev. B}
  {\bfseries 62} (2000) 8777}.

\bibitem{Chippari:2023vnx}
F.~Chippari, M.~Picco and R.~Santachiara, \emph{{Two-dimensional Ising and
  Potts model with long-range bond disorder: A renormalization group
  approach}},
  \href{https://doi.org/10.21468/SciPostPhys.15.4.135}{\emph{SciPost Phys.}
  {\bfseries 15} (2023) 135}
  [\href{https://arxiv.org/abs/2306.01887}{{\ttfamily 2306.01887}}].

\bibitem{Lecce:2024hnd}
I.~Lecce, M.~Picco and R.~Santachiara, \emph{{Magnetic exponent for the
  long-range bond-disordered Potts model}},
  \href{https://doi.org/10.1103/PhysRevE.110.064154}{\emph{Phys. Rev. E}
  {\bfseries 110} (2024) 064154}
  [\href{https://arxiv.org/abs/2407.13456}{{\ttfamily 2407.13456}}].

\bibitem{Dotsenko:1994sy}
V.~Dotsenko, M.~Picco and P.~Pujol, \emph{{Renormalization group calculation of
  correlation functions for the 2-d random bond Ising and Potts models}},
  \href{https://doi.org/10.1016/0550-3213(95)00534-Y}{\emph{Nucl. Phys. B}
  {\bfseries 455} (1995) 701}
  [\href{https://arxiv.org/abs/hep-th/9501017}{{\ttfamily hep-th/9501017}}].

\bibitem{Cardy:1996xt}
J.~Cardy, \emph{{Scaling and renormalization in statistical physics}},
  Cambridge Lecture Notes in Physics, Cambridge University Press (1996).

\bibitem{Gaberdiel:2008fn}
M.R.~Gaberdiel, A.~Konechny and C.~Schmidt-Colinet, \emph{{Conformal
  perturbation theory beyond the leading order}},
  \href{https://doi.org/10.1088/1751-8113/42/10/105402}{\emph{J. Phys. A}
  {\bfseries 42} (2009) 105402}
  [\href{https://arxiv.org/abs/0811.3149}{{\ttfamily 0811.3149}}].

\bibitem{Montvay}
I.~Montvay and G.~Munster, \emph{Quantum fields on a lattice}, Cambridge
  Unversity Press, Cambridge (1997).

\bibitem{Benedetti:2020yvb}
D.~Benedetti, R.~Gurau and K.~Suzuki, \emph{{Conformal symmetry and composite
  operators in the $O(N)^{3}$ tensor field theory}},
  \href{https://doi.org/10.1007/JHEP06(2020)113}{\emph{JHEP} {\bfseries 06}
  (2020) 113} [\href{https://arxiv.org/abs/2002.07652}{{\ttfamily
  2002.07652}}].

\bibitem{Kleinert:2001hn}
H.~Kleinert and V.~Schulte-Frohlinde, \emph{{Critical Poperties of $\phi^4$
  Theories}}, World Scientific (2001),
  \href{https://doi.org/10.1142/4733}{10.1142/4733}.

\bibitem{Cardy:1986gw}
J.L.~Cardy, \emph{{Effect of Boundary Conditions on the Operator Content of
  Two-Dimensional Conformally Invariant Theories}},
  \href{https://doi.org/10.1016/0550-3213(86)90596-1}{\emph{Nucl. Phys. B}
  {\bfseries 275} (1986) 200}.

\bibitem{Cappelli:1986hf}
A.~Cappelli, C.~Itzykson and J.B.~Zuber, \emph{{Modular Invariant Partition
  Functions in Two-Dimensions}},
  \href{https://doi.org/10.1016/0550-3213(87)90155-6}{\emph{Nucl. Phys. B}
  {\bfseries 280} (1987) 445}.

\bibitem{Ruelle:1998zu}
P.~Ruelle and O.~Verhoeven, \emph{{Discrete symmetries of unitary minimal
  conformal theories}},
  \href{https://doi.org/10.1016/S0550-3213(98)00639-7}{\emph{Nucl. Phys. B}
  {\bfseries 535} (1998) 650}
  [\href{https://arxiv.org/abs/hep-th/9803129}{{\ttfamily hep-th/9803129}}].

\bibitem{cgm25}
M.~Correia, M.~Giroux and S.~Mizera, \emph{{SOFIA: Singularities of Feynman
  Integrals Automatized}},  \href{https://arxiv.org/abs/2503.16601}{{\ttfamily
  2503.16601}}.

\end{thebibliography}\endgroup
